\documentclass[12pt, preprint]{elsarticle}

\journal{Joule}

\biboptions{numbers,sort&compress,super}

\usepackage{libertine}
\usepackage{libertinust1math}
\usepackage[a4paper, total={7in, 10in}]{geometry}
\usepackage{afterpage}
\usepackage{changepage}
\usepackage{amsmath}
\usepackage{bbold}
\usepackage{graphicx}
\usepackage{eurosym}
\usepackage{mathtools}
\usepackage{url}
\usepackage{booktabs}
\usepackage{epstopdf}
\usepackage{xfrac}
\usepackage{tabularx}
\usepackage{bm}
\usepackage{subcaption}
\usepackage{blindtext}
\usepackage{longtable}
\usepackage{multirow}
\usepackage{threeparttable}
\usepackage{pdflscape}
\usepackage[export]{adjustbox}
\usepackage[version=4]{mhchem}
\usepackage[parfill]{parskip}
\usepackage[colorlinks=true, linkcolor=blue, urlcolor=blue, citecolor=blue]{hyperref}
\usepackage[nameinlink,sort&compress,capitalise,noabbrev]{cleveref}
\usepackage[prependcaption,textsize=footnotesize]{todonotes}
\usepackage{placeins}
\usepackage{siunitx}
\usepackage{lineno}
\modulolinenumbers[5]

\DeclareSIUnit{\tco}{t_{\ce{CO2}}}
\DeclareSIUnit{\sieuro}{\mbox{\euro}}
\DeclareSIUnit{\twh}{{\tera\watt\hour}}
\DeclareSIUnit{\mwh}{{\mega\watt\hour}}
\DeclareSIUnit{\kwh}{{\kilo\watt\hour}}
\DeclareSIUnit\year{a}

\newcommand{\bneuro}[1]{#1\,bn\euro{}$_\text{2020}$/a}

\renewcommand{\ttdefault}{\sfdefault}

\newcommand{\customlabel}[2]{%
  \begingroup%
  \def\next{\label{#1}}%
  \hypersetup{pdfinfo={Title={#2}}}%
  \phantomsection%
  \gdef\@currentlabelname{#2}%
  \next%
  \endgroup%
}

\usepackage{lipsum}

\usepackage[resetlabels,labeled]{multibib}
\newcites{S}{Supplementary References}
\bibliographystyleS{elsarticle-num}

\newcommand{\co}{\ce{CO2}~}

\begin{document}

\begin{frontmatter}

    \title{A Minimal Methanol Backstop for High Electrification Scenarios}
	\author[TUB]{Philipp Glaum}
	\author[TUB]{Fabian Neumann}
	\author[RISE]{Markus Millinger}
	\author[TUB]{Tom Brown\corref{correspondingauthor}}
	\ead{t.brown@tu-berlin.de}
        \cortext[correspondingauthor]{Lead contact and corresponding author.}

	\address[TUB]{Department of Digital Transformation in Energy Systems, Technische Universität Berlin, Berlin, Germany}
	\address[RISE]{Built Environment: System Transition: Energy Systems Analysis, RISE Research Institutes of Sweden, Göteborg, Sweden}

    \abstracttitle{Summary}
	\begin{abstract}
          Electrification of sectors such as land transport and building heating is a cost-effective pathway to deep decarbonization. However, some sectors still require energy-dense fuels — including aviation, shipping and backup power — or chemical feedstocks. While a `hydrogen economy' is often proposed to fill these hard-to-electrify gaps, it faces challenges in transport, storage, and infrastructure coordination.
        We introduce a `minimal~methanol~backstop' to supply residual demand in highly-electrified systems. As a liquid fuel, methanol is easy to store and transport, and avoids infrastructure lock-in. Produced from hydrogen and carbon monoxide, it can help integrate biogenic carbon from decentralized biomass wastes and residues. Using a European energy system model constrained to be carbon-neutral, we show that methanol-based systems increase total system costs by 2.4\% relative to hydrogen-based systems, an increase that remains below 6\% across sensitivities.
        We argue that this modest cost premium is justified by reduced infrastructure complexity.
	\end{abstract}




\end{frontmatter}


\section*{Highlights}
\begin{itemize}
\item Green methanol can decarbonize energy sectors where electrification is unsuitable
\item As a liquid it is easier to handle than hydrogen, and it helps integrate biomass
\item A methanol-based net-zero system is 2-6\% more expensive than a hydrogen-based one
\item For this modest expense, difficulties scaling up hydrogen infrastructure are avoided
\end{itemize}

\section*{Graphical Abstract}

\begin{center}
\includegraphics[width=0.8\textwidth]{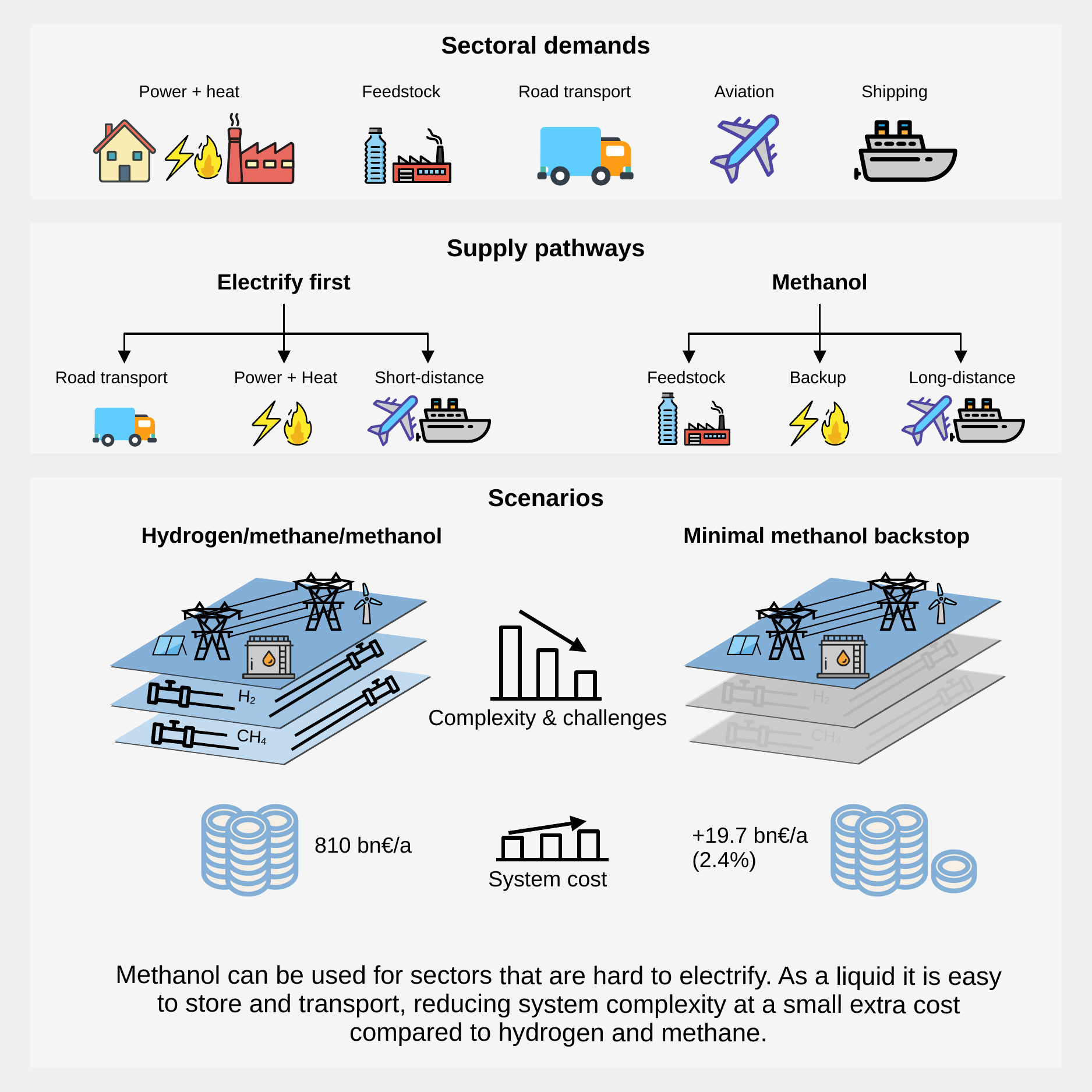}
\end{center}

\section*{Introduction}
\label{sec:intro}

Achieving deep decarbonization will require profound changes to how energy is consumed and supplied. There are varying levels of consensus in the literature regarding the best approach for each sector. Electrifying demand and simultaneously decarbonizing electricity has been identified by the International Panel on Climate Change as a key strategy~\cite{IPCC_2022_WGIII_TS}, and it forms the cornerstone of the International Energy Agency's scenario for net-zero emissions~\cite{IEA24}. The European Commission stated in its communication for its 2040 climate target that `[e]lectrification with a fully decarbonised power system by 2040 is the main driver of the energy transition'\cite{com2040}. While electrification can be highly effective, it is accompanied by challenges regarding the cost of end-user devices, battery weight for long-distance transport, as well as grid integration. It is widely accepted that electrification is a cost-effective strategy in land transport~\cite{Link2024,Bogdanov2024,Zeyen2025,IEA-EV-25}, short-distance shipping and aviation~\cite{Schaefer2019,Bogdanov2024}, building heating~\cite{brownSynergiesSectorCoupling2018}, and low- to medium-temperature process heat~\cite{Madeddu_2020}. There is some consensus that green electrolytic hydrogen is the most effective emissions abatement strategy for ammonia production and iron ore reduction for steelmaking~\cite{bockris1972,VOGL2018736,vanderzwaanElectricityHydrogendrivenEnergy2025}. For the remaining sectors, such as long-distance shipping and aviation, high value chemicals (HVC), high-temperature heat as well as backup power for dark wind lulls, several ideas are currently being discussed.

One prominent idea is the `hydrogen economy', whereby hydrogen is produced using electrolysis where renewables are abundant, then transported to demand centres where it is either used directly for power or heat, used chemically, or turned into other molecules like liquid fuels~\cite{bockris1972,haldane1923}. Hydrogen can be transported using pipelines and then stored in underground solution-mined salt caverns. While hydrogen emits no greenhouse gases when oxidized, several potential problems accompany the hydrogen economy~\cite{johnsonRealisticRolesHydrogen2025}: hydrogen is a small molecule, which tends to escape through valves and embrittle steel~\cite{martinReviewChallengesUsing2024}; it needs to be compressed for transport and storage; it has a low volumetric density, which increases storage needs and severely complicates its use for long-distance shipping and aviation; cheap underground storage requires suitable salt deposits for solution mining, which are not available everywhere, and would require a substantial build-out in Europe~\cite{frischmuth2024}; as a gas, it is costly to transport by vehicle and is moved most economically in GW-scale pipelines, which require lumpy investments and close coordination of storage, transport, supply and demand during the build-up phase, while risking stranded assets; its leakage contributes indirectly to global warming~\cite{Sand2023}.

\begin{figure}[!ht]
    \centering
    \includegraphics[width=0.8\textwidth]{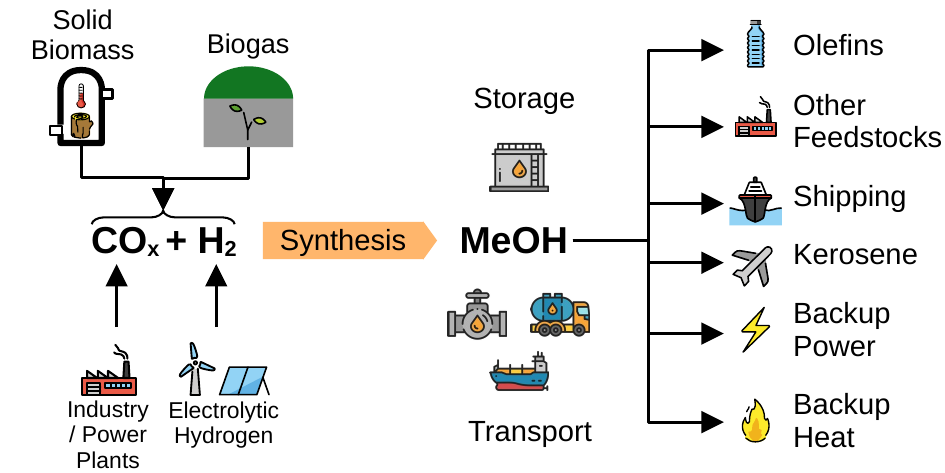}
    \caption{Methanol production and consumption pathways. Own figure created using icons designed by OpenMoji~\cite{HfggmuendOpenmoji2025}.}
    \label{fig:methanol_pathways}
\end{figure}

Our contention is that for most of the use cases of hydrogen, it is easier to use the hydrogen locally and transport the hydrogen derivatives (like ammonia, methanol or hydrocarbons) instead. In particular methanol, which is liquid at room temperature and pressure, can be used to serve most of these needs, while being much simpler to store and move around. As shown in~\cref{fig:methanol_pathways}, many sectors can use methanol, either directly in the case of shipping or via conversion processes like methanol-to-aromatics/olefins (MtA/O) for HVC or methanol-to-kerosene (MtK) for aviation. These conversion processes are already competitive with alternative routes based on Fischer–Tropsch synthesis because of the higher product selectivity of methanol synthesis~\cite{bubeKeroseneProductionPowerbased2024}, and are technologically mature, with technology readiness levels (TRLs) of 7 and above~\cite{ETPCleanEnergy}. Usage of hydrogen or methane for backup power and backup heat in district heating could also be replaced by methanol, with the option to capture the carbon dioxide emissions using innovative generation technologies like the oxyfuel Allam cycle~\cite{brownUltralongdurationEnergyStorage2023} or fuel cells~\cite{Bertau2014}. Since the other uses of methanol in shipping, aviation and industry are large, the use for backup would be small in comparison.

While there is a cost and energetic penalty to producing methanol compared to hydrogen, the benefits are manifold: it can be stored in steel tanks at extremely low cost regardless of the local geology (in contrast, cost-effective hydrogen storage relies on highly-localized salt deposits); it can reuse existing fossil liquid storage and pipeline facilities; it is less explosive; it is already traded globally; it can power gas turbines with minimal retrofitting~\cite{cuiComparativeStudyMethanol2025} (conversion costs are estimated at \$3 million per turbine~\cite{Bertau2014}, less than 1\% of the investment cost of a 500~MW turbine); it can be transported at a low cost by truck, rail, barge, ship or pipeline; it is biodegradable; it scales down to small MW-scale production or consumption use cases because of the easy handling~\cite{Bertau2014}.

The use of methanol from gasified coal to replace oil imports was already suggested in the 1980s by Friedrich Asinger~\cite{Asinger1986}, but it was Nobel Prize laureate George Olah in the early 2000s who first called for a `methanol economy' that would use renewable `green' methanol to replace fossil fuels across a range of applications~\cite{olah2003,olah2005,Bertau2014}. We propose a new twist to these concepts: methanol should be used rather as a last resort for the gaps left in a highly-electrified economy, where the use of molecules is kept to a minimum: a `minimal methanol backstop'. Methanol is ideally suited to the role of a gap-filler since it can be transported easily at small scales by vehicle, without the need for large discrete pipelines that are lumpy in the economic sense. If usage grows, transport can switch to pipelines. If, however, use cases are out-competed in the course of electrification, stranded assets in the form of pipelines or underground storage are avoided.
Finally, since methanol infrastructure can adapt flexibly to demand, it is a hedge against other uncertainties affecting demand, such as the exact amounts of carbon capture and biomass that can be expected in the future~\cite{Blanco2018}.

Since carbon oxides are needed for its synthesis and then released on combustion, a `minimal methanol backstop' requires careful carbon accounting. Using carbon oxides from biogas, biomass gasification or other bio-processes, or \co captured upon combustion in the power and heat sectors, at waste-to-energy plants, or onboard ships~\cite{JOHNSEN20241000}, ensures that the carbon in methanol is either biogenic or recycled. This allows the carbon cycle to be closed for scenarios with net-zero emissions; any carbon entering the system from fossil sources has to be compensated by carbon dioxide removal (CDR) involving the sequestration of biogenic carbon. Small amounts of \co captured directly from the air (DAC) can also top up the supply, although DAC is currently expensive and the scale-up is uncertain~\cite{brazolla2025}.

As methanol is suited to small-scale applications, it can also help absorb widely-distributed sustainable carbon from decentral biogenic waste and residue sources at small (MW) scales, with the option to bind excess carbon in them by supplementing with electrolytic hydrogen. Methanol can then easily be transported from small sites to central facilities for further use or distribution.  This contrasts with biogas, which when upgraded to biomethane either needs to be liquefied for transport~\cite{gustafssonCentralizedDecentralizedHow2024} or requires a gas grid connection, which may not be cost-effective if the biogas plant is far from the pipeline network~\cite{matschossConsolidatedPotentialAnalysis2020}.  Where methanol is synthesized directly from biomass without additional hydrogen we refer to it as `biomethanol'; where hydrogen is added to top up the carbon-rich syngas mixture before methanol synthesis we refer to it as `e-biomethanol'; where captured carbon dioxide is combined with electrolytic hydrogen we call it `e-methanol' (even if the \co is of biogenic origin). Biomethanol is likely to be lower cost to start with, requiring minimal government support~\cite{korberg2021}. With increasing fossil fuel phase-out, the more expensive use of excess CO/\co with electrolytic hydrogen in e-biomethanol and e-methanol may become competitive.

Today both the uses of methanol and green production capacity are rising fast. Global fossil-based methanol production reached around 110 million tonnes per year (110~Mt/a) in 2023 and demand is forecast to grow at 3.5\% per year~\cite{mi2024,methanex2024}. Since its commercialization of methanol-to-olefin technology in 2010, China has been using methanol synthesised from coal for a large share of its olefin production~\cite{liang2023}. Given its promise as a potentially clean fuel, container shipping companies like Maersk and COSCO have since 2024 been retrofitting ships from fuel oil to methanol (which requires adjustments to engines, fuel delivery and tanks) and commissioning dual-fuel ships that can switch between fuel oil and methanol, so that as of August 2025 there were 60 methanol-capable ships on the water and more than 300 on order~\cite{gmf2025}. In 2025 renewable methanol production will rise to 1.7 Mt/a, of which the majority is biomethanol plants~\cite{migena2025}. Prominent e-methanol plants that opened in 2025 include the Kassø facility in Denmark, which produces 42 kt/a using green hydrogen and \co from a local biogas plant, and a 50~kt/a facility in Taonan, China, using green hydrogen and \co from biomass gasification. The pipeline of planned projects for biomethanol and e-methanol amounts to 18.5 Mt/a and 23.4 Mt/a respectively by 2030~\cite{migena2025}.

In this paper, we build on previous work for the power sector~\cite{brownUltralongdurationEnergyStorage2023} to explore the implications of using methanol instead of hydrogen or methane as the backstop energy carrier in a highly-electrified carbon-neutral energy scenario for Europe. While costs are expected to increase slightly, we argue that the savings in other areas, often poorly accounted for by energy system models, may outweigh the costs: savings from having the cost-effective option of vehicular transport rather than hard-to-regulate lumpy monopolistic infrastructure like pipelines, lower dangers of stranded assets, simplicity of transport and distribution, and avoiding the greenhouse gas effects of methane and hydrogen leakage.

We model 100 regions in Europe using the open-source model PyPSA-Eur to analyze future European energy systems, covering all energy demand and non-energy feedstocks, constrained to have net-zero carbon dioxide emissions. The default \textit{settings} that are then varied in the sensitivity analysis include: electrification is implemented wherever possible, including all road transport, domestic aviation, domestic shipping, and industry process heat where biomass is not used today and where fuels are not required for their chemical properties or radiant heat (the challenges of deep electrification are discussed below); carbon dioxide sequestration is constrained to 200 Mt\co per year, which is similar to the European Commission's Carbon Management Strategy's target of 250~Mt\co per year in 2050~\cite{comcms} as well as other studies looking at the feasibility of Carbon Capture and Storage~(CCS) scale-up~\cite{Schreyer2025,Gidden2025}; the biomass resource is limited to the medium potential of domestic wastes and residues from the JRC ENSPRESO database~\cite{ruizENSPRESOOpenEU282019} (347~TWh/a wet biomass for biogas, 1050~TWh/a solid biomass for gasification, 151~TWh/a biogenic municipal solid waste); imports of fuels from outside Europe are not allowed; no relocation of energy-intensive industries inside Europe is allowed; electricity transmission can be expanded by up to 25\% in MWkm volume, in line with the transmission system operators' Ten Year Network Development Plan~\cite{TYNDP2024}; a pipeline network to transport liquefied \co can be built.

For our analysis, we define five main \textit{scenarios} to contrast the roles of three possible gap-fillers for power and heat when prices are high: hydrogen, methane and methanol. In the scenarios, we sequentially deactivate the transport options for the gaseous carriers hydrogen and methane.
All scenarios allow the transport of oil, methanol, biomass, carbon dioxide and electricity. In each scenario the following gaseous pipeline networks are activated:
\begin{itemize}
    \item \textbf{All Networks (all)}: both hydrogen and methane transmission
    \item \textbf{Only Methane Network (CH$_4$)}: only methane transmission
    \item \textbf{Only Hydrogen Network (H$_2$)}: only hydrogen transmission
    \item \textbf{No Gaseous Fuel Networks (none)}: neither hydrogen nor methane transmission, but local distribution of hydrogen and methane inside the regions is still allowed
    \item \textbf{Minimal Methanol Backstop (min. MeOH)}: neither hydrogen nor methane transmission, and local distribution of hydrogen and methane inside the regions is also forbidden. Hydrogen may only be used captively inside industrial facilities for ammonia, steel, methanol and kerosene production. Hydrogen may only be stored locally in overground tanks. No methane is consumed. Only solid biomass and methanol may be used for backup heat and power plants. Methanol must be used for high-temperature heat demand in flat glass production that cannot be electrified~\cite{rehfeldtDirectElectrificationIndustrial2024}.
\end{itemize}

\section*{Results}
\label{sec:results}

\begin{figure}[!ht]
    \centering
    \includegraphics[width=.7\textwidth]{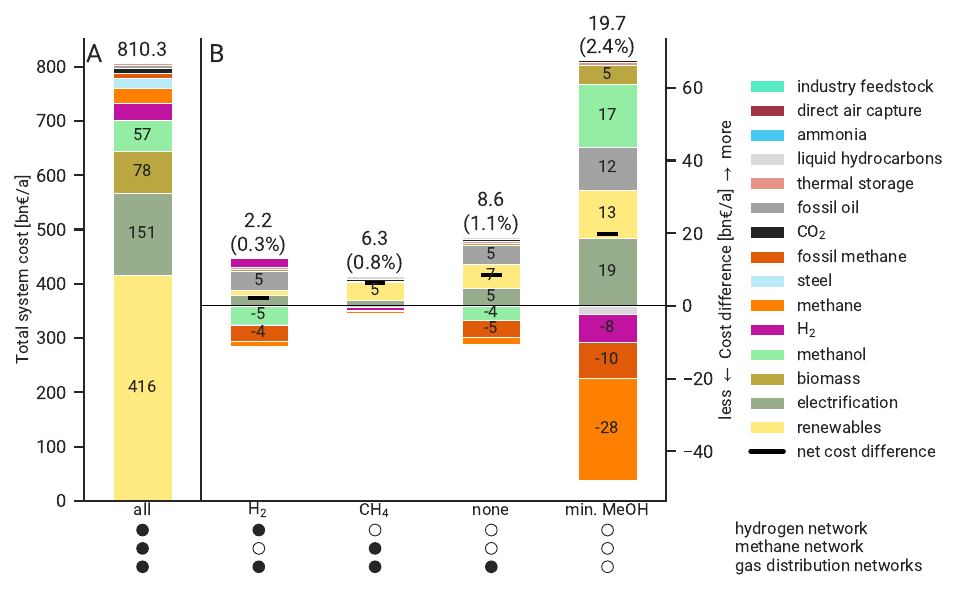}
    \caption{\textbf{Comparison of total annual system costs for the different scenarios.} The panel A shows the absolute cost of the `all networks' scenario. The panel B shows the cost increases and decreases of the other scenarios by component relative to the `all networks' scenario. The net absolute and relative cost difference is shown at the  top of each bar. A breakdown of the cost groups is given in~\cref{tab:cost_groups}.}
    \label{fig:co2_networks_total_cost}
\end{figure}

\begin{figure}[!ht]
    \centering
    \includegraphics[width=1\textwidth]{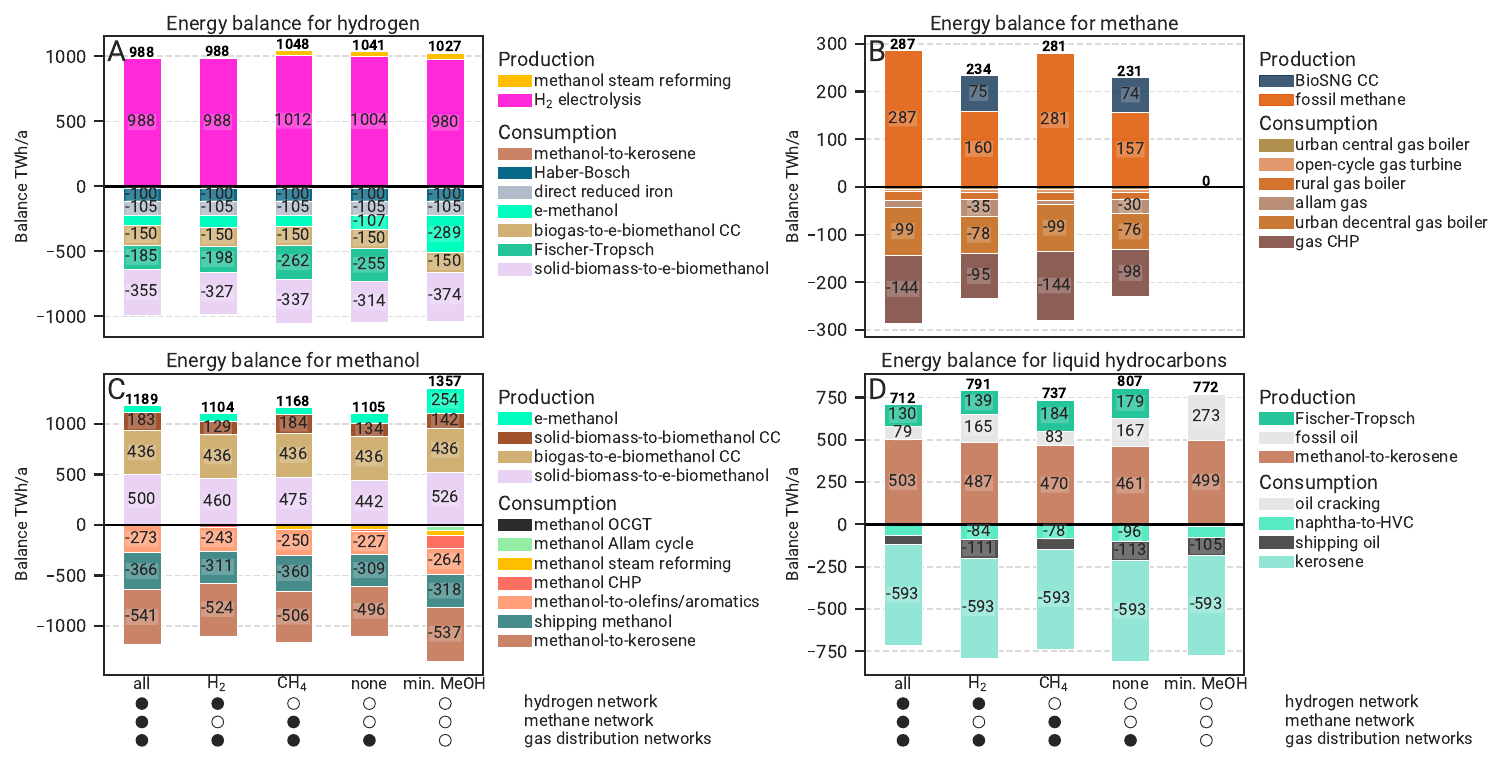}
    \caption{\textbf{Energy balances of the different scenarios for hydrogen, methane, methanol and oil.} The positive values show supply and the negative values show consumption. The bold number above each bar shows the total supply or consumption in TWh/a.}
    \label{fig:co2_networks_balances}
\end{figure}

\subsection*{Default scenario with all networks}

\cref{fig:co2_networks_total_cost}~A shows total annual system costs of \bneuro{810} for the default configuration of the model, where all energy conversion pathways are allowed, and the model is able to use and expand the following inter-regional networks: electricity, hydrogen, methane and carbon dioxide.
Methanol and oil can be traded between regions. Most of the cost in the system come from cost for renewable generation~(\bneuro{416}), electrification~(\bneuro{151}) of heat and transport (vehicle costs are excluded), biomass utilization~(\bneuro{78}), and the production and utilization of hydrogen~(\bneuro{32}) and its derivatives–methanol~(\bneuro{57}) and ammonia~(\bneuro{2}).

In the default scenario the usage of methanol is substantial: from~\cref{fig:co2_networks_balances}~C we see 1189~TWh/a of methanol usage, of which 94\% comes from biogenic sources and the rest from e-methanol (where the CO$_2$ comes from a mixture of biomass and other captured sources).
Methanol is mainly used to serve aviation, shipping and high value chemicals (HVC) demand (\cref{fig:co2_networks_balances}~C), either directly in the case of shipping, or through methanol-to-olefins (MtO), methanol-to-aromatics (MtA) or methanol-to-kerosene (MtK) processes.

MtO, MtA and MtK are competitive with the green naphtha to steam cracker route for HVC and Fischer-Tropsch for kerosene production because of the higher product selectivity.
Moreover, the Fischer-Tropsch process has a higher minimal part load constraint~(50\%) compared to methanol synthesis~(30\%), making it less flexible.
Consequently,  Fischer-Tropsch is only used if all products (kerosene, shipping oil, naphtha) can be fully used and sufficient cheap hydrogen is available to supply its minimal part load.
Methane and solid biomass rather than methanol are used for backup power or heat in this scenario~(\cref{fig:co2_networks_backup}).
The supply of methanol from biogas (which includes utilization of excess CO$_2$) is exhausted, and the rest is provided by solid biomass conversion~(\cref{fig:co2_networks_solid_biomass_balances}) and methanolisation of \co and electrolytic hydrogen.

The hydrogen production of 988~TWh/a comes exclusively from electrolysis (\cref{fig:co2_networks_balances}~A). It primarily serves the e-biomethanol~(505~TWh/a) and e-methanol (80~TWh/a) production, the Fischer-Tropsch production (185~TWh/a), the direct reduction of iron~(105~TWh/a) and the synthesis of ammonia via the Haber-Bosch process~(100~TWh/a).

The 200 MtCO$_2$/a sequestration budget is used to sequester captured unavoidable process emissions~(121~Mt/a), captured \co from waste incineration~(11~Mt/a), from Allam Cycle methane~(3~Mt/a) and from biomass conversion processes~(65~Mt/a), see~\cref{fig:co2_networks_carbon_flows_all}. This allows unabated emissions of 65~Mt/a from fossil methane (54~Mt/a) and shipping oil (11~Mt/a). 287~TWh/a of fossil methane is used for heat or power in Combined Heat and Power plants~(CHPs), Allam cycle gas and gas boilers~(\cref{fig:co2_networks_balances}~B). A total of 79~TWh/a of fossil oil and 130~TWh/a of Fischer-Tropsch oil are used for the production of kerosene, shipping oil, and plastics~(\cref{fig:co2_networks_balances}~D). At the same time, 503~TWh/a of methanol is used for the production of kerosene.

\subsection*{Dynamics towards a minimal methanol backstop}

Now we explore the competing roles of methane, hydrogen and methanol as gap-fillers in power and heat for periods of the year when renewables are scarce. As we remove gaseous options, we expect costs to increase, but not so drastically since methanol is always the substitute of last resort. If we separately remove either the hydrogen or methane network option, costs barely increase due to cost-effective substitutes in other energy carriers (\cref{fig:co2_networks_total_cost}~B).

Without a methane transmission network, costs rise by \bneuro{2.2}~(0.3\%), and methane and methanol usage decrease while fossil oil increase (\cref{fig:co2_networks_total_cost}~B). This is because fossil methane from European extraction sites can no longer be transported to district heating areas in our model.
Backup power and heat systems respond by reducing the use of fossil methane overall by around 44\% and covering the remainder with locally converted biogenic synthetic natural gas (BioSNG,~\cref{fig:co2_networks_balances}~B).
Additionally, methanol and hydrogen CHPs cover heat and electricity demand (\cref{fig:co2_networks_backup}).
The avoided use of fossil methane frees up part of the \co sequestration allowance.
In consequence, 86~TWh of additional fossil oil replace some of the synthetic methanol used for shipping, HVC and kerosene (\cref{fig:co2_networks_balances}~D).

Without a hydrogen network, costs rise by \bneuro{6.3}~(0.8\%). This is mainly attributed to higher costs for renewable energy generation.
Costs for renewables rise because hydrogen has to be produced locally at steel and ammonia plants, which requires more wind and solar capacity in low-yield areas, limiting the exploitation of the best resources.
In addition, 36~TWh/a of hydrogen is delivered to steel and ammonia plants by steam reforming methanol, a process that adds considerable conversion and reconversion losses~(\cref{fig:co2_networks_balances}~A).
In the \hyperlink{sec:sensitivity}{Sensitivity Section}, we explore a sensitivity analysis where ammonia and steel plants can relocate to avoid these costs.
Methanolisation, the largest consumers of hydrogen, can be flexibly sited where hydrogen is cheap and abundant, with the required \co captured from industrial applications and transported there by pipeline.

Removing both hydrogen and methane transmission networks yields an effect that is largely additive, increasing costs by \bneuro{8.6}~(1.1\%).
The effect is relatively limited since methane distribution is still possible within each region and can provide backup power and heat.
To supply regions without local methane production, methanol steps in~(\cref{fig:co2_networks_backup}).

In the `minimal methanol backstop' scenario, the distribution of methane and hydrogen as well as the use of these for process heat, electricity production and district heating is completely prohibited~(\cref{fig:co2_networks_balances}~A,B), forcing methanol into these use cases as biomass is limited.
The advantage of this scenario is that no distribution networks for methane or hydrogen are required.
All hydrogen is used captively inside the facilities where it is electrolyzed for the production of ammonia, methanol, kerosene and iron.
Only hydrogen derivatives that are liquid -- ammonia, methanol and kerosene -- are transported outside of industrial facilities.
In the absence of hydrogen distribution, underground hydrogen storage cannot be used and only aboveground storage tanks located onsite at industrial facilities are available.
As a result, Fischer-Tropsch becomes uncompetitive compared to methanol production, as it relies on low-cost underground storage to buffer cost-effective hydrogen for its high minimum part-load operation.

For this scenario, we see a cost increase of \bneuro{19.7}~(2.4\%).
Cost savings come from eliminated uses of methane technologies~(\bneuro{-28}), fossil methane~(\bneuro{-10}) and hydrogen~(\bneuro{-8}), which are offset by higher expenditures in electrification~(\bneuro{19}), methanol~(\bneuro{17}), renewables~(\bneuro{13}), fossil oil~(\bneuro{12}), and biomass~(\bneuro{5}).
Because there is no fossil methane in this scenario, more fossil oil can be consumed (273~TWh/a) and emissions offset.
Fossil oil substitutes some methanol demand for HVC, kerosene and shipping oil.
However, since there is no Fischer-Tropsch production, the total methanol consumption in the `minimal methanol backstop' scenario increases to 1357~TWh/a (\cref{fig:co2_networks_balances}~C), being the largest among the regarded scenarios.
The effect that methanol is used as a transport medium to supply small local hydrogen demands in fertilizer production and steelmaking persists in this scenario.
We assumed for these scenarios that methane and hydrogen pipelines capacities can be expanded as continuous variables; if only lumpy discrete pipeline sizes are allowed, the cost increase of the `minimal methanol backstop' scenario decreases slightly to \bneuro{19.3} compared to the `all networks' scenario (\cref{fig:co2_network_total_cost_discretized}), because the pipelines tend to be over-dimensioned compared to the transport need.

Consistent between all the scenarios is that methanol demand stays rather constant (1104-1357~TWh), see~\cref{fig:co2_networks_balances}~C.
For further inspection, we depict the electricity balances of all scenarios in~\cref{fig:co2_networks_electricity_balances}.

\subsection*{Operational aspects of the minimal methanol backstop}

We now explore the usage of methanol in the most extreme `minimal methanol backstop' scenario.

\begin{figure}[!ht]
    \centering
    \includegraphics[width=0.7\textwidth]{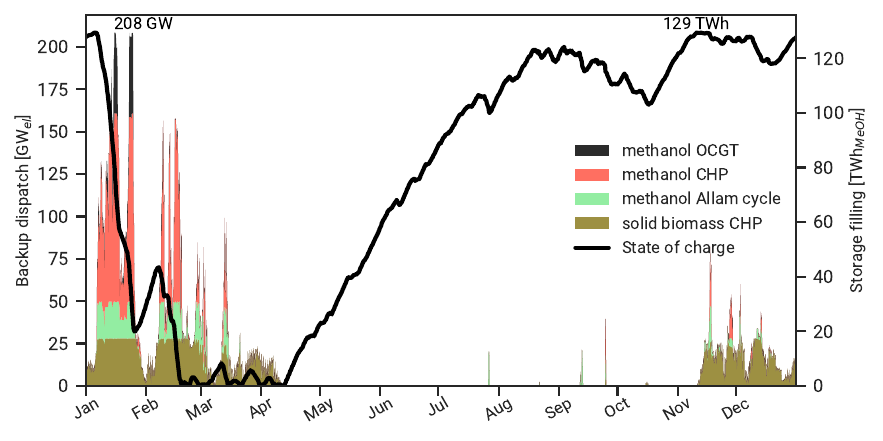}
    \caption{\textbf{Methanol storage levels and backup power plant dispatch.} Storage profile is shown as a solid line. The stacked area chart shows the dispatch of methanol CHP and backup power plants.}
    \label{fig:co2_networks_backup_and_storage}
\end{figure}

The storage behavior of methanol is shown in~\cref{fig:co2_networks_backup_and_storage}. Methanol storage of 129~TWh$_\text{MeOH}$ is mainly depleted during winter months for power and district heating supply during dark wind lulls, and then replenished in the summer when there is abundant solar energy. During the winter period, district heating methanol CHPs (without carbon capture) run with average full load hours (FLH) of 450~h and have an electricity output of up to 111~GW. Methanol in Allam cycle turbines run with FLH 950~h with capacity of 20~GW, while 48~GW unabated Open Cycle Gas Turbine~(OCGT) are only used for peak power generation (FLH 120~h) when prices are high because of their lower investment costs compared to CHPs. Besides methanol, solid biomass CHPs provide another 28~GW of backup capacity, running with FLH of 2240~h.

\begin{figure}[!ht]
    \centering
    \includegraphics[width=.8\textwidth]{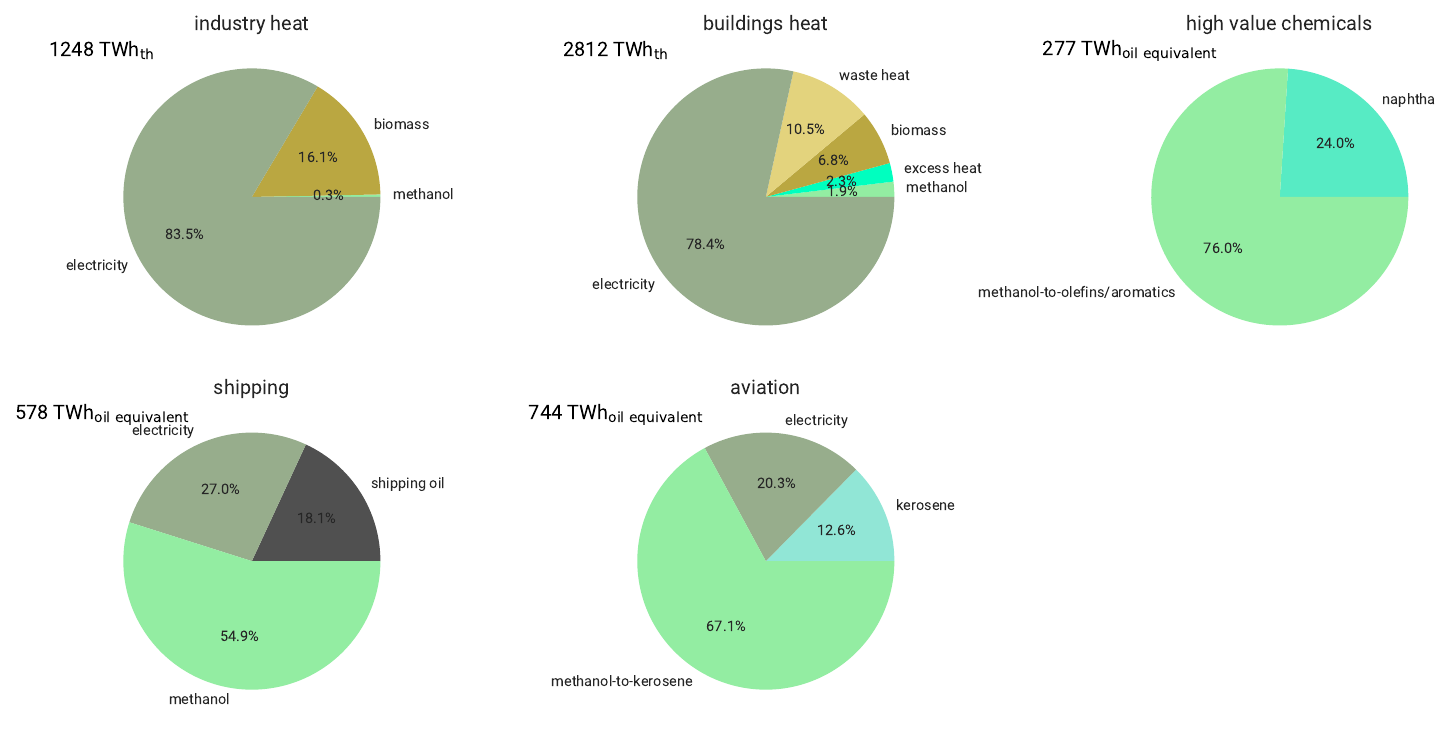}
    \caption{\textbf{Supply fractions of secondary/final demand in residential and industry heat, road transport, aviation, shipping, and high-value chemicals (HVC).} The demands for residential and industry heat are in TWh thermal, while the demands for transport and HVC are in TWh oil equivalent.}
    \label{fig:co2_networks_methanol_supply}
\end{figure}

\cref{fig:co2_networks_methanol_supply} shows the different energy carriers used in each sector. Apart from backup services in power and district heating provision, methanol is used to serve high-temperature process heat in flat glass production where electrification is not possible because of the need for radiant heat to melt the glass~\cite{rehfeldtDirectElectrificationIndustrial2024}. Neither hydrogen nor methanol are perfect substitutes for methane in this application, because they burn with a much lower emissivity, but this could be compensated by adapting the burners~\cite{zier2021}. Finally, the majority of the methanol is used in shipping (55\%), the production of aviation fuels (67\%) and high-value chemicals (76\%).

Methanol is used for these applications only if they cannot be cost-effectively electrified, or supplied with biomass, waste and excess heat (due to limited potential), or cost-effectively be replaced by fossil oil (with given emission and sequestration limits).
Of the demand for industrial process heat, 84\% is electrified, 16\% uses solid biomass, and less than 1\% uses methanol. Since hybrid process heat solutions are not allowed in the model (i.e., using fuel as a backup for an otherwise electrified process in high price periods), methanol consumption is limited to supplying heat of 4~TWh/a for flat glass production.
In buildings heat, methanol CHPs provide 2\% of total demand (all in district heat), while 78\% is electrified, 7\% uses biomass, another 10\% comes from waste incineration and 2\% uses excess heat from exothermic processes.
High value chemical production uses methanol as a feedstock for 76\% of its demand, while the remaining 24\% is served by fossil naphtha.
In shipping and aviation only international transport uses methanol or fossil oil either directly or in form of kerosene, because they cannot be electrified, while for domestic transport the model electrifies.

\begin{figure}[!ht]
    \centering
    \includegraphics[width=.9\textwidth]{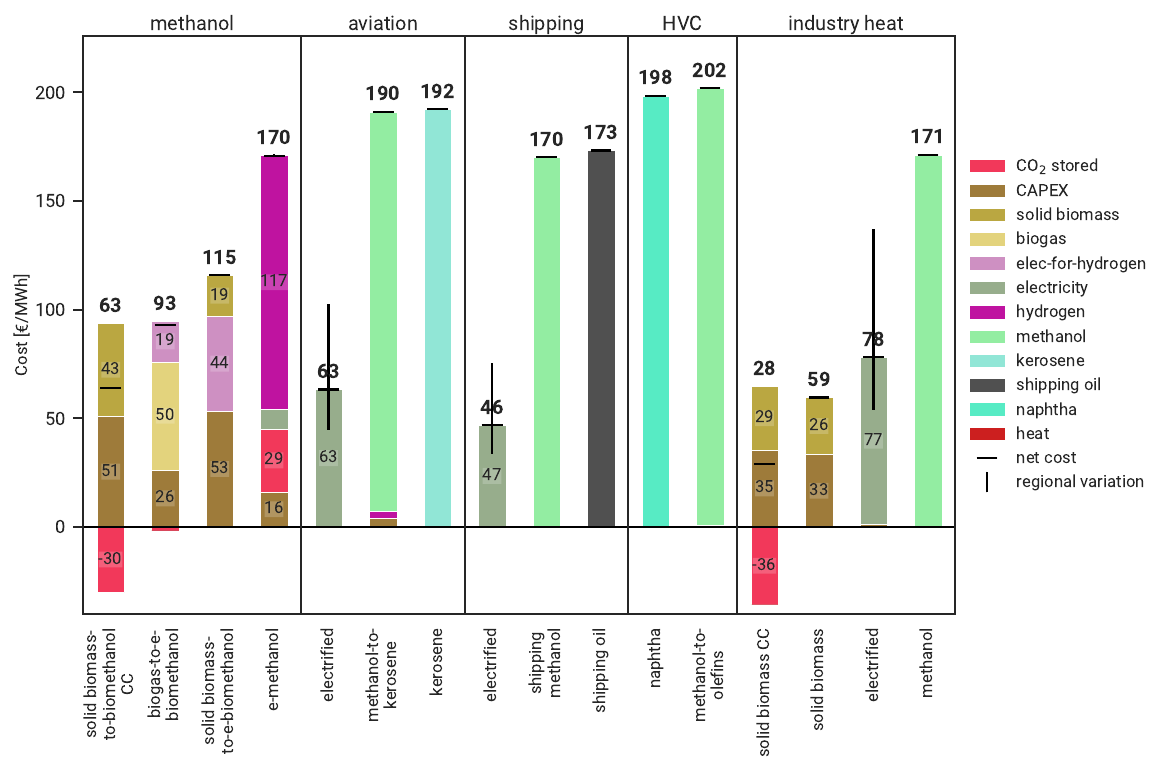}
    \caption{\textbf{Merit order curve of levelized costs for different methanol routes and final demands.} The costs are disaggregated into cost elements and consider the endogenous \co price for fossil routes. Costs are reported as demand-weighted averages across regions and time. For technologies that store captured carbon, resulting revenue streams are represented as negative cost elements. The dashed line indicates the net cost obtained by aggregating all positive and negative elements. Whiskers denote the range of demand-weighted costs across the 100 modelled regions.}
    \label{fig:co2_networks_methanol_merit_order}
\end{figure}

The levelized costs for different methanol production routes and for fuel options to serve the individual final demands in aviation, shipping, chemicals and industry heat are shown as a merit order in~\cref{fig:co2_networks_methanol_merit_order}. For the levelized cost calculation, we consider the model-endogenous prices for the fuels, electricity and \co which include the cost for production, storage and transport. The hydrogen for e-biomethanol production is reported as the electricity demand of the integrated electrolyzers.

For methanol production, the biomethanol and e-biomethanol routes are the cheapest, with costs between 63–115~\euro/MWh$_\text{MeOH}$, while synthetic e-methanol costs around 170~\euro/MWh$_\text{MeOH}$ owing to a large hydrogen cost element.
In the solid biomass to biomethanol with Carbon Capture~(CC) route, only the hydrogen and carbon contained in the biomass are used for methanol synthesis, while the excess \co is captured. This captured \co can either be sequestered or converted into additional methanol through the addition of hydrogen.
As a result, carbon capture provides a revenue stream that reduces the levelized cost of biomethanol with CC to 63~\euro/MWh$_\text{MeOH}$.
The solid biomass to e-biomethanol route on the other hand mixes additional hydrogen into the syngas mixture to use up the excess CO/\co for more methanol production, consuming electricity for the electrolysis. Similarly the biogas to e-biomethanol uses electricity to generate hydrogen to use up some of the excess carbon for methanol.
For short-distance aviation and shipping, electrification offers the lowest final energy costs for all regions -- even though variations in wind and solar resource availability make electricity prices vary widely across Europe.
Liquid hydrocarbons from Fischer-Tropsch process and methanol can be used interchangeably with small cost differences for long-distance aviation, shipping and high value chemicals production.
Depending on the region, industry heat is supplied at the lowest cost by biomass or electricity.
Both routes are cheaper than the methanol route. However, peak electricity prices can be considerably higher than the costs of the methanol route, which could make hybrid heating solutions attractive during times of electricity prices above 171~\euro/MWh.
Similar to methanol, the lowest levelized cost for industry heat is achieved with biomass CC, provided that there is an offtaker for the captured CO$_2$.

\begin{figure}[!th]
    \centering
    \includegraphics[trim=0 1cm 0 0cm,width=0.9\textwidth,clip=true]{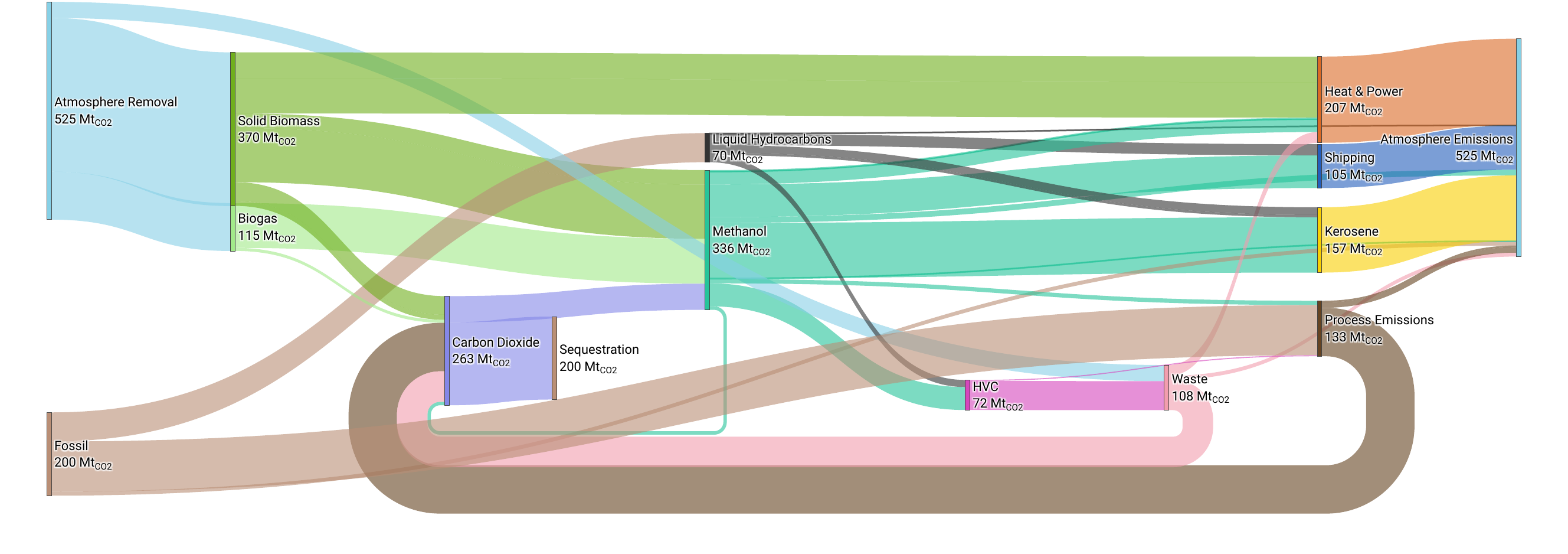}
    \caption{\textbf{Sankey diagram of annual carbon flows in the `minimal methanol backstop' scenario.}}
    \label{fig:co2_networks_carbon_flows}
\end{figure}

Finally, we examine the carbon flows in the `minimal methanol backstop' scenario (\cref{fig:co2_networks_carbon_flows}).
Biomass absorbs 524~MtCO$_2$/a from the atmosphere and enters the model as solid biomass, biogas and biogenic municipal solid waste.
A substantial part of solid biomass (147~MtCO$_2$/a) is used for heat and power applications where it is combusted and goes back to the atmosphere.
Carbon capture from these processes is not economical, because they are only used for backup with FLH of 2240~h.
Another 165~Mt/a of biogenic \co is used for biomethanol and e-biomethanol production, and 57~Mt/a is captured.
Biogas enters the system with 115~Mt/a biogenic \co, mostly converted to e-biomethanol.
In total, 200~Mt/a of fossil \co enters the system, 70~Mt/a from fossil oil and 130~Mt/a through fossil process emissions in industry like cement production.
Besides the direct fossil process emissions there are 3~Mt/a coming from the high value chemical production that uses liquid hydrocarbons or methanol as a feedstock.
The system captures 117~Mt/a of the total 133~Mt/a process emissions, while the remaining emissions escape to the atmosphere.
Most of the carbon in high value chemicals is captured in waste-to-energy plants.
Overall, 263~Mt/a of \co is captured of which 200~Mt/a is sequestered (matching the fossil carbon entering the system) and 63~Mt/a is used for e-methanol production.
To complete the carbon cycle, both methanol and liquid hydrocarbons are then used for shipping, aviation fuels, high value chemicals, power and heat. 525~Mt/a ends up in the atmosphere again, balancing the amount absorbed by the biomass in the first place.
The Sankey diagrams of the carbon flows for the other default scenarios are available in~\cref{fig:co2_networks_carbon_flows_all,fig:co2_networks_carbon_flows_h2,fig:co2_networks_carbon_flows_ch4,fig:co2_networks_carbon_flows_none}.

\subsection*{Spatial aspects of the minimal methanol backstop}

\begin{figure}[!ht]
    \centering
    \includegraphics[width=1\textwidth]{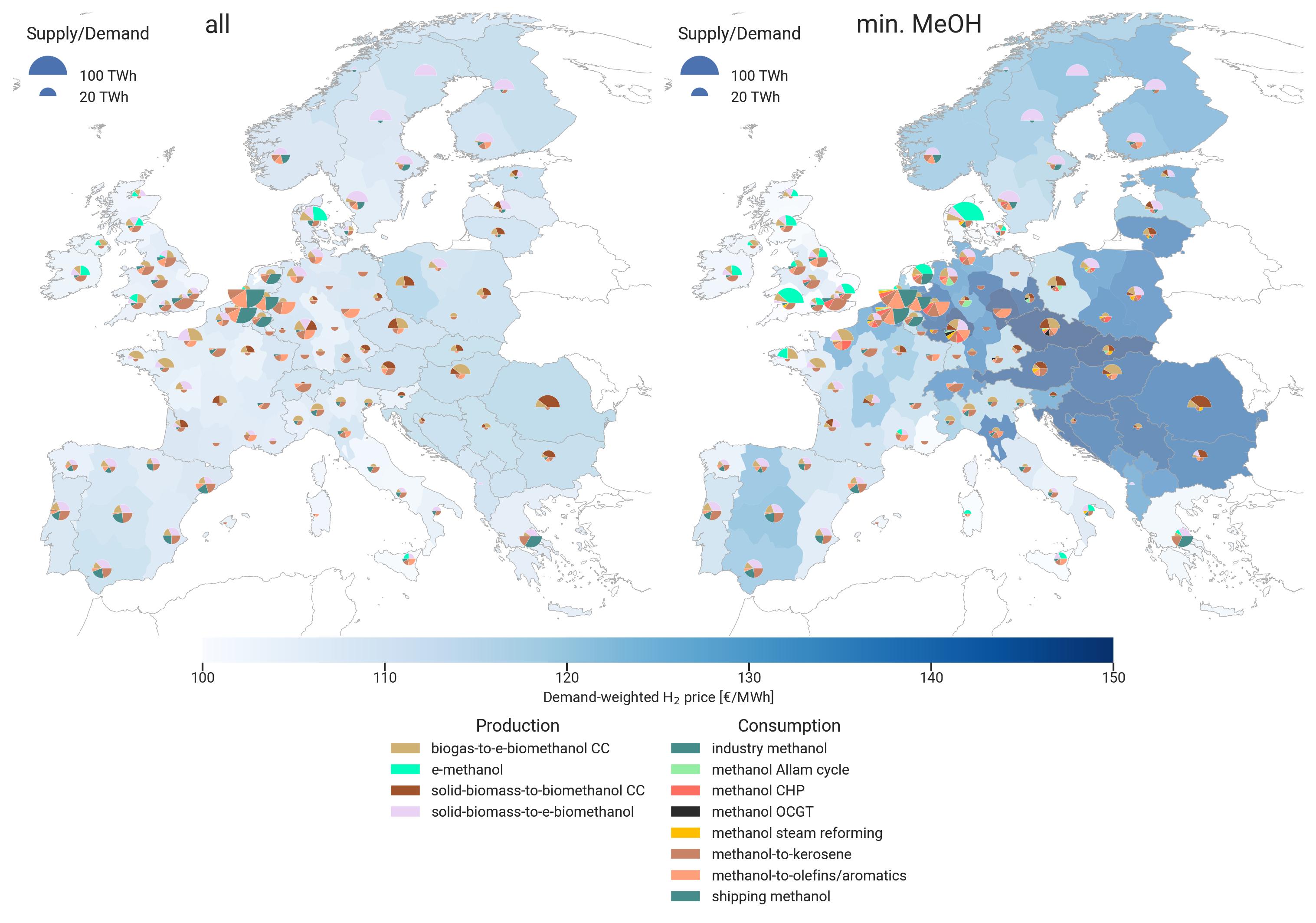}
    \caption{\textbf{Geographical patterns of methanol production and consumption in the `all networks' and `minimal methanol backstop' scenario.} Upper semi-circles show the production of methanol by technology route, while the lower semi-circles show the consumption of methanol by application. The choropleth layer shows demand-weighted average hydrogen prices per region.}
    \label{fig:methanol_map}
\end{figure}

\cref{fig:methanol_map} shows the geographical production and usage patterns of methanol for the `all networks' and `minimal methanol backstop' scenario.

The production of methanol is split between synthetic and biogenic production sites. Methanol synthesis tends to be centered around the periphery of Europe, either where there are good wind resources along the coastlines, or in the South where solar resources are plentiful for green hydrogen production. Lower hydrogen prices indicate regions with good renewable resources.
Biogenic methanol production is broadly distributed across Europe, reflecting the spatial distribution of biomass potentials. While e-biomethanol tends to be co-located in regions with low-cost electricity, biomethanol production also occurs in areas with limited renewable energy potential and higher electricity prices.

The consumption locations of methanol are partially exogenously set and partially endogenously determined by the model.
Fuel demands for shipping and aviation are set by sea- and airport locations, and feedstock demands are set by chemical industry locations.
Methanol for backup generation is only used in the `minimal methanol backstop'.
OCGT and Allam cycle as well as CHPs are endogenously sited in the center of Europe where the renewable resource endowment is worse and heating demand is high.
One advantage of these locations is the ease of exporting power to the periphery through Europe's extensive power transmission grid.
Furthermore, the periphery needs less local backup power because of the presence of large flexible loads in the form of electrolyzers.

\co pipelines transport the \co from inland capture points to sites with methanol synthesis or \co sequestration sites (\cref{fig:co2_stored_map}).
Regardless of the route, the carbon for methanol production comes solely from biogenic sources, i.e., biogenic waste and solid biomass, and captured \co of unavoidable process emissions. No direct air capture is required in any of the scenarios with default setting.

Geographical production and usage maps for hydrogen and methane are provided in~\cref{fig:hydrogen_map,fig:methane_map}.

\subsection*{\hypertarget{sec:sensitivity}{}Sensitivity on various input parameters}

In this section, we explore how the cost of the `minimal methanol backstop' scenario reacts to altering the input parameters with high uncertainty and high impact on the system. These include measures that represent possible bottlenecks, such as difficulties electrifying end uses or building enough transmission, as well as positive developments, like the availability of more sequestration or sustainable biomass:

\begin{itemize}
    \item \textbf{low electrification:} lower electrification than anticipated, where high and medium temperature heat for industry, heavy duty vehicles for road transport, domestic aviation and navigation cannot be electrified;
    \item \textbf{no power transmission:} a thought experiment where we remove the option for power transmission altogether, but still allow electricity distribution within each region;
    \item \textbf{high demand:} road, aviation and shipping transport, as well as high value chemical production, increase by 30\% compared to current levels;
    \item \textbf{low biomass potential:} the biomass availability is reduced from the medium potential (1550~TWh/a) to the low potential (840~TWh/a) given in the JRC ENSPRESO database~\cite{ruizENSPRESOOpenEU282019};
    \item \textbf{today's transmission:} using the power transmission capacities of today instead of allowing a 25\% expansion;
    \item \textbf{no biomass backup:} removing the option to use solid biomass directly for backup in power and heat;
    \item \textbf{no \co transport:} removing the option for \co transport between regions by pipeline;
    \item \textbf{relocation:} considering the relocation of steel and ammonia production within Europe;
    \item \textbf{imports:} allowing imports of green energy and materials from outside Europe with prices from~\cref{tab:import_prices};
    \item \textbf{400 Mt/a \& infinite sequestration limit:} increasing yearly \co sequestration limits from 200~Mt/a in the default to 400~Mt/a and unlimited;
    \item \textbf{high biomass potential:} increasing biomass availability to the high potential (3320~TWh/a) given in the JRC ENSPRESO database;
    \item \textbf{techno-economic assumptions for 2050:} taking techno-economic assumptions for the year 2050 instead of 2030;
    \item \textbf{\co reduction targets 90/95\%:} less ambitious \co reduction targets with reduction of 90\% and 95\% compared to 1990 emissions;
    \item \textbf{high biomass potential and inf. sequestration:} considering a combination of high biomass potential (3320~TWh/a) and infinite \co sequestration.
\end{itemize}

\begin{figure}[!ht]
    \centering
    \includegraphics[width=1\textwidth]{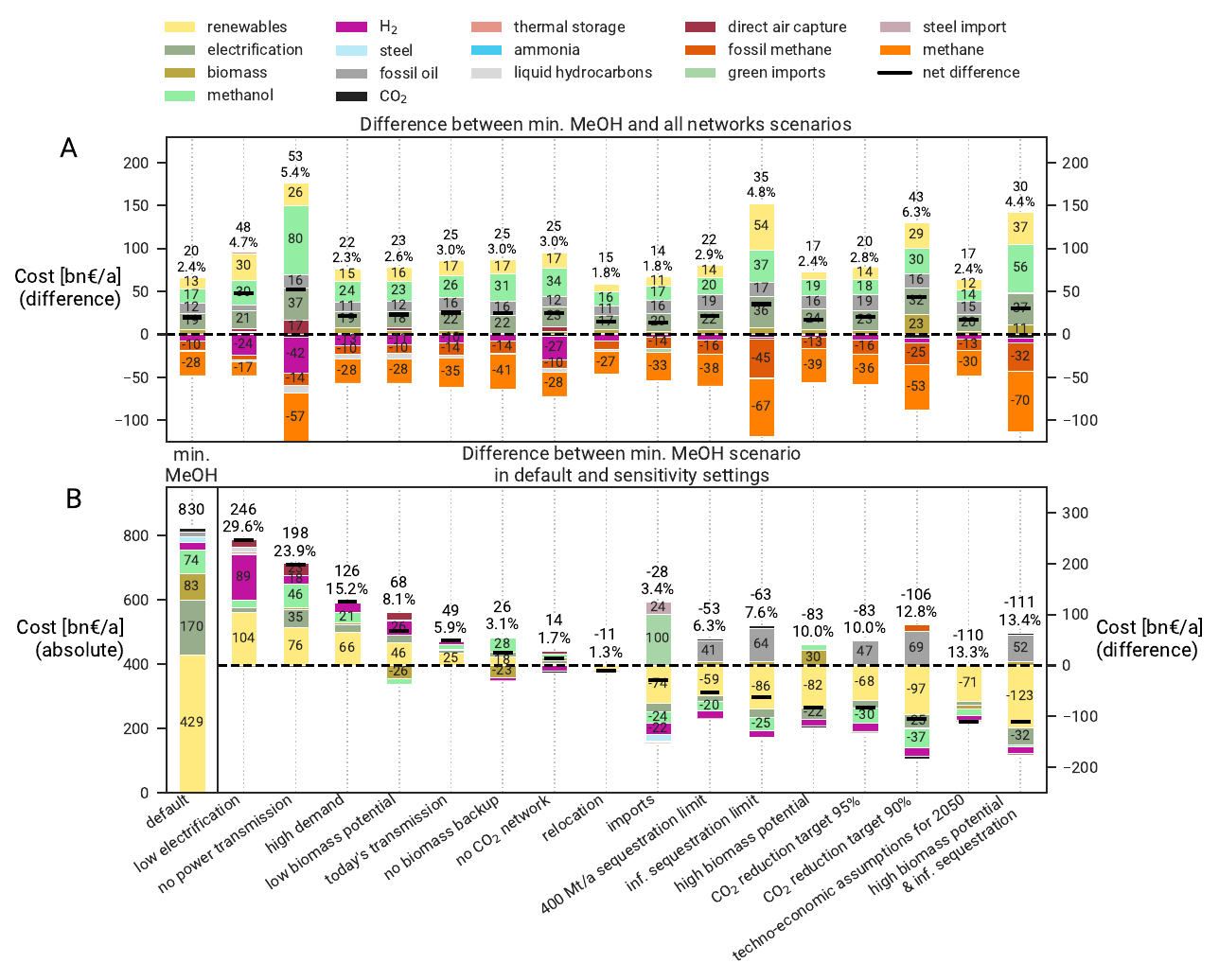}
    \caption{\textbf{Cost sensitivity for the different sensitivity settings and the scenarios.} The top panel shows the cost difference between the `all network' and the `minimal methanol backstop' scenario for the different sensitivity settings. The lower panel compares the absolute cost difference for the `minimal methanol backstop' scenario between the default setting (with \co network) and the other sensitivity settings.}
    \label{fig:sensitivity_cost_diff_abs_rel}
\end{figure}

\cref{fig:sensitivity_cost_diff_abs_rel}~top shows the cost premium in the `minimal methanol backstop' compared to the `all networks' scenario for each input parameter setting.
\cref{fig:sensitivity_cost_diff_abs_rel}~bottom explores how the cost of the `minimal methanol backstop' scenario itself reacts to altering the input parameters.
The outer left bar depicts the default setting with the sequestration limit of 200~Mt$_\text{CO2}$/a and the medium biomass potential of 1550~TWh/a.

In the default setting, the total cost of the `minimal methanol backstop' is \bneuro{830}~(\cref{fig:sensitivity_cost_diff_abs_rel}~bottom) and the cost premium of the `minimal methanol backstop' amounts to \bneuro{20}~(2.4\%) (\cref{fig:sensitivity_cost_diff_abs_rel}~top).

Removing electrification of high and medium temperature industry heat, domestic aviation and navigation as well as heavy duty vehicles forces the model to use synthetic fuels in these sectors.
This drives up the total system cost of the `minimal methanol backstop' by \bneuro{246}~(29.6\%), since methanol demand rises by 62\%~(\cref{fig:sensitivity_energy_carrier_consumption}) and the only production route with free capacity is the expensive e-methanol route which requires direct air capture (DAC).
The cost premium of the `minimal methanol backstop' compared to `all networks' rises to \bneuro{48}~(4.7\%) because the share of fuels in final demand increases, which is magnified by the cost premium of methanol over hydrogen or methane.
The increase in the cost premium reveals that the smaller cost premium in the default setting of 2.4\% relies to some extent on a high electrification rate that minimizes fuel usage.

In a thought experiment that removes the power transmission grid, the cost in the `minimal methanol backstop' increases by \bneuro{198}~(23.9\%) because only local renewable generation can satisfy electric demands in each region.
Backup capacities increase from 208~GW in the default to 400~GW, which increases the demand for methanol by 33\%.
As a consequence, more renewable generation (+\bneuro{76}) is cost-effective. Additionally, we observe more battery storage capacities.
The cost premium of the `minimal methanol backstop' in this setting is the highest among all sensitivity settings with \bneuro{53}~(5.4\%).
This is because with gasous networks, the system can partly evade the lack of power transmission by transporting gaseous fuels. Additionally, the system can import more fossil methane.

If we have 30\% higher demand for road, aviation and shipping transport as well as high value chemical production, the cost of the `minimal methanol backstop' increases by \bneuro{126}~(15.2\%). This cost rise is mainly caused by the higher demand for methanol and liquid hydrocarbon (\ref{fig:low_electrification_balances}) leading to small amount of DAC and more renewable generation and hydrogen production.
The cost premium of the `minimal methanol backstop' is \bneuro{22}~(2.3\%) which is similar to the default setting. This similarity illustrates that the `minimal methanol backstop' is robust to the considered demand increase.

Reducing the biomass availability from the medium to low potential leads to a cost increase of \bneuro{68}~(8.1\%) in the `minimal methanol backstop'. Biogenic methanol production decreases by 38\% so that the model is forced to use DAC and e-methanol, for which the model builds more renewable generation and hydrogen production.
The cost premium of the `minimal methanol backstop' increases to \bneuro{23}~(2.6\%).

If instead of allowing the electricity transmission grid to grow by 25\%, we limit it to today's size, then the cost of the `minimal methanol backstop' increases by \bneuro{49}~(5.9\%). Similar but less pronounced than the case of `no power transmission', these changes are driven by an increase in the need for fuels, which is driven by the local backup needs when renewable fluctuations cannot be balanced spatially by international trade. The cost premium of the `minimal methanol backstop' is at \bneuro{25}~(3\%).

Comparably, if biomass is not allowed for backup power and CHPs, where it covers 45\% of demand in the default setting~(see Figure \ref{fig:co2_networks_backup}), it must be substituted by fuels that drive up the cost of the `minimal methanol backstop' by \bneuro{26}~(3.1\%).
The the cost premium of `minimal methanol backstop' increases to \bneuro{25}~(3\%).

Removing the option to build \co pipelines increases the cost in the `minimal methanol backstop' by \bneuro{14}~(1.7\%) since the hydrogen for methanol synthesis has to be produced where \co is available, which typically raises the cost of the hydrogen. Fossil emissions cannot be brought from inland locations to the coast for sequestration, so have to be compensated with negative emissions at sequestration sites. The model is also forced to use DAC in locations where biogenic or industrial \co is not available. These results align with Hofmann et al.~\cite{Hofmann2025}
The cost premium of the `minimal methanol backstop' increases slightly from \bneuro{20}~(2.4\%) in the default setting to \bneuro{25}~(3\%), because `all networks' still has the ability to transport carbon through the methane network.

Allowing the relocation of steel and ammonia production within Europe leads to cost savings of \bneuro{11}~(1.3\%). By relocating these industries' hydrogen demand to regions with low hydrogen costs, the model requires less renewable generation and can avoid the conversion losses of using methanol as a transport medium for hydrogen (see~\cref{fig:relocation_balances}).
With relocation, the cost premium decreases to \bneuro{15}~(1.8\%) because the hydrogen transmission network is not as essential anymore to transport cost-effective hydrogen to industry sites. The small benefit of relocation inside Europe has to be weighed against the significant practical and political challenges, such as local job losses, training  workforces at new sites, supply chain interruption and geopolitical security~\cite{Verpoort2024}.

Allowing imports of green fuels and products, as well as the relocation of industries, leads to cost savings of \bneuro{28}~(3.4\%), primarily due to lower green production costs abroad compared to Europe. The import shares are 100\% for hot briquetted iron (steel), 37\% for methanol, and 100\% for ammonia.
Some methanol is still produced in Europe because of the availability of low-cost biogenic carbon sources (it is assumed that the exporting regions rely on direct air capture). These green imports contribute to significant reductions in the spending on renewable generation and hydrogen production, by \bneuro{-74} and \bneuro{-22}, respectively. These findings are consistent with previous studies highlighting the advantages of green imports~\cite{neumannGreenEnergySteel2025}.
The cost premium of the `minimal methanol backstop' scenario decreases to \bneuro{14}~(1.8\%), because Fischer-Tropsch and methanol can be imported without the need for hydrogen or methane networks. Although many of the concerns alluded to above about relocation of steel and ammonia apply here too, we note that Europe is already dependent on imports for most of its fossil gas and oil imports.

Increasing the \co sequestration allowance from 200 ~Mt$_\text{CO2}$/a in the default setting to 400~Mt$_\text{CO2}$/a and unlimited sequestration leads to cost decreases of \bneuro{53}~(6.3\%) and \bneuro{63}~(7.6\%), respectively.
In the unlimited case, 499.8~Mt$_\text{CO2}$/a are sequestered.
This creates space for more fossil oil usage in the system, substituting the use of synthetic fuels from hydrogen and captured carbon (CCU).
Any fossil fuel emissions that can be captured are sequestered, while unabated emissions, such as those from aviation, are compensated by carbon dioxide removal (CDR), sequestering the biogenic \co that would otherwise have been used for CCU.
Methanol production decreases from 1358~TWh/a under a 200~Mt$\text{CO2}$/a sequestration allowance to 770~TWh/a and 590~TWh/a with increased allowances of 400 and infinite~Mt$\text{CO2}$/a, respectively.
Green methanol is prioritized in shipping, where it is most cost-effective, and substituted with fossil sources for chemicals and aviation.
If we compare the cost premium to the `all network' scenario (\cref{fig:sensitivity_cost_diff_abs_rel}~top), we observe a cost increase of 22–\bneuro{35}~(2.9–4.8\%) because the model is no longer able to use as much fossil methane with CCS or compensated by CDR.

Increasing biomass availability from 1550~TWh/a to 3320~TWh/a leads to a cost decrease of \bneuro{83}~(10\%) in the `minimal methanol backstop'.
This is because more of the synthetic e-methanol production (with captured carbon) is replaced by less expensive biomethanol production (~\cref{fig:co2_networks_methanol_merit_order}). Almost all (95\%) of the available biomass is used in the model.
The price of methanol drops from 173~€/MWh to 115~€/MWh.
The cost premium of the `minimal methanol backstop' compared to `all networks' reduces slightly compared to the default setting to \bneuro{17}~(2.4\%), since both systems can utilize more biomass resources.

If the overall net-zero CO$_2$ emission target is relaxed to a 95\% or 90\% reduction compared to 1990 levels, the total costs decrease by \bneuro{83}~(10\%) and \bneuro{106}~(12.8\%) respectively. This shows the sharp increase in marginal abatement costs for the final 5\%.
Since the relaxation of the CO$_2$ target allows more fossil gas to be used throughout the system, the high cost of replacing this fossil gas with methanol raises the cost premiums of a `minimal methanol backstop' to \bneuro{20}~(2.8\%) and \bneuro{43}~(6.3\%) respectively. This setting shows that replacing methane with methanol in backup use cases only makes economic sense as we approach net-zero emissions.

Considering cost assumptions for the year 2050 instead of 2030 leads to cost savings of \bneuro{110}~(13.3\%) in the `minimal methanol backstop'. Cost savings mainly come learning in renewable generation assets (-\bneuro{71}), as well as hydrogen electrolysis, and synthesis processes like methanol production. The price of methanol drops here to 120~€/MWh.
The cost premium of the `minimal methanol backstop' decreases to \bneuro{17}~(2.4\%) because both scenarios benefit from lower costs assumptions.

Finally, the setting with the highest cost savings in the `minimal methanol backstop' scenario is the combination of high biomass potential and infinite \co sequestration with \bneuro{111}~(13.4\%) lower costs compared to the default setting. This is because more of the synthetic methanol production (with captured carbon) is replaced by less expensive biomethanol production (see~\cref{fig:co2_networks_methanol_merit_order}). Additionally, capturing biogenic carbon offsets more fossil oil usage.
The cost premium of the `minimal methanol backstop' compared to `all networks' increases to \bneuro{30}~(4.4\%) because the `all networks' scenario can use more fossil fuels combined with CCS and compensated by BECCS.

Common across all analyzed settings is the cost premium of 14–\bneuro{53}~(1.8–5.4\%) of the `minimal methanol backstop' compared to `all networks'.
Excluding the settings for low electrification, infinite sequestration, lower emission reductions and no power transmission, the cost premium range is reduced to 1.8-3\% (14–\bneuro{25}).
Renewable electricity production is required for local hydrogen and methanol production and adds cost between \bneuro{+11} and \bneuro{+54}. Additionally, the higher methanol demand increases the cost for methanol production facilities from at least \bneuro{+14} up to \bneuro{+80}.

Across all scenarios at least 567~TWh/a (with \co reduction target of 90\%) and up to 2198~TWh/a (with low electrification) of methanol are used, which is largely supplied from biogenic sources (50-100\%), with imports or synthetic methanol production making up any differences in demand.

\begin{figure}
    \centering
    \includegraphics[width=0.5\textwidth]{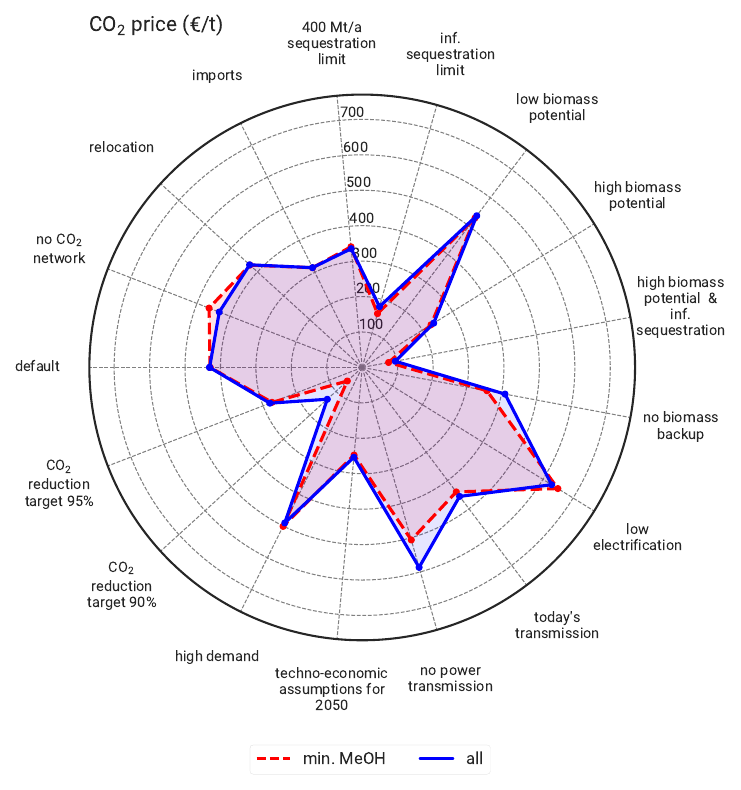}
    \caption{\textbf{Sensitivity of \co prices for sensitivity settings.} The spider web charts show the variation in \co prices for the `all networks' and the `minimal methanol backstop' scenario in the different sensitivity settings.}
    \label{fig:sensitivity_co2_prices}
\end{figure}

The \co price necessary to reach these scenarios can be derived from the shadow price of the carbon dioxide emissions constraint in the models.
The \co prices of the `all networks' and `minimal methanol backstop' scenarios vary significantly across the different sensitivity settings (\cref{fig:sensitivity_co2_prices}), ranging between 57 and 650~€/t$_\text{CO2}$, with the lowest price occurring in `minimal methanol backstop' for the setting with 90\% \co reduction target and the highest price with low electrification.
In the default setting, the `minimal methanol backstop' has a \co price of 429~€/t, dropping to 248~€/t with 2050 techno-economic assumptions, 228~€/t with high biomass potential and 158~€/t with unlimited sequestration.
Combining high biomass potential and unlimited sequestration leads to a \co price of 76~€/t. When assessing some of the low \co prices, it must be borne in mind that the model has already priced in the exogenous assumptions, such as the high levels of electrification.

What might be counter-intuitive at first is that the `minimal methanol backstop' scenario often has lower \co prices than the `all networks' scenario.
This is because the systems in `minimal methanol backstop' scenario often builds more renewables for methanol production, which help decarbonize the total system.
One example where the `minimal methanol backstop' has a higher \co price than the `all networks' scenario is the setting without \co network. Here, the `minimal methanol backstop' not only builds more renewable capacities for methanol production, but also relies on more expensive \co from direct air capture. This drives up the cost, because all cost effective CDR options are already exhausted.

Total cost comparisons and energy balances of the different settings are given in the appendix (\cref{fig:sensitivity_total_cost,fig:no_co2_network_total_cost,fig:no_co2_networks_balances,fig:relocation_total_cost,fig:relocation_balances,fig:imports_total_cost,fig:imports_balances,fig:seq_400_total_cost,fig:seq_400_balances,fig:inf_seq_total_cost,fig:inf_seq_balances,fig:low_bio_total_cost,fig:low_bio_balances,fig:high_bio_total_cost,fig:high_bio_balances,fig:high_bio_inf_seq_total_cost,fig:high_bio_inf_seq_balances,fig:todays_transmission_total_cost,fig:todays_transmission_balances,fig:no_grid_total_cost,fig:no_grid_balances,fig:low_electrification_total_cost,fig:low_electrification_balances,fig:high_demand_total_cost,fig:high_demand_balances,fig:no_bio_total_cost,fig:no_bio_energy_balances,fig:reduced_aims_total_cost,fig:reduced_aims_energy_balances,fig:2050_cost_total_cost,fig:2050_cost_energy_balances,fig:sensitivity_energy_carrier_consumption}).

\FloatBarrier

\section*{Discussion}
\label{sec:discussion}

The final energy supply in net-zero scenarios tends to be dominated by electricity, followed by carbonaceous liquids for long-haul transport and the chemical industry.
This result is robust across different scenarios, including those where some hydrogen-based fuels are replaced by unabated fossil fuels compensated by carbon dioxide removal.
The extent of electrification emerges as the dominant cost driver across all considered scenarios.
Besides electrification, high biomass potential helps to reduce system costs significantly providing more cheap, carbon-neutral liquids.
Combined with sufficient sequestration capacities, capturing and storing biogenic carbon could offset more unabated emissions from fossil fuels.

While hydrogen has dominated discussions of hard-to-electrify sectors, it does not need to extend beyond dedicated industrial clusters for the production of ammonia, steel and methanol.
As we show, other uses of hydrogen in backup power and heat could also be covered by a carbonaceous liquid like methanol.
Methanol supplied by fossil fuels is already cost-effective in applications such as chemicals and attractive for shipping, so it could provide a bridge between the fossil and net-zero worlds, while also offering easy handling and resilience against infrastructure uncertainty.
The bulk of methanol production comes from biomass supplemented by electrolytic hydrogen, with more expensive synthetic e-methanol covering the rest.
We will now discuss implementation and stakeholder aspects, as well as alternative options to methanol.

\subsection*{Practical conclusions for planners}

Each network scenario represents an extreme solution for the energy system, but in reality a mixed strategy of parallel infrastructures, differing from region to region, would allow to hedge against unforeseen circumstances. While some countries like Germany have decided for a full transition to the hydrogen economy like our `only hydrogen network' scenario~\cite{klimaschutzFortschreibungNationalenWasserstoffstrategie,OneStopShopWasserstoffWasserstoff}, we present here alternatives with similar cost that allow a middle way: building some regional networks for hydrogen (the `no gaseous fuel networks' scenario), e.g., from coastal regions with good wind conditions to steel-making locations, phasing out the methane network slowly, and in parts making a transition to methanol once methane demand has declined. This mixed strategy simplifies the difficult coordination and regulation problem of simultaneously scaling up hydrogen transport, storage, production and demand. It essentially de-links components of the system, allowing each component to be pursued separately.

First steps on the way to a minimal methanol backstop could include support mechanisms for biomethanol to bridge the gap between supply cost and demand willingness to pay, similar to the double auctions for the H2Global programme~\cite{H2GlobalsFoundation}. Initial producers could be biogas plants that have fallen out of electricity support programmes, as discussed below.

\subsection*{Resilience aspects}

In future scenarios where much of the demand is electrified, the resilience of the power system to disturbances and its ability to recover quickly, including from full blackouts, becomes ever more important.
Liquids offer advantages in this regard, since they can be bunkered next to power plants in large volumes in aboveground tanks, they do not need to rely on the functioning of a pipeline network  and they do not require any special geology like the salt deposits needed to build underground salt caverns for hydrogen. As was argued in Brown and Hampp~\cite{brownUltralongdurationEnergyStorage2023} and Göke et al~\cite{göke2025}, liquid fuel storage provides resilience against unusual events, such as low wind years, volcano eruptions, and physical or cyber attacks on infrastructure.
Storing liquid fuels in tanks next to power plants is an approach often followed for black-start-capable power plants today. While the extra cost of our extreme sensitivity run without any power transmission is substantial, a middle ground could be found where enough power capacity is kept in each region, along with reserves of methanol, so that the regions can run in a disconnected state for a few days or weeks  in an emergency. In the event of a blackout, this would allow island grids to be formed before the full system is reconnected.

\subsection*{Political economy aspects}

We now assess the idea of a `minimal methanol backstop' in light of its impact on groups with political power.

{\bf Fossil gas industry:} One purported benefit of a hydrogen economy is that hydrogen would provide an alternative business model to the fossil gas industry, offering opportunities for hydrogen production and the chance to build a parallel pipeline network. While a minimal methanol backstop would still require plenty of green hydrogen production, now captive inside industrial plants, it does assume a complete removal of gaseous fuel pipelines. On the positive side for the gas industry, there is an economic case for transporting \co by pipeline between regions, which could provide a new revenue stream.

{\bf Biogas industry:} Biogas today is mostly used in Germany for baseload electricity and heat production at small scales on farms; in other countries a majority share is upgraded to biomethane and used in land transport or shipping~\cite{StateBiogasIndustry}. Both of these use cases recede in renewables-dominated net-zero scenarios: backup power and heat is only cost-effective for a few hundred hours per year, while land transport is electrified. The possibility to produce e-biomethanol, which has many different use cases in the future, provides an alternative business model to biogas plants. Already today, shipping companies like Maersk are securing long-term contracts for green methanol~\cite{MaerskSignsLongTerm}. Challenges include the need to link up smaller biogas plants with a local biogas network to get the economies of scale for methanol production~\cite{gustafssonCentralizedDecentralizedHow2024}, as well as transitioning substrates away from today's energy crops towards more sustainable sources like agricultural wastes (e.g., straw, manure), residues (e.g., from forestry) and cover crops.

{\bf Regulation:} The fossil gas industry has tended towards oligopolistic structures: its large-scale production is concentrated in a few countries and companies, and the pipeline network is close to a natural monopoly. Hydrogen transport and storage would have similar concentrations of ownership as for fossil gas. In contrast, methanol production has many technical pathways to production, as well as enabling a spatially diverse set of actors to participate. The many different ways it can be transported (ship, barge, train, truck, pipeline) make it more resilient to mono- and oligopolistic market power in a pipeline network.

{\bf Smoother transition from fossil gas:} Methanol is a much easier drop-in fuel for methane than hydrogen in applications such as turbines or simple burners~\cite{Bertau2014}. This allows planners to focus on electrifying the economy and reducing fossil gas usage, before replacing fossil gas with methanol. Hydrogen, in contrast, needs careful planning to replace usage devices, pipeline networks and storage to account for its different physical properties. This planning must be done decades in advance under high levels of uncertainty regarding the final consumption volumes.

{\bf Prolonging the use of fossil gas:} Since methanol can be used as a drop-in replacement for fossil gas in turbines and burners, one concern is that it may lead companies to claim to want to switch to methanol as a cover for continued use of fossil gas, and thus delay more radical change. For example, in the power sector it could lead to continued use of gas power plants with a promise to switch to methanol that is never realized, either because of higher cost or because climate change is de-prioritized politically. Or for industrial heat, the promise to switch to methanol in the future could be used to argue for keeping gas burners and delaying electrification strategies. While these are legitimate concerns, they also apply to any dual-fuel strategies for hydrogen and methane as well. It should be clear to all participants that any combustion processes will be expensive in the future: fossil fuels because of \co pricing, and green fuels because of high production costs. This holds especially for scenarios with lower sequestration and biomass potentials.

\subsection*{Challenges of deep electrification}

The starting point for our analysis is a high level of
electrification, including all road transport, domestic aviation and
shipping, and most of heating in buildings and industry. This brings
considerable cost savings compared to the `low electrification'
sensitivity as well as simplifications from reducing the number of
energy carriers that are delivered to end consumers. While there is
strong agreement in the literature that electrification is the leading
strategy for decarbonisation~\cite{IPCC_2022_WGIII_TS,IEA24}, there are
several challenges. For example, while the total cost of ownership of
electric passenger cars is lower than their fossil-fuelled equivalents even
today, the higher upfront cost may deter some
consumers~\cite{IEA-EV-25}.  Efficiency measures for buildings that
might be necessary for some heat pump installations are not directly
costed in the model. While our model estimates both transmission and
distribution grid costs, there has been considerable inflation in grid
component costs since 2022 that is not accounted for~\cite{ludererEnergiewendeKosteneffizientGestalten2025}. Furthermore,
the coarse grid resolution may underestimate grid upgrade costs~\cite{frysztacki2021strong}, and the costs for ancillary services are
not included, although they are estimated to be low once there is
sufficient storage and demand response in the system~\cite{burdenresponse}.

Further challenges as well as estimated costs of the exogenous
assumptions on electric vehicle charging infrastructure, building
renovations and district heating expansion are given in Section
\nameref{sec:SI_discussion}.

\subsection*{Concepts with alternative liquid energy carriers}

Much of the advantage of methanol stems from its liquid state in ambient conditions. In Section \nameref{sec:SI_discussion} we consider alternative liquid energy carriers: ammonia, liquefied methane, DME, ethanol, liquefied hydrogen, Fischer-Tropsch fuels and liquid organic hydrogen carriers.

Further discussion on challenges, technical uncertainties, model limitations, pipelines for \co and methanol, and leakage of methane and hydrogen can also be found in Section \nameref{sec:SI_discussion}. A regularly-updated collection of frequently asked questions about the concept, along with answers, can be found online~\cite{meohfaq}.

\subsection*{Conclusion}
\label{sec:conclusion}

In a highly-electrified net-zero world, much of the remaining non-electrified demand can be met with hydrogen derivates rather than hydrogen itself. In particular, methanol, which is cost-effectively used in these scenarios for shipping, aviation kerosene and the chemicals industry, could be used as a `gap filler' for the remaining demands for fuels in backup power and heat. The extra cost of using methanol over hydrogen or methane in this `minimal methanol backstop' scenario for a net-zero Europe is just \bneuro{20}~(2.4\% of system costs), compared to an ideal deployment of hydrogen and methane infrastructure. For this low extra cost we argue that there are considerable advantages: no scale-up of complex new infrastructure to coordinate like for hydrogen; a smoother transition away from methane for gas turbines than for hydrogen with drop-in methanol; an easy way to absorb decentral biogenic wastes and residues into the energy system; a more flexible transport and storage infrastructure in the uncertain period while technologies for net-zero compete.

\section*{Methods}
\label{sec:methods}

\subsection*{Overview of European energy system model PyPSA-Eur}

For our analysis, we employ the open-source, sector-coupled energy system optimization model PyPSA-Eur~\cite{horschPyPSAEurOpenOptimisation2018,brownPyPSAEurOpenSectorcoupled2024}. The model covers the European Union plus Norway, Switzerland, the United Kingdom and the Balkan countries. It is built with Snakemake~\cite{molderSustainableDataAnalysis2021} and PyPSA~\cite{brownPyPSAPythonPower2018}, which uses Linopy~\cite{hofmannLinopyLinearOptimization2023} for model formulation and solver communication.
The optimization process aims to determine a cost-optimal energy system, accounting for both capital and operational expenditures, without considering the transition pathway–i.e., assuming an overnight scenario. It incorporates constraints such as \co limits, energy conservation, and technology-specific restrictions, including minimal part-load operation.

PyPSA-Eur encompasses multiple sectors, including electricity, heat, industry, agriculture, shipping, aviation, and land transport. Sector-specific energy demands are sourced from various databases, primarily using the energy balances from the JRC IDEES database~\cite{jointresearchcentreeuropeancommissionJRCIDEESIntegratedDatabase2017}, with additional national databases used for countries not covered by JRC IDEES.
To meet these demands, the model incorporates various energy carriers, including electricity, gas, oil, methanol, hydrogen, biomass, heat, and ammonia. Many energy demands, such as those for electricity, heat, transport, and steel production, are determined endogenously within the model, allowing it to dynamically optimize the allocation of energy carriers. Exogenously defined demands are illustrated in~\cref{fig:sectorial_demand}.

The model co-optimizes the capacity and operation of generation, storage, transport, and conversion technologies.
For generation, we model renewable energy sources such as solar, wind, hydro, and dispatchable generators like combined heat and power (CHP) plants or hydrogen turbines.
To obtain the time series for renewable generation, the model uses Atlite to generate renewable potentials and generation profiles for a given weather year~\cite{hofmannAtliteLightweightPython2021}.
Storage options include salt caverns for hydrogen, overground storage tanks for all fuels, seasonal and short-term heat storage, batteries, and pumped hydro storage.
For transmission, the model includes the high-voltage electricity grid and pipelines for hydrogen, methane and \co to transport energy and \co over longer distances, i.e., between regions.
The electricity transmission grid is derived from processed data of the OpenStreetMap project~\cite{openstreetmapcontributorsPlanetDumpRetrieved2017,xiongModellingHighvoltageGrid2025}.
For the transport of methanol, oil and solid biomass across Europe, we do not model a dedicated transport infrastructure but assume cost where they compete with other energy carriers.
For conversion, the model includes both energy and non-energy conversion processes.
Non-energy conversion refers to processes such as high-value chemical production from naphtha or steel production from hot briquetted iron.
Energy conversion includes processes such as electrolysis, which converts electricity and water into hydrogen, and gas boilers, which convert methane into residential heat.
Thereby, the model can produce hydrogen, methane, methanol, and oil from various sources, including renewable electricity, fossil fuels, and biomass. An overview of relevant conversion pathways for our analysis is given in~\cref{tab: overview_techs}.
All these components have distinct technological and cost parameters, which are collected in a technology database~\cite{aarhusuniversityPyPSATechnologyData2023}, primarily sourced from the Danish Energy Agency Technology Data~\cite{danishenergyagencyTechnologyData2022}. For our analysis, we use techno-economic assumptions for the year 2030~\cite{TechnoecomicAssumptions20302025}.

\subsection*{Temporal and spatial resolution}

For our analysis, we resolve 100 representative regions and 2190 time steps obtained from a segmentation clustering on the weather year 2013 using TSAM~\cite{hoffmannParetooptimalTemporalAggregation2022}.
The 100 representative regions are selected based on regional electricity demand, using the k-means algorithm~\cite{frysztackiComparisonClusteringMethods2022}. High-demand regions such as Germany, France, and the UK, have more regions resolved.
The time series aggregation clusters by wind, solar and load series to sequential periods of varying durations. This allows clustering, for instance, nighttime hours with little time series variations together while resolving midday solar feed-in peaks at higher temporal resolution. The use of 2190 time steps was found to keep a high level of accuracy (compared to the hourly solution) and significantly reduce computational time~\cite{NEUMANN20231793}. Computational resources are discussed below.

\subsection*{Biomass supply and consumption}

Biomass potentials are restricted to residual products, such as agricultural residues, forest residues, and manure. These residual products are not in direct competition with food production. The regional potentials are taken from the medium scenario for 2030 of the JRC ENSPRESO database~\cite{ruizENSPRESOOpenEU282019}, and then varied between the low and high scenarios for the sensitivity analysis.

Solid biomass can be used directly for specific applications, such as steam and  medium-temperature heat provision in industry or boilers in rural areas. It can also be converted into methane, hydrogen, methanol, or oil. Biogas can be converted into methane or methanol.

\subsection*{Methanol supply and consumption}
Our model encompasses two biogenic methanol routes, namely solid biomass-to-methanol and biogas-to-methanol. Solid (mostly woody) biomass is gasified to syngas in large 200 ktMeOH/a (around 140 MW-MeOH) plants, since the solid biomass is easier to aggregate and transport over long distances. Biogas from anaerobic digesters is run through a steam methane reformer to produce syngas at smaller scale (around 30~MW-MeOH), since the substrates have a high water content and transportation is more expensive. There is a trade-off between economies plant size and the logistic costs for transporting biomass over the longer distances  that a larger collection area would require, which we did not explore here. In both pathways, biogenic carbon and hydrogen present in the syngas are used for methanol synthesis. Since the carbon to hydrogen ratio in biomass exceeds that in methanol, in both pathways electrolytic hydrogen can be added to the syngas mixture to enhance the methanol yield, producing e-biomethanol. Assumptions for producing e-biomethanol from biogas were taken from the literature~\cite{bubePowerBiogasMethanol2024}. Alternatively, the model can also rely only on the hydrogen intrinsic in the biomass itself and produce biomethanol without the addition of hydrogen. The surplus carbon can then be captured and either sequestered or combined with additional hydrogen for further utilization (e-methanol in the case of methanol).

In addition to biogenic sources, methanol can also be produced from hydrogen and captured CO$_2$, or from fossil fuels. For fossil methanol, we consider both grey and blue methanol, using syngas from steam methane reforming of fossil methane. Blue methanol differs by incorporating carbon capture to reduce excess \co emissions during synthesis. For both grey and blue methanol, the embedded carbon in the methanol is fossil.

On the consumption side, methanol can be used either as a feedstock or for energetic applications (\cref{fig:methanol_pathways}). Energetic uses include power and heat provision, shipping and aviation fuels, and hydrogen production from methanol reforming. For shipping we do not allow onboard capture of carbon dioxide. Methanol can also be used for non-energetic purposes as a feedstock for high-value chemicals (e.g.,~olefins and aromatics). For power and heat applications, we assume the same efficiency and cost parameters as methane-based processes.

\subsection*{Oil refining and Fischer-Tropsch synthesis}

Oil products are distinguished by three categories based on their carbon chain length: naphtha (light gasoline), kerosene, and diesel. The Fischer-Tropsch synthesis and crude oil refining yield different product distributions.
For crude oil, we assume an average product split based on European refinery yields, i.e., 20\% naphtha, 10\% kerosene, 60\% diesel, and 10\% self-consumption~\cite{grupalotosYieldStructuresSelected2018}.
For Fischer-Tropsch synthesis, we assume final product shares of 25\% naphtha, 44\% kerosene, and 31\% diesel~\cite{konigSimulationEvaluationProcess2015}. The overall product conversion efficiency from hydrogen to FT products is assumed to be 70\%.
Additionally, the model can achieve greater product selectivity through cracking and recycling. When converting long-chain hydrocarbons into shorter-chain products, such as diesel to kerosene or kerosene to naphtha, we assume a conversion loss of 10\%, which is based on electric steam cracking losses~\cite{lechtenbohmerDecarbonisingEnergyIntensive2016}.

\subsection*{Transmission and distribution costs}

Our analysis evaluates methanol, hydrogen, and methane as competing energy carriers in a strongly electrified system. They serve diverse applications -- residential and industrial heating, backup power during low renewable generation -- and must be assessed considering both transmission and distribution costs.  An overview is provided in~\cref{tab:transport_cost}.

For methane and hydrogen, we include both investment costs for pipelines and operational costs for compression at the transmission level.
At the distribution level,  we apply a uniform 20~\euro{}/MWh cost for power and heat applications on the consumer side, such as CHPs and industrial processes~\cite{ludererEnergiewendeKosteneffizientGestalten2025}. Costs for building and operating new \co pipelines are also included.

For methanol, we apply a uniform cost of 4~\euro{}/MWh for transmission and distribution on the consumer side. This value represents the consumption-weighted average from a supplementary model run where methanol transport costs were accounted for endogenously~(\cref{fig:methanol_transport_map}).
In these runs, we consider transport of methanol by truck at 80\euro{}/t$_\text{MeOH}$/1000~km~\cite{diasEnergyEconomicCosts2020}, neglecting potentially less costly alternatives such as transport by pipeline, train, barge or ship.
Oil competes with methanol in the transport sector and for producing high-value chemicals. We assume transport and distribution costs to be the same as for methanol.

We allow the transport of solid biomass across regions with average cost of 0.1 \euro/km/MWh. Biogas must be upgraded to methane in order to cross model regions. For distribution costs within regions, we assume that solid biomass, such as forest residues, is collected and transported to a central conversion facility, whereas manure and straw are first digested and converted into biogas at decentralized locations.
Decentralized digestion of agricultural biomass has environmental benefits because it retains essential nutrients needed for soil health~\cite{gustafssonCentralizedDecentralizedHow2024}. The resulting biogas is then transported via pipelines to a central biogas upgrading facility.
Assuming an average transport distance of 10~km to the central conversion facility and transportation cost of 0.1~\euro{}/MWh/km for solid biomass, the resulting total transport cost amounts to 1~\euro{}/MWh~\cite{pabloruizJRCEUTIMESModelBioenergy2015}. For biogas  distribution via pipelines, costs are around 7~\euro{}/MWh for a 10 km distance~\cite{gustafssonCentralizedDecentralizedHow2024}.

\begin{table}
    \centering
    \begin{tabular}{l|c}
        \toprule
        Energy carrier & Transport cost \\
        \midrule
        Overhead transmission & 442 [€/km/MW] \\
        Submarine transmission & 1008 [€/km/MW] \\
        New methane (pipeline) & 87 [€/km/MW] \\
        New hydrogen (pipeline)  & 282 [€/km/MW] \\
        \co (pipeline) & 2116 [€/km/(t$_{\text{CO}_2}$/h)] \\
        Solid biomass (truck)& 0.1 (mean) [€/km/MWh]  \\
        Methanol (truck) & 0.027 [€/km/MWh] \\
        \bottomrule
    \end{tabular}
    \caption{Transport cost per km for different energy carriers}
    \label{tab:transport_cost}
\end{table}

\subsection*{Electrification of heating and transport}

Since we explore gap-fillers in high electrification scenarios, we assume that substantial shares of heat and transport demand that have been deemed viably electrifiable in recent literature are electrified, either exogenously assumed (land transport, domestic aviation and shipping) or in competition with other options.

In buildings and district heating networks the option of heat pumps is always available in competition with other sources like gas boilers, biomass boilers and CHPs; it is assumed that the demand for space heating declines by 20\% by 2050 compared to 2025 due to building renovations~\cite{Zeyen2021}, which are assumed exogenously and not included in the total cost.  It is assumed that 60\% of urban areas in Northern European countries are supplied with district heating, based on potentials presented in Persson et al.~\cite{Persson2011}. For industrial heat, existing biomass usage is retained, while best-in-class electrification options are used to cover much of the low and medium-temperature demand~\cite{NEUMANN20231793}. The model can then choose whether to electrify or use fuels for the remaining medium- and high-temperature demand, assuming that electrification is possible in nearly all sectors; fuels are retained where necessary for the high-temperature process heat in flat glass production~\cite{Madeddu_2020, rehfeldtDirectElectrificationIndustrial2024}.
If industry heat processes are electrified, we do not allow hybrid heating solutions, i.e., using a fossil fuel as a backup for an otherwise electrified process. The costs of converting industrial processes are not included in the model, because we focus here on the energy supply.

Besides heat, we exogenously assume the full electrification of road transport~\cite{Zeyen2025}, domestic shipping and aviation~\cite{Schaefer2019,bogdanovEffectsDirectIndirect2024}. The electrification of shorthaul aviation relies on batteries with a gravimetric density at least 800~Wh/kg, which should be achievable by mid-century~\cite{Schaefer2019}. Since we assume these changes exogenously and focus here on the energy supply, we do not include costs for vehicles or charging infrastructure in the model.

\subsection*{Fossil fuels and green energy imports}

Our analysis exclusively considers European fossil gas and oil extraction, which primarily originate from Norway and the United Kingdom~\cite{statistaInfographicEuropesBiggest2022}.

When considering imports from overseas, we only include green energy carriers and products. These include green hydrogen, Fischer-Tropsch oil, methanol, ammonia, hot briquetted iron and crude steel. The costs associated with green carriers and products given in~\cref{tab:import_prices} are taken from Neumann et al.~\cite{neumannGreenEnergySteel2025}.

\begin{table}
    \centering
    \begin{tabular}{ll}
        \toprule
        Fuel & Price\\
        \midrule
        Methanol & 140 \euro/MWh \\
        Hydrogen & 100 \euro/MWh \\
        Fischer-Tropsch oil & 160 \euro/MWh \\
        Ammonia & 110 \euro/MWh \\
        Hot briquetted iron & 480 \euro/t \\
        \bottomrule
    \end{tabular}
    \caption{Overview of the green fuel and product import prices}
    \label{tab:import_prices}
\end{table}

\subsection*{Computational resources}

Each scenario was solved with Gurobi v12.0.1 on a high-performance cluster node with two AMD EPYC 7543 CPUs and 16 DDR4-2933 64GB ECC RAM units. With 8 threads available, a single scenario required between 16 and 19 hours and up to 130~GB of memory.

\section*{Acknowledgements}

M.M. acknowledges funding from the Swedish Energy Agency, project numbers P2022-01082 and 2023-00888, through the CETPartnership (https://cetpartnership.eu/) project RESILIENT via the European Union's Horizon Europe research and innovation programme under grant agreement no. 101069750. We thank Johannes Hampp, Gunnar Luderer, Falko Ueckerdt, Michael Liebreich, Hannah Daly, Barry McMullin, Steve Green, Philipp Stöcker, Jerry Murphy, Hanno Böck, Elisabet Liljeblad, Bobby Xiong and many others for useful discussions. The responsibility for the contents lies with the authors.

\section*{Author Contributions}


\textbf{Philipp Glaum}: Conceptualization – Data curation – Formal Analysis – Investigation – Methodology – Software – Validation – Visualization – Writing - original draft\\
\textbf{Tom Brown}: Conceptualization – Formal Analysis – Investigation – Project administration – Supervision – Validation – Writing - original draft – Writing - review \& editing\\
\textbf{Fabian Neumann}: Conceptualization – Formal Analysis – Investigation – Supervision – Validation – Writing - review \& editing\\
\textbf{Markus Millinger}: Conceptualization – Formal Analysis – Investigation – Validation – Writing - review \& editing


\section*{Resource Availability}
\subsection*{Lead Contact}
Requests for further information and resources should be directed to and will be fulfilled by the lead contact, Tom Brown (\href{mailto:t.brown@tu-berlin.de}{t.brown@tu-berlin.de}).
\subsection*{Materials Availability}
This study did not generate new physical materials or unique reagents.
\subsection*{Data and Code Availability}
The code to reproduce the experiments and visualizations is archived on Zenodo: \href{https://doi.org/10.5281/zenodo.17108229}{https://doi.org/10.5281/zenodo.17108229}.
A dataset of the modeling results has been deposited to Zenodo: \href{https://doi.org/10.5281/zenodo.15388950}{https://doi.org/10.5281/zenodo.15388950}.
Technology data assumptions were taken from \url{https://github.com/p-glaum/technology-data/tree/bio_methanol/}.
We also refer to the documentation of PyPSA (\href{https://pypsa.readthedocs.io}{https://pypsa.readthedocs.io}) and PyPSA-Eur (\href{https://pypsa-eur.readthedocs.io}{https://pypsa-eur.readthedocs.io}), for technical instructions on how to install and run the model.

\section*{Declaration of Interests}

The authors declare no competing interests.

\section*{Declaration of Generative AI and AI-assisted technologies in the writing process}

During the preparation of this work the authors used chatGPT~\cite{ChatGPT} in order to improve readability. After using this tool/service, the authors reviewed and edited the content as needed and take full responsibility for the content of the published article.

\addcontentsline{toc}{section}{References}
\renewcommand{\ttdefault}{\sfdefault}
\bibliography{references}


\newpage

\makeatletter
\renewcommand \thesection{S\@arabic\c@section}
\renewcommand\thetable{S\@arabic\c@table}
\renewcommand \thefigure{S\@arabic\c@figure}
\makeatother
\renewcommand{\citenumfont}[1]{S#1}
\setcounter{equation}{0}
\setcounter{figure}{0}
\setcounter{table}{0}
\setcounter{section}{0}

\section{Supplementary Information}
\label{sec:si}

\subsection*{Supplementary material of the methods section}

\begin{figure}[!ht]
    \centering
    \includegraphics[width=0.8\textwidth]{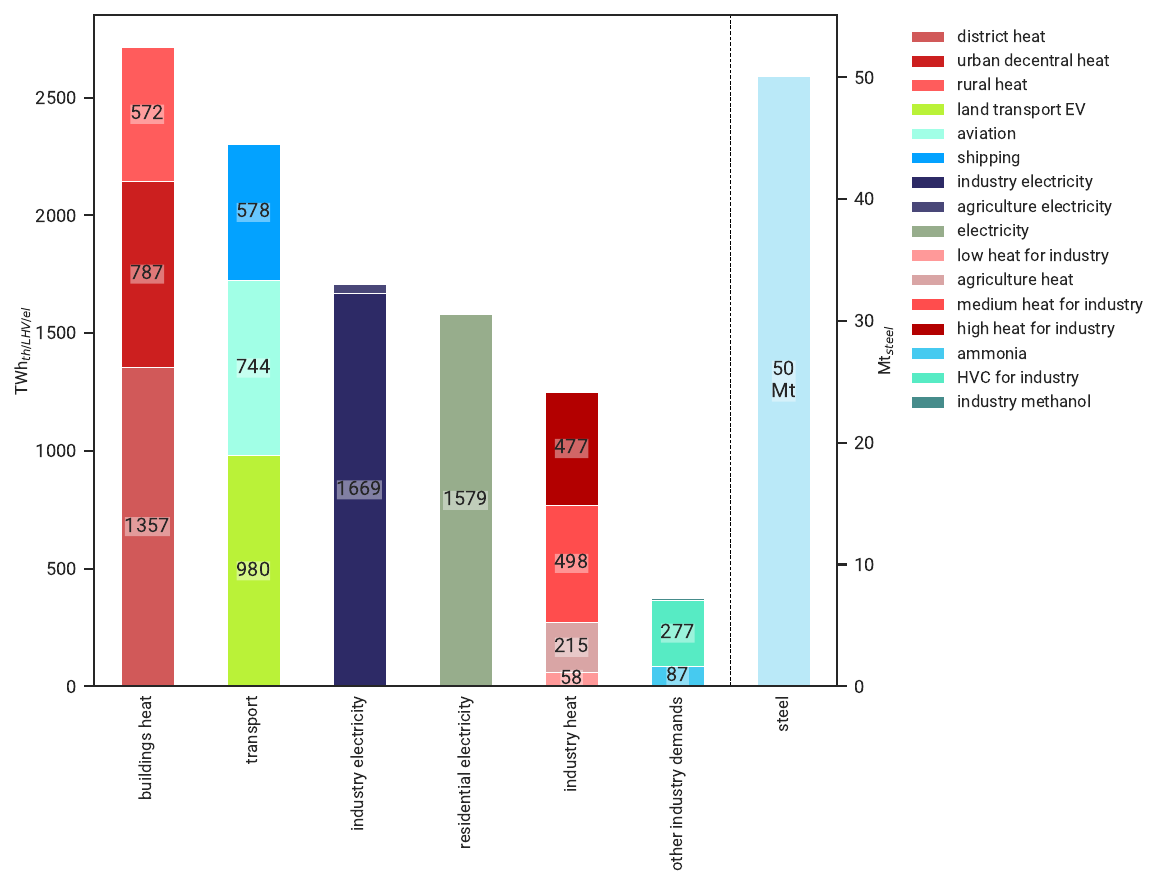}
    \caption{\textbf{Annual exogenously given sectorial demands.}}
    \label{fig:sectorial_demand}
\end{figure}

\begin{table}
    \begin{tabular}{lll}
        \toprule
        \midrule
        Air-source heat pump & Allam cycle with methanol & Ammonia cracker \\
        Bio synthetic natural gas (CC) & Biogas to methane (CC) & Biogas to methanol (CC) \\
        Biomass to oil products (CC) & Blue methanol & Combined-cycle methanol turbine (CC) \\
        Direct air capture & Fischer-Tropsch & Floating offshore wind (DC) \\
        Grey methanol & Hydrogen electrolysis & Hydrogen fuel cell \\
        Haber-Bosch & Hydrogen CHP & Hydrogen for industry heat \\
        Methane boiler & Methane CHP (CC) & Methane for industry heat \\
        Methanol CHP (CC) & Methanol for industry heat (CC) & Methanol steam reforming (CC) \\
        Methanol synthesis & Methanol to gasoline & Methanol to kerosene \\
        Methanol to olefins/aromatics & Offshore wind (AC) & Offshore wind (DC) \\
        Onshore wind & Open-cycle gas turbine & Open-cycle hydrogen trubine \\
        Open-cycle methanol turbine & Resistive heater & Run-of-river \\
        Sabatier & Solid biomass CHP (CC) & Solid biomass to hydrogen \\
        Solid biomass to methanol (CC) & Steam methane reforming (CC) & Utility-scale solar PV (tracking) \\
        Solar heater & Waste CHP (CC) &  \\
        \bottomrule
    \end{tabular}
    \caption{\textbf{Overview of available conversion technologies in the model.} The techno-economic assumptions for these technologies can be found in the analysis specific technology database~\protect\citeS{TechnoecomicAssumptions20302025S}. (CC: with Carbon Capture)}
    \label{tab: overview_techs}
\end{table}

\begin{table}
    \centering
    \begin{tabular}{ll}
        \toprule
        Fossil fuel & Price \\
        \midrule
        Fossil methane & 35 [€/MWh] \\
        Fossil oil & 52 [€/MWh] \\
        \bottomrule
    \end{tabular}
    \caption{Overview of the fossil fuel prices assuming the lower heating value.}
    \label{tab:fossil_prices}
\end{table}

\FloatBarrier

\subsection*{Additional figures of the main results}
\begin{figure}[!ht]
    \centering
    \includegraphics[width=0.9\textwidth]{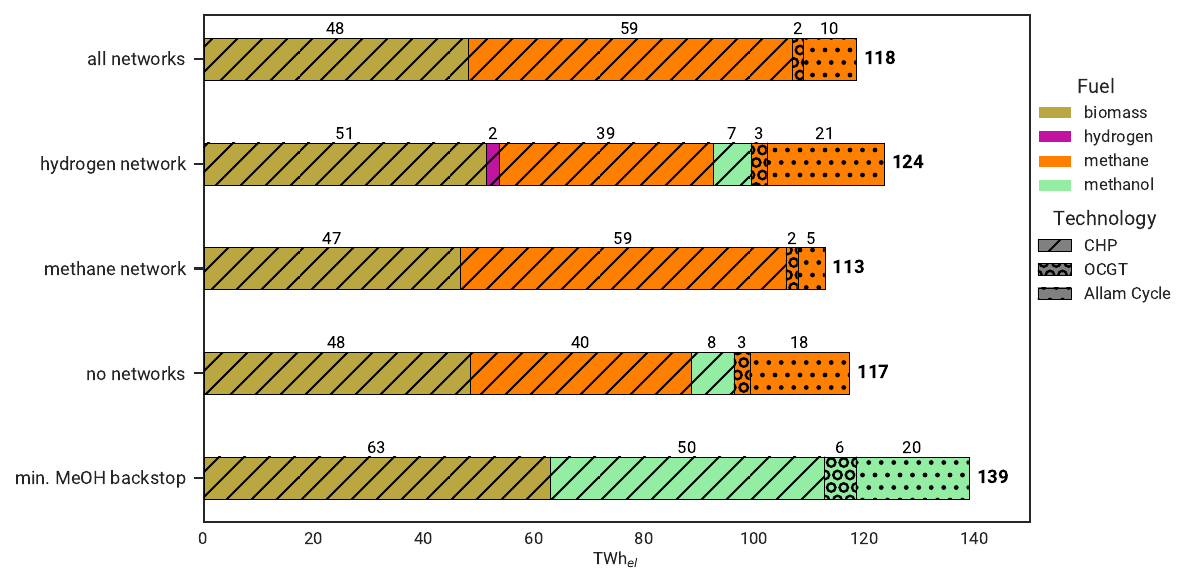}
    \caption{\textbf{Backup power and heat generation in the different scenarios for the default setting with \co network.} The colors show the different fuel types and the hatches show the different technologies used for backup generation.}
    \label{fig:co2_networks_backup}
\end{figure}
\begin{figure}[!ht]
    \centering
    \includegraphics[width=.7\textwidth]{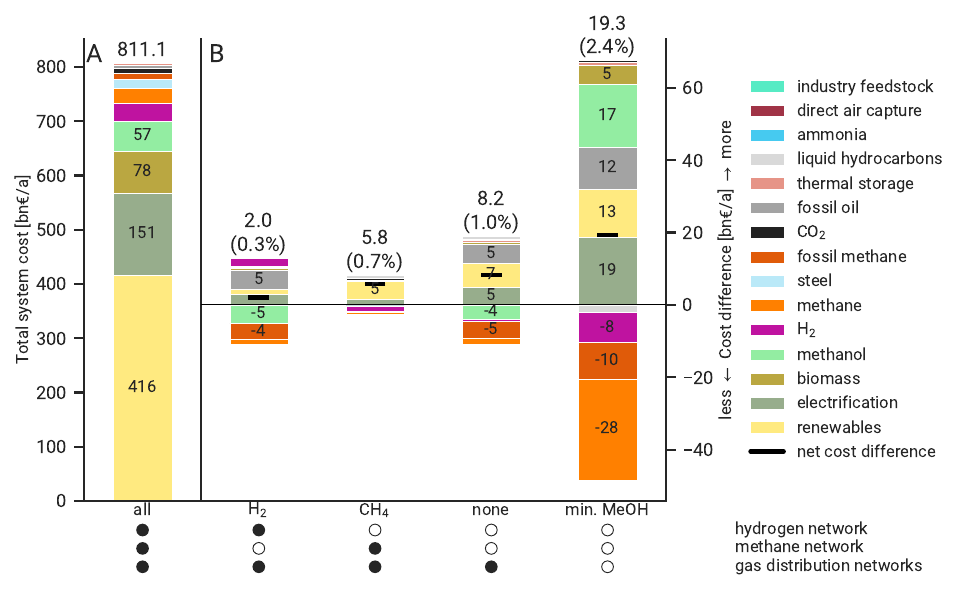}
    \caption{\textbf{Comparison of total system costs with post-discretization of H$_2$, methane and \co pipelines for the different scenarios in the default setting with \co network.} The panel A shows the absolute cost of the `all networks' scenario. The panel B shows the cost increases and decreases of the other scenarios by component relative to the `all networks' scenario. The net absolute and relative cost difference is shown at the top of each bar. A breakdown of the cost groups is given in~\cref{tab:cost_groups}.}
    \label{fig:co2_network_total_cost_discretized}
\end{figure}
\begin{figure}[!ht]
    \centering
    \includegraphics[width=0.8\textwidth]{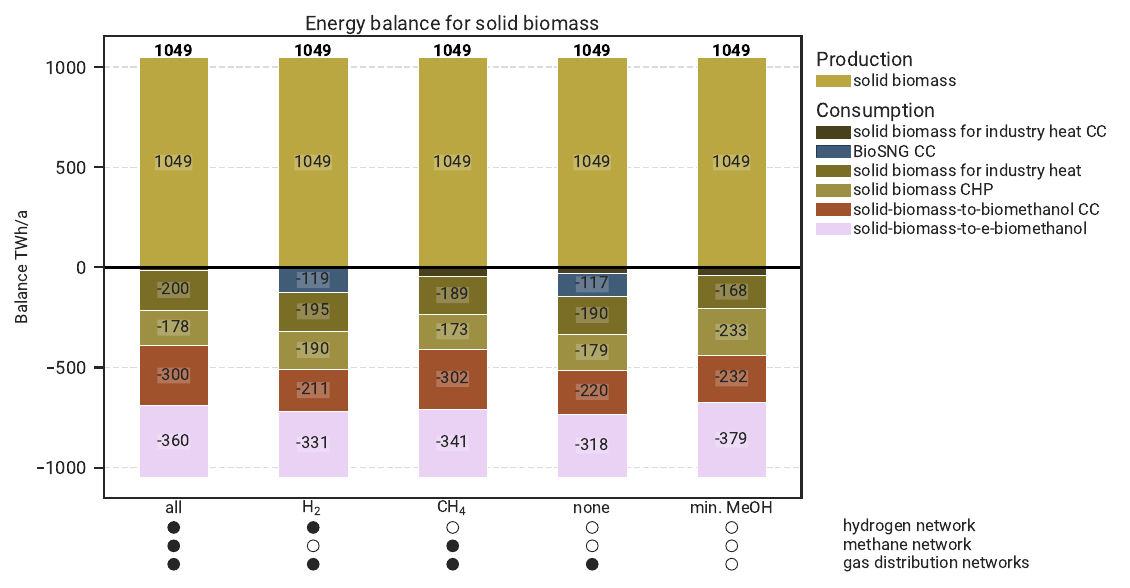}
    \caption{\textbf{Energy balances of the different scenarios for solid biomass.} The positive values show supply and the negative values show consumption. The bold number above each bar shows the total supply or consumption in TWh/a.}
    \label{fig:co2_networks_solid_biomass_balances}
\end{figure}
\begin{figure}[!ht]
    \centering
    \includegraphics[width=0.8\textwidth]{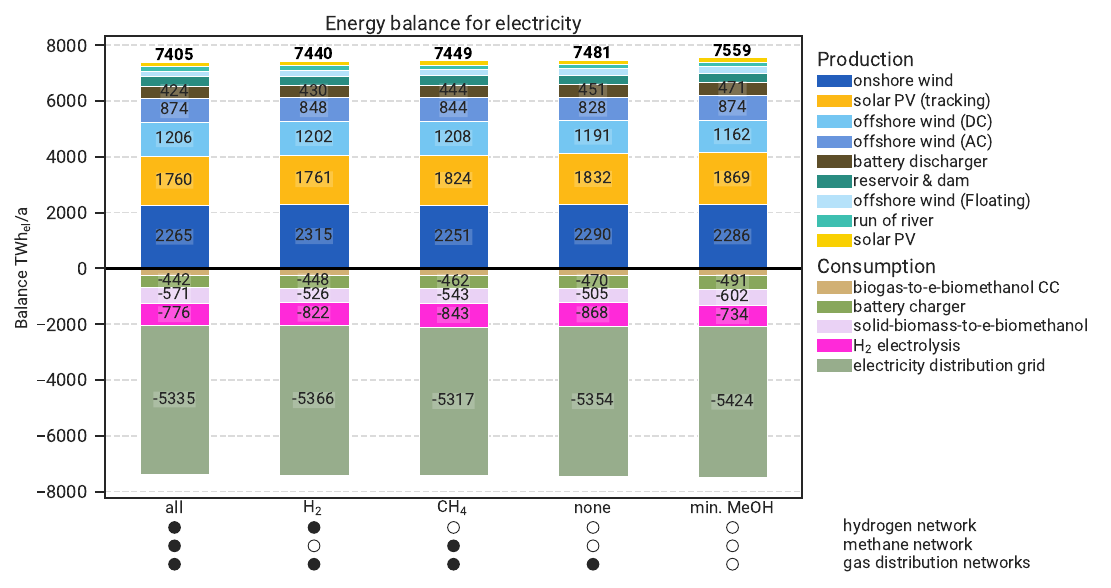}
    \caption{\textbf{Energy balances of the different scenarios for electricity.} The positive values show supply and the negative values show consumption. The bold number above each bar shows the total supply or consumption in TWh/a.}
    \label{fig:co2_networks_electricity_balances}
\end{figure}
\begin{figure}[!ht]
    \centering
    \includegraphics[width=1\textwidth]{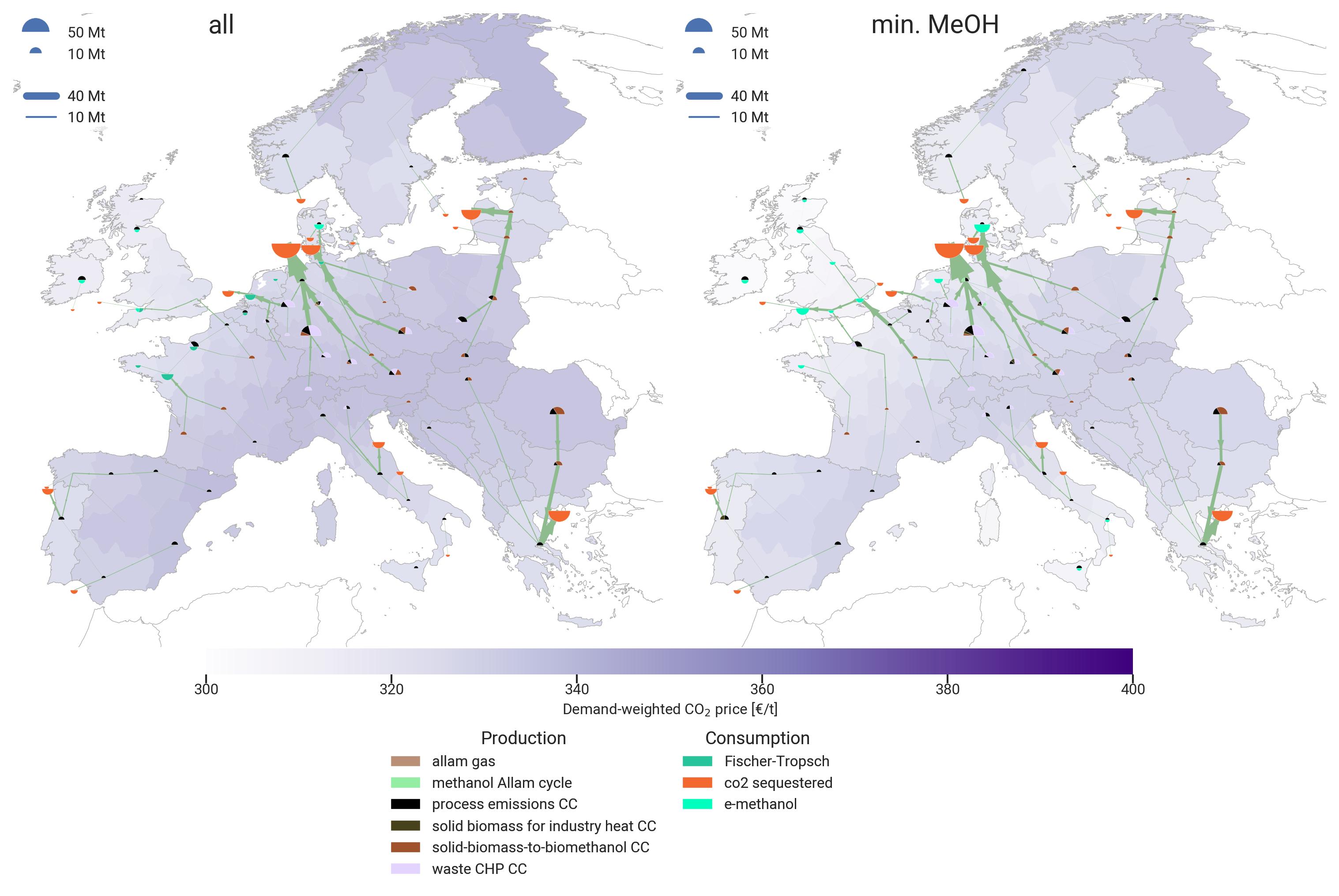}
    \caption{\textbf{Geographical patterns of carbon sources and utilization/storage in the `all networks' and `minimal methanol backstop' scenario.} Upper semi-circles show the source of \co by technology route, while the lower semi-circles show where the carbon goes. The choropleth layer shows demand-weighted average prices for the captured \co per region.}
    \label{fig:co2_stored_map}
\end{figure}
\begin{figure}[!ht]
    \centering
    \includegraphics[width=1\textwidth]{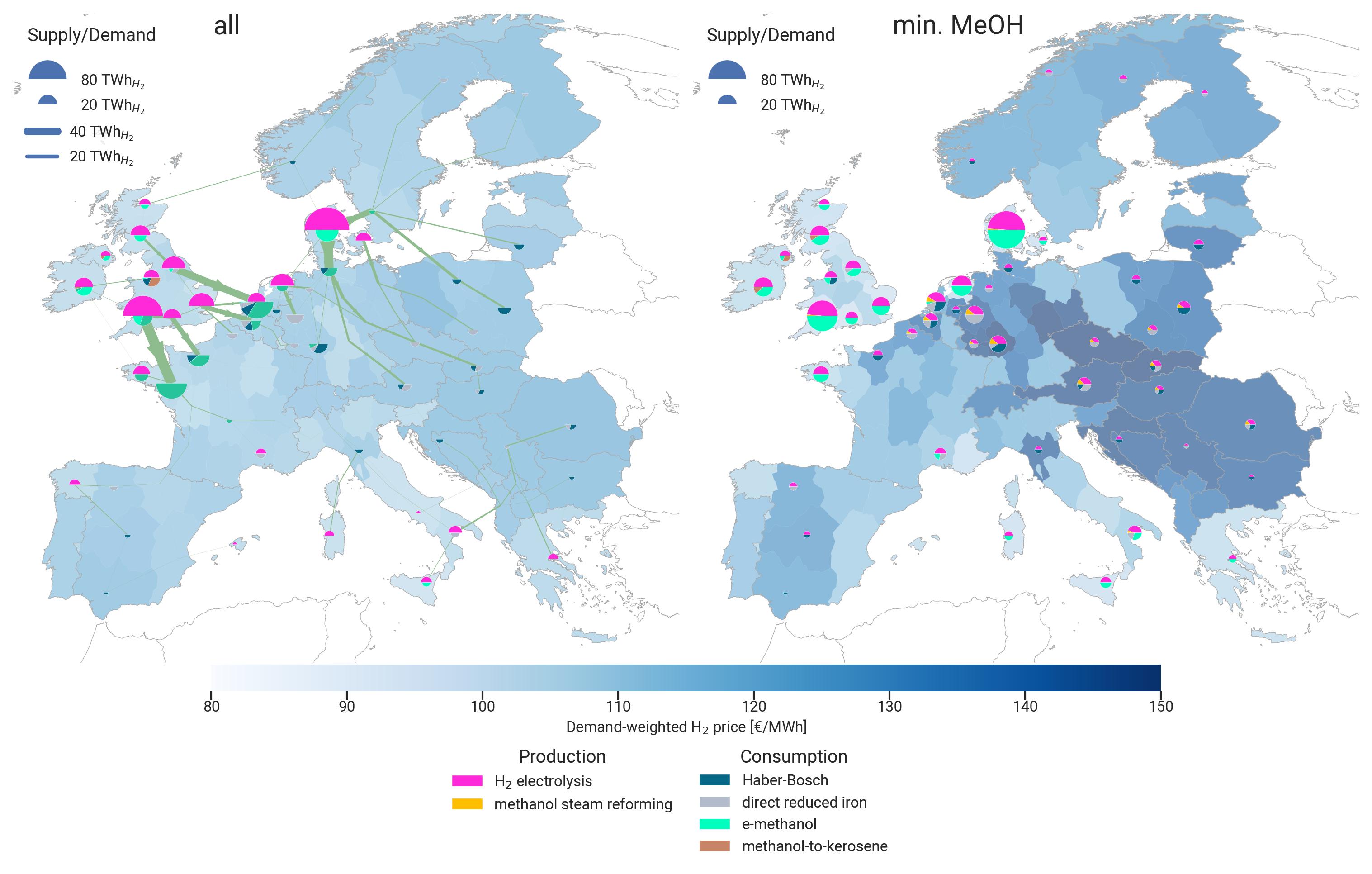}
    \caption{\textbf{Geographical patterns of hydrogen production and utilization/storage in the `all networks' and `minimal methanol backstop' scenario.} Upper semi-circles show the production of hydrogen by technology route, while the lower semi-circles show where the hydrogen goes. The choropleth layer shows demand-weighted average prices for the hydrogen per region.}
    \label{fig:hydrogen_map}
\end{figure}
\begin{figure}[!ht]
    \centering
    \includegraphics[width=.5\textwidth]{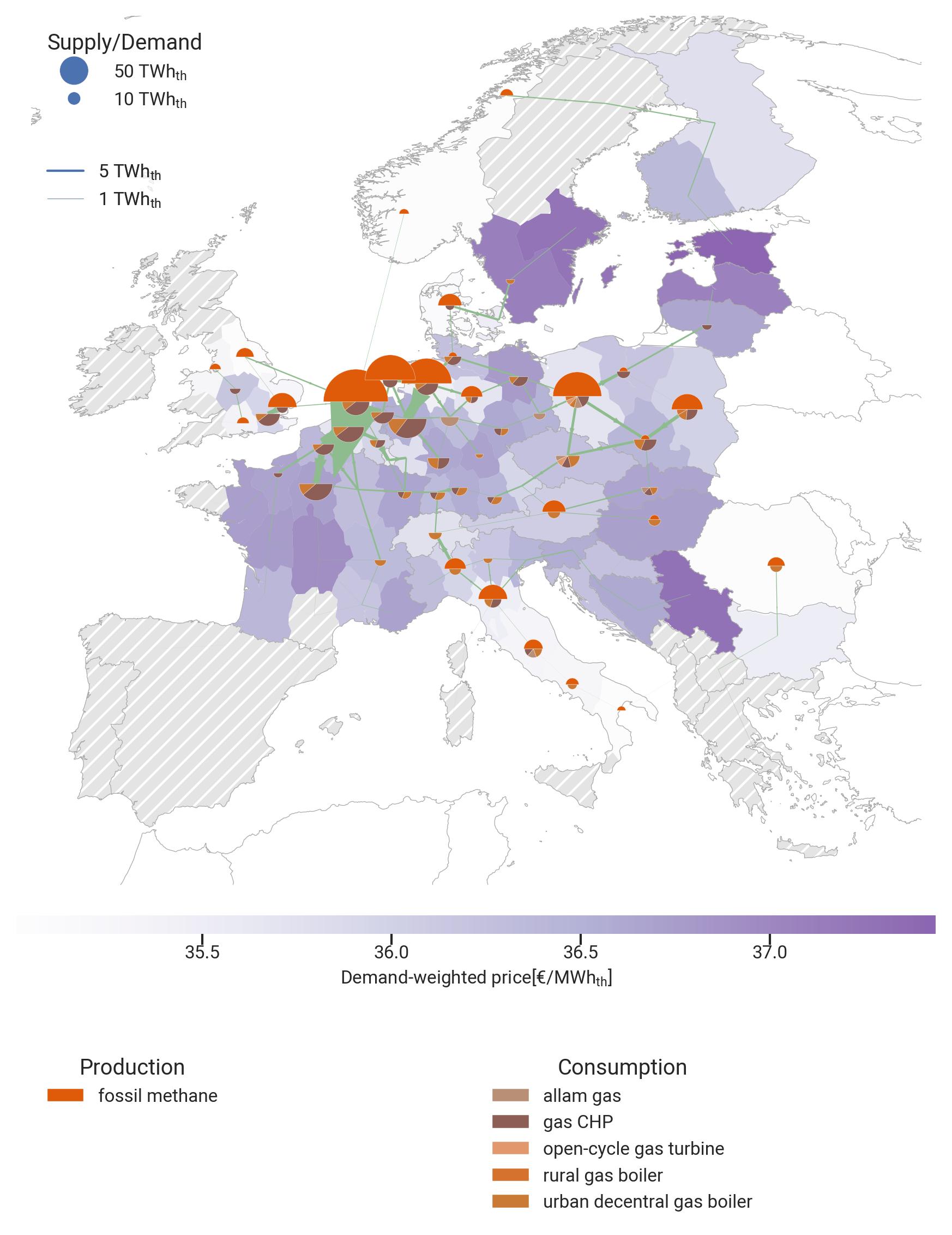}
    \caption{\textbf{Geographical patterns of methane sources and utilization/storage in the `all networks' scenario.} Upper semi-circles show where methane comes from, while the lower semi-circles show where the methane goes. The choropleth layer shows demand-weighted average prices for the methane per region.}
    \label{fig:methane_map}
\end{figure}

\begin{figure}[!ht]
    \centering
    \includegraphics[width=0.8\textwidth]{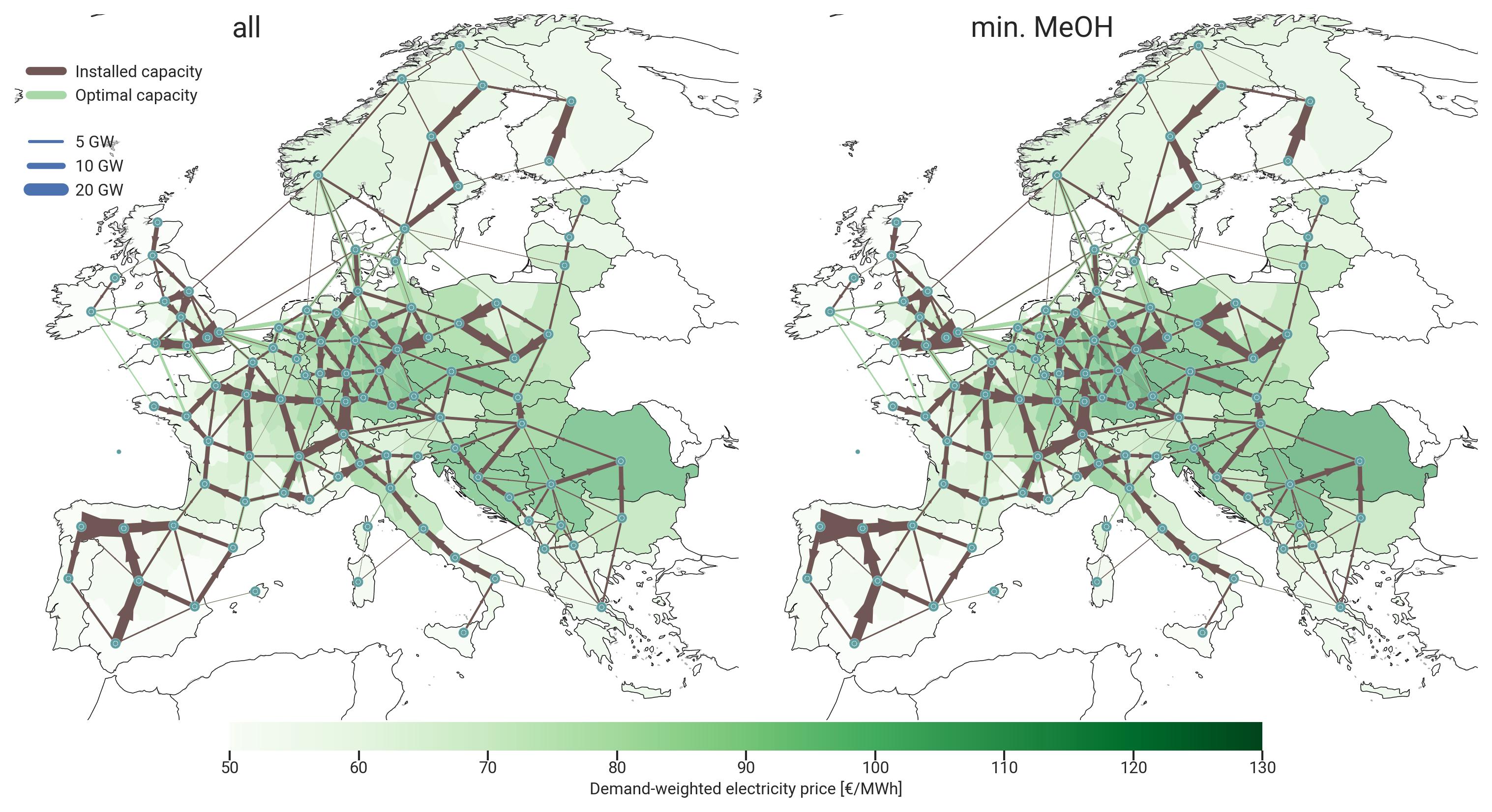}
    \caption{\textbf{Transmission line expansion map in the `all networks' and `minimal methanol backstop' scenario.}  The installed capacity is the darker color in the foreground, while the existing capacity is the lighter color in the background. The units of the lines are in GW. For both scenarios, the current transmission grid has a capacity of 265~TWkm, while the opimized transmission grid has 332~TWkm. This is an increase of 67~TWkm~(25\%).}
    \label{fig:line_expansion_map}
\end{figure}

\begin{figure}[!ht]
    \centering
    \includegraphics[trim=0 1cm 0 0cm,width=0.9\textwidth,clip=true]{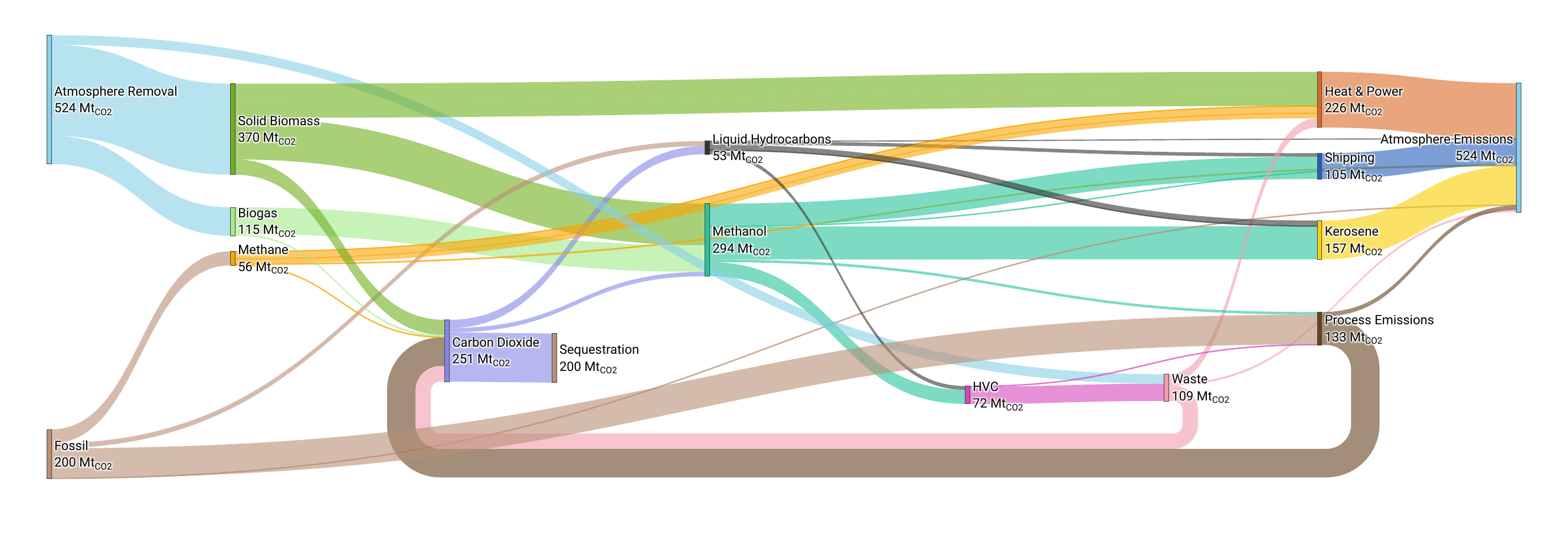}
    \caption{\textbf{Sankey diagram of carbon flows in the `all networks' scenario.}}
    \label{fig:co2_networks_carbon_flows_all}
\end{figure}

\begin{figure}[!ht]
    \centering
    \includegraphics[trim=0 1cm 0 0cm,width=0.9\textwidth,clip=true]{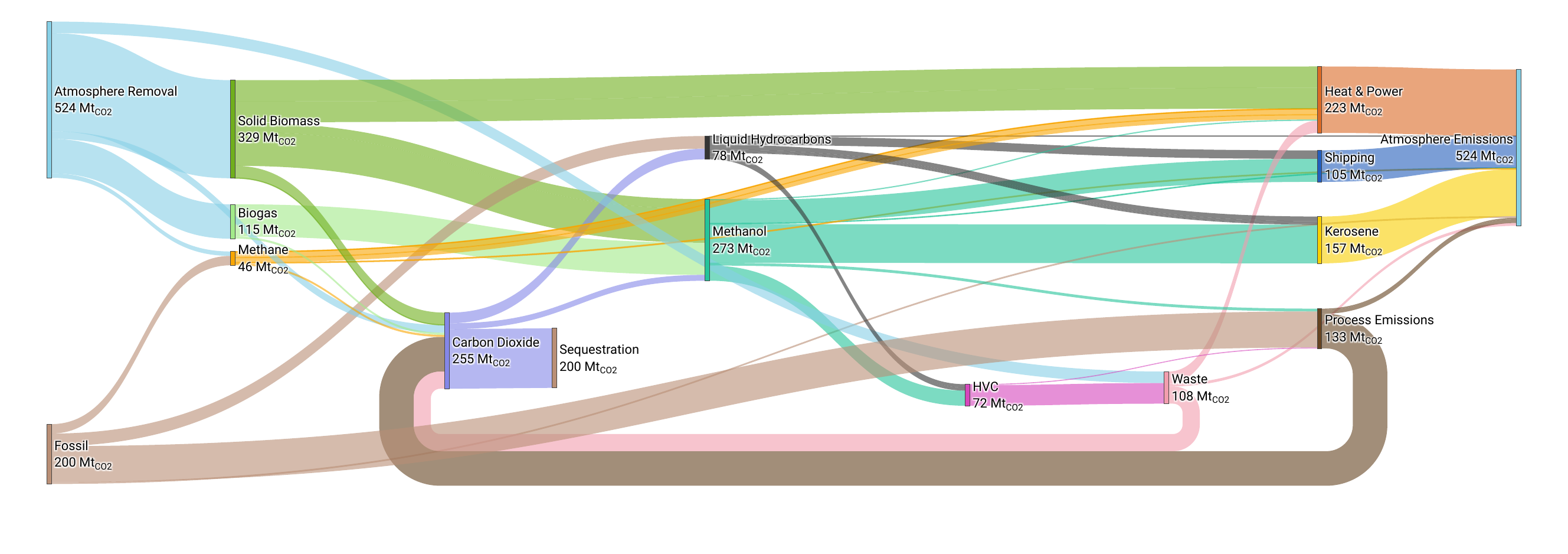}
    \caption{\textbf{Sankey diagram of carbon flows in the `hydrogen network' scenario.}}
    \label{fig:co2_networks_carbon_flows_h2}
\end{figure}

\begin{figure}[!ht]
    \centering
    \includegraphics[trim=0 1cm 0 0cm,width=0.9\textwidth,clip=true]{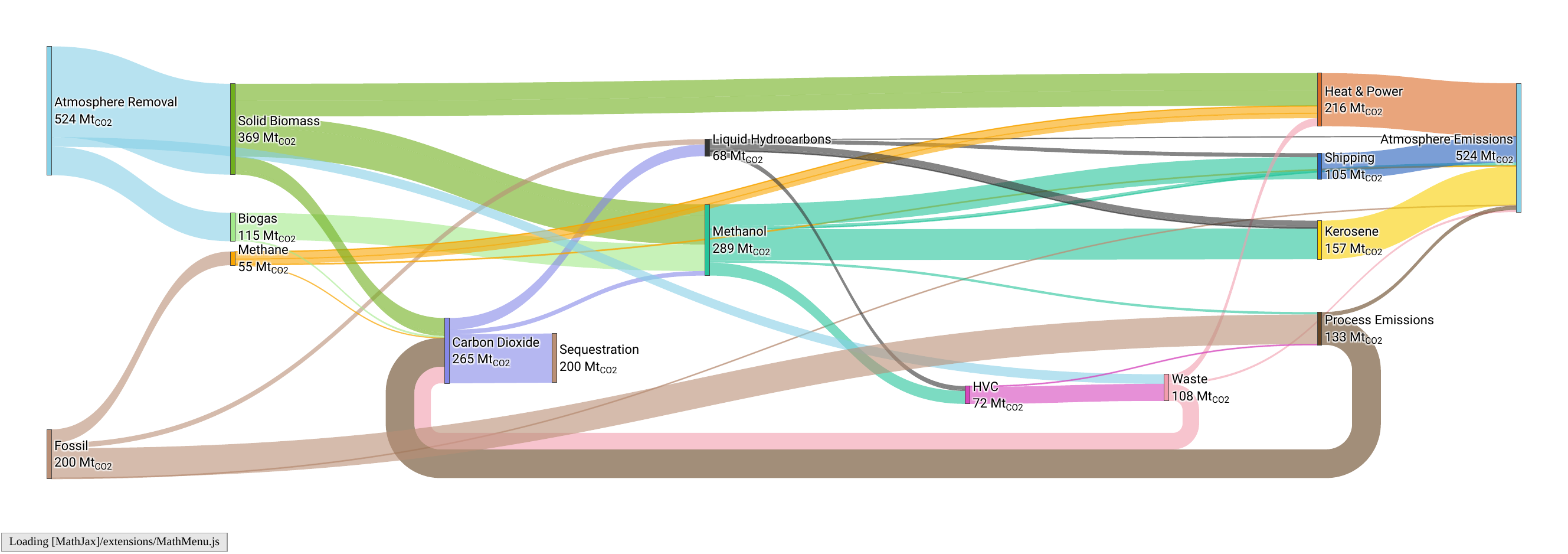}
    \caption{\textbf{Sankey diagram of carbon flows in the `methane network' scenario.}}
    \label{fig:co2_networks_carbon_flows_ch4}
\end{figure}

\begin{figure}[!ht]
    \centering
    \includegraphics[trim=0 1cm 0 0cm,width=0.9\textwidth,clip=true]{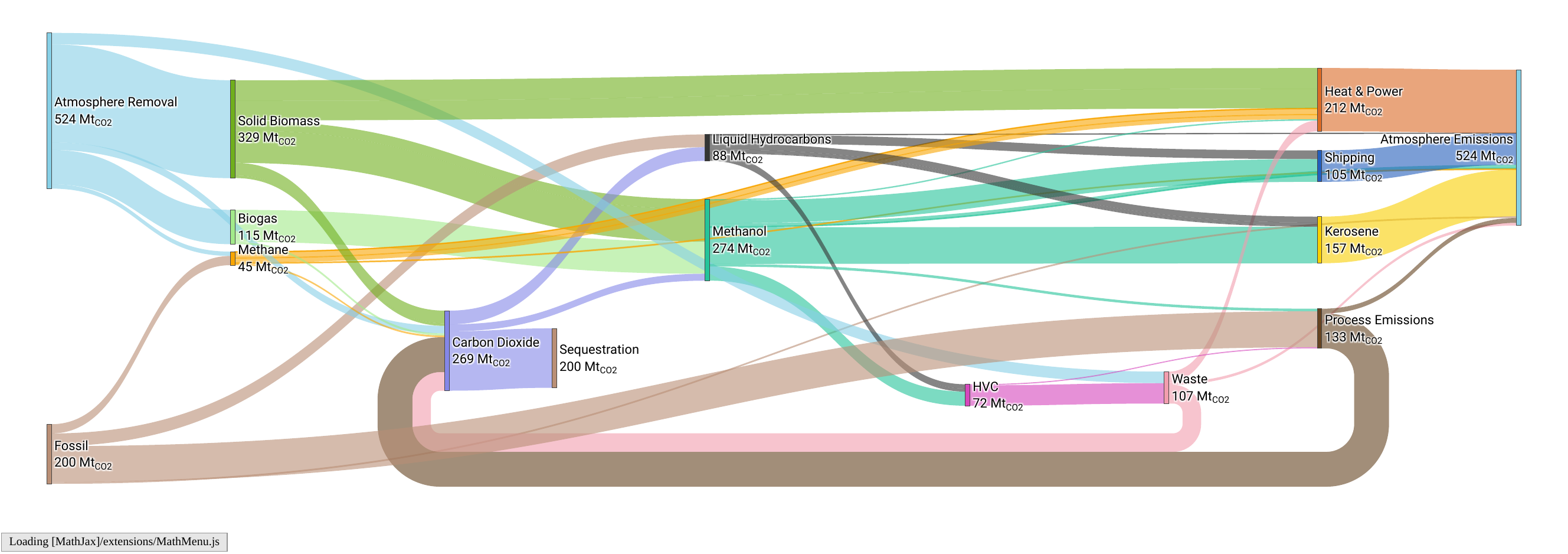}
    \caption{\textbf{Sankey diagram of carbon flows in the `no networks' scenario.}}
    \label{fig:co2_networks_carbon_flows_none}
\end{figure}

\begin{table}
    \centering
    \begin{tabular}{ll}
    \toprule
     Cost group & Members \\
    \midrule
    CO$_2$ & Direct air capture, CO$_2$ sequestration, CO$_2$ pipeline \\

    H$_2$ & Hydrogen store, Ammonia cracker, Hydrogen pipeline, Steam methane reforming (CC),\\
    & Hydrogen electrolysis, Open cycle hydrogen turbine, Hydrogen for industry heat,\\
    & Hydrogen fuel cell, Hydrogen CHP, Solid biomass to hydrogen, Methanol steam reforming\\

    ammonia & Ammonia store, Haber-Bosch \\

    biomass & Solid biomass for industry heat (CC), Biomass boiler, Biomass CHP (CC) \\

    electrification & Distribution grid, DC lines, AC lines, Battery Storage, Heat pump, Resistive heater\\
    & Electricity for industry heat\\

    fossil methane & Fossil methane \\

    fossil oil & Fossil oil \\

    green imports & Fischer-Tropsch import, Methanol import, Ammonia import, Hydrogen import \\

    methane & Bio synthetic natural gas (CC), Biogas to gas (CC), Methane pipeline, Gas CHP (CC),\\
    & Open-cycle gas turbine, Gas boiler, Gas for industry heat (CC), Sabatier, Allam cycle gas \\

    methanol & Blue/grey methanol, Solid biomass to biomethanol (CC), Biogas to e-methanol, Solid biomass\\
    & to e-methanol, Open cycle methanol turbine, Allam cycle methanol, Combined cycle\\
    & methanol turbine (CC), e-methanol, Methanol for industry heat (CC), Methanol CHP \\
    & (CC), Methanol to gasoline, Methanol to olefins/aromatics, Methanol to kerosene \\

    oil & Solid biomass to oil products (CC), Fischer-Tropsch \\

    renewables & Solar thermal, Solar PV, Utility-scale solar PV (tracking), Reservoir \& dam,\\
    & Pumped hydro storage, Onshore wind, Offshore wind, Run-of-river \\

    steel & Direct reduced iron, Electric arc furnace \\

    steel import & Hot briquetted iron import, Steel import \\

    thermal storage & Water tank \\
    \bottomrule
    \end{tabular}
    \caption{\textbf{Overview of cost groups and their members.} The cost groups are used to aggregate costs in the cost plots. (CC: with Carbon Capture)}
    \label{tab:cost_groups}
\end{table}

\FloatBarrier
\subsection*{Further discussion}
\label{sec:SI_discussion}

\textbf{Alternative liquid energy carrier concepts}

{\bf Ammonia (NH$_3$):} Green ammonia is a liquid under moderate pressure~(10 bar) or cryogenic conditions~(-33$^\circ$C), and does not require a sustainable carbon source, leading some to suggest using it as an energy carrier~\citeS{davisNetzeroEmissionsEnergy2018,serviceLiquidSunshine2018,mucci2025}. However, ammonia is much harder to handle than methanol because of its higher toxicity; it does not combust very well and combustion results in high nitrogen oxide emissions (nitrous oxide, N$_2$O is a potent greenhouse gas, while NO$_x$ are air pollutants); converting ships from fuel oil to ammonia is arduous and there can be engine slippage~\citeS{bertagniMinimizingImpactsAmmonia2023,gmf2025S}; and it requires more renewable electricity~\citeS{bertagniMinimizingImpactsAmmonia2023, kanchirallaHowVariationsShip2023}. Furthermore, it does not address the need for carbonaceous fuels in aviation or feedstocks in the chemical industry.

{\bf Liquefied methane (LCH$_4$):} Fossil gas is regularly transported in liquid form below its boiling point of -162$^\circ$C, making liquid methane LCH$_4$ a possible alternative to methanol. It can be transported by truck and be flexibly combined with the existing methane pipeline and storage infrastructure. LCH$_4$ has already been suggested as a carrier to gather biomethane from decentral sources in the absence of a gas grid~\citeS{gustafssonCentralizedDecentralizedHow2024S}. On the other hand, it requires considerable energy to liquefy; LCH$_4$ storage facilities must reliquefy the constant but slow boil-off, which is faster for smaller volumes; the cryogenic conditions result in trucking costs nearly ten times those of methanol~\citeS{diasEnergyEconomicCosts2020S}; slip emissions from engines can be substantial~\citeS{Ushakov2019}; converting ships from fuel oil to LCH$_4$ is more challenging than to methanol~\citeS{gmf2025S}. Perhaps most decisively, there is no clear long-term demand that absolutely requires methane, unlike the high volume demands for methanol in MtA/O/K processes.

\textbf{Dimethyl ether (DME)}: Direct DME production from \co can be more thermodynamically favourable than methanol, and it is less toxic. On the negative side, it is a gas at ambient conditions, and is usually liquefied under pressures of around 5 bar, much like propane.

\textbf{Liqueified hydrogen (LH$_2$)}: The temperature (-253$^\circ$C) and energy requirement for hydrogen liquefaction are even more extreme than methane, which is why gaseous hydrogen is the dominant concept for its usage as an energy carrier. Widespread use of LH$_2$ is therefore prohibitive.

\textbf{Ethanol}: The next highest alcohol, ethanol, has many of the attractive properties of methanol (liquid at ambient conditions, existing infrastructure, existing production routes from biomass). Ethanol is extensively used today as a first generation biofuel mixed with gasoline, but the production from maize and sugarcane requires extensive land use and is therefore not sustainable. The synthesis of ethanol, both from \co and H$_2$ or from residual cellulosic biomass is less mature, as are process pathways using it as a platform chemical for producing HVC and kerosene. While the use of marine vessels operating on methanol is growing, both through new orders and retrofits, ethanol is not being considered~\citeS{AlternativeFuelsInsight}.

\textbf{Fischer-Tropsch fuels}: The Fischer-Tropsch process can be used to produce kerosene for aviation, diesel for ships, and naphtha for steam crackers to produce HVC. However, the production process is less flexible than methanol~\citeS{stefanbubeFischerTropschMethanolbasedKerosene2024} and the variety of products it produces makes it less suitable as a platform solution.

\textbf{Liquid Organic Hydrogen Carriers (LOHC)}: Compared to methanol, LOHCs are more expensive and have a lower technological readiness level. Methanol would still be needed for MtO/A/K processes. On the positive side, the waste heat from power generation could be used to dehydrogenate LOHCs.

\textbf{Challenges}

Here we bring together the challenges for reaching a `minimal methanol
backstop' that have been discussed throughout the paper. We note that
these challenges apply in varying degrees to all the scenarios and
thus are to a large extent the challenges of reaching net-zero
emissions in a cost-effective way.

{\bf Electrification:} As the high cost impact of the `low
electrification' setting show, the `minimal methanol' concept relies
on a high level of electrification to minimise fuel use and thus lower
its cost premium versus the `all networks' scenario. The complete
electrification of land transport, the high share of heat pumps in
buildings, the widespread electrification of process heat in industry
as well as of domestic aviation and shipping requires concerted action
by consumers and the public.  While the total cost of ownership of
electric passenger cars is lower than their fossil-fuelled equivalents even
today, the higher upfront cost may deter some
consumers~\citeS{IEA-EV-25S}. A similar story is expected for heat
pumps~\citeS{ludererEnergiewendeKosteneffizientGestalten2025S}.  There
are technical challenges for reaching the highest temperatures in
industry~\citeS{rehfeldtDirectElectrificationIndustrial2024S} as well as
the 800~Wh/kg battery density required for short-haul
flights~\citeS{Schaefer2019S}. The increased electricity distribution
grid capacity is estimated in the model, but the feasibility of its
expansion is not assessed. The additional costs of vehicle charging
infrastructure are estimated below.

{\bf Management of biomass:} All scenarios see a simultaneous shift of
biomass feedstocks as well as of the use cases for biomass compared to the situation in 2025. Feedstocks
shift away from energy crops towards sustainable wastes and residues
that do not increase land use. Instead of being used for land
transport, baseload electricity and heating, biomass is used for
chemicals and fuels for long-haul transport. While these changes would
significantly reduce the land use of the energy system, and be
disruptive for agriculture, they would give guaranteed future revenue
streams for farmers.

{\bf Stranded assets:} For all the scenarios there is a risk of
stranded assets in some industry sectors as well as for gas
distribution due to declining use of fossil fuels. However, if managed
well, all assets can be retired before they reach their end of life or
are fully depreciated.  In the steel and basic chemicals sectors in
Europe, 90\% of blast furnaces need relining between 2025 and 2040
while 41\% of steam crackers reach their end of life before
2030~\citeS{Neuwirth2024}. This allows a complete replacement of the
plant fleet by following their reinvestment cycles~\citeS{Neuwirth2024}.  For gas networks,
which are dominated by distribution pipes to buildings for heating,
many of the assets are also decades old and expected to be fully
depreciated by 2050~\citeS{bouacida2022}. For remaining newer assets,
many European countries have recently allowed faster depreciation so
that gas grid operators can recover their costs before the pipelines
are decommissioned~\citeS{rosenow2025}. Decommissioning pipes to make
them safe, however, does represent a substantial one-off cost, estimated
to be 25 billion euros in Germany and 29 billion euros in the United
Kingdom~\citeS{rosenowElephantRoomHow2024}.

{\bf Fast scale-up of green methanol production:} Current green
methanol production in Europe amounts to around 0.6~TWh/a. Projects in
the pipeline would raise this capacity 100-fold to 60~TWh/a in 2030,
but it would still have to increase 20-fold by 2050 for our scenarios.

\textbf{Costs of exogenous infrastructure assumptions}

There are several exogenous infrastructure assumptions whose costs
are not included in the sum of total system costs, since they
are not influenceable by the optimization and stay constant across the
scenarios. We give some rough estimates from the literature for the
costs of these exogenous assumptions:
\begin{itemize}
\item {\bf Building efficiency measures}: We assume a 20\% reduction in the space heating needs by 2050 compared to 2025. Zeyen et al.\citeS{Zeyen2021S} had annualised costs for building retrofits of 144~bn€/a to achieve up to a 59\% reduction in space heating demand. Retrofitting costs increase non-linearly the deeper the renovation, so we estimate a cost of around 40~bn€/a for the annualised costs in our case. Renovations to buildings on a regular basis are necessary anyway, and provide inhabitants with more comfort, so some of these costs would be incurred regardless of the energy transition.
\item {\bf District heating expansion}: Our model assumes a substantial increase in the number of homes heated by district heating. Our reference~\citeS{Persson2011S} from 2011 for the district heating potentials had marginal costs of up to 2~\euro/GJ for the annualised capital expenditure for expanding district heating networks divided by the annual heating demand. Adjusting for inflation of 12\% between 2010 and 2020 as well as taking the average rather than marginal cost, we come to around 8~\euro/MWh. A more recent study~\citeS{FALLAHNEJAD2024122154} by some of the same authors with much more aggressive building renovation (which raises specific costs, because the demand is spread more thinly) has average costs of 32~\euro/MWh. If we take 20~\euro/MWh as a value and increase district heating demand by around 900 TWh/a, this would have annualised additional costs of 18 bn€/a.
\item {\bf Electric vehicle charging infrastructure}: Charging infrastructure studies show wide ranges, reflecting uncertainty around where vehicles will charge (public versus private charging points) and system boundaries (whether upstream distribution grid upgrades, stationary batteries and renewables are included). A 2022 study estimated investment costs of up to 168 bn€ in the EU by 2030 for private and public charging infrastructure, which annualises to around 16 bn€/a~\citeS{acea2022}.
Another study that only reaches 4.5\% share of electric cars by 2030 and 62\% by 2050 finds that `investments range between 8.7 and 16.2~bn€ in 2021–2030' and ` an additional 44.3 to 80.3~bn€ of investments in 2031–2050'~\citeS{TSIROPOULOS2022119446}, which is much lower.  A back of the envelope calculation for full electrification with investment costs per kW of 250~€/kW, average 50~kW per charging point, 260 million vehicles in 2050 and 0.1 charging points per vehicle gives 325 bn\euro total investment, which annualises to around 30~bn€/a. A study on megawatt charging for battery electric trucks (BETs) in 2025 found that a `fleet-wide public and depot infrastructure for Europe would cost about €6.6-10.8 billion (or 2.9-4.7 €cents/km) per year for BETs'~\citeS{ploetz2025}. Including cars and trucks, an annual upper bound cost of 40 bn€/a is a reasonable estimate, which is around 30 bn€/a more than the costs of maintaining petrol and diesel stations.
\item {\bf Electric vehicle capital costs}: Although the total cost of ownership of electric vehicles is expected to be lower than fossil vehicles for cars and trucks by 2030~\citeS{IEA-EV-25S}, the upfront costs are currently higher. By 2050 this situation will likely change given dramatic falls in battery costs. Given the high uncertainty and the likelihood that even the upfront costs of electric vehicles will be lower, we have not estimated these costs.
\end{itemize}
The additional transport costs from electrification (the 30~bn€/a for charging) are dwarfed by the 246~bn€/a additional costs of the `low electrification' sensitivity that largely arises from the production of e-fuels for internal combustion engines.

\textbf{Use of pipelines for \co and methanol}

The `minimal methanol backstop' scenario still assumes pipelines are used for \co transport, while methanol transport could use pipelines if volumes are large enough~\citeS{meohpipeline1985}. This might appear to undercut the benefits of moving away from pipelines for hydrogen and methane transport. However, both \co and methanol can be transported as liquids (for \co at -35 to -15$^\circ$~C and 12 to 25 bar), which gives the flexibility to use vehicles like barges, ships, trains and trucks for smaller volumes and during the scale-up. Pipelines for liquids have a much narrower diameter and lower energy requirements compared to gases. Methane and hydrogen can be liquefied but need much lower temperatures for liquefaction, which is reflected in much higher vehicular transport costs. Furthermore, we showed above in a sensitivity that the removal of the \co network has little impact on the results, just making them 2-3\% more expensive overall and barely increasing the cost premium of the `minimal methanol backstop'. Trucking costs were used for methanol transport as a worst case; pipelines would make the scenarios cheaper.

\textbf{Leakage of methane and hydrogen}

Leakages of methane and hydrogen, which exacerate global warming, were not evaluated here and could favour the minimal methanol backstop concept. Methane has a global warming potential which is 28 times worse than carbon dioxide over 100 years (GWP100). Hydrogen is not directly a greenhouse gas, but alters the duration of other greenhouse gases in the atmosphere, resulting in a GWP100 of 11.6$\pm$2.8~\citeS{Sand2023S}. Methane leakages are in principle relatively easy to detect with infrared cameras and thus regulate, but in practice leakage rates over the whole supply chain could be up to 5\% in some regions~\citeS{wellerNationalEstimateMethane2020}. In a highly electrified world much of the leakage from gas production facilities and distribution to buildings could be avoided, since methane would mainly be used in industrial facilities.

We will now show an example calculation of the impact. Assume that 1\% of methane leaks for a supply of 200~TWh/a. At GWP100 this has an impact of 4~MtCO$_2$eq/a. Priced at 400~€/t\co this would add costs of \bneuro{1.6}, which while large, is not sufficient on its own to tip the cost analysis in favour of methanol.

\textbf{Technical uncertainties}

\textbf{Allam cycle}: There is currently only one major manufacturer of Allam cycle turbines, NETPower, which is currently facing cost overruns and delays in its new projects~\citeS{anchondoNetPowerDelays2025}.
We note that very little Allam capacity is built in our simulations, which could easily be substituted by unabated gas turbines, gas turbines with carbon capture, or in the future by solid oxide fuel cells that allow easy \co capture without oxyfuel.

\textbf{Lower volumetric density of methanol for shipping}: Methanol has around half the volumetric density of fuel oil, necessitating larger tanks in ships, or more regular refuelling stops.

\textbf{Hydrogen pipeline cost uncertainty}: Since large-diameter hydrogen pipelines have rarely been built, there is some uncertainty around the cost of new and repurposed pipelines. We have used industry estimates, but the gas grid operator Gasunie has seen in 2025 that costs escalate by nearly a factor of three from 1.5~bn\euro{} to 3.8~bn\euro{}~\citeS{KabinetsaanpakKlimaatbeleid2025}.

\textbf{Modelling limitations}

The model is built as a continuous linear optimization problem, so that many locations, technologies and temporal behaviours can be explored. The downside is that many non-linearities in the operation of assets or the cost scalings with size are neglected. To address this, we make a mixture of conservative assumptions and apply post-processing. For example, pipeline capacities are post-discretised once the continuous capacities reach certain thresholds; smaller biogas plants are aggregated within a 10~km radius with local biogas pipelines so that methanol synthesis achieves sufficient economies of scale; blanket factors are applied for liquid and solid transport.
Besides, the model has perfect forsight for the whole year which means that the operation of all assets is cost optimal.
We do not model the transport of methanol and oil between regions which could lead to simultaneous production and consumption in regions far apart.
For our industry demands we consider constant demand profiles, which are based on the JRC IDEES energy balance from 2019 and may not reflect the future demand and flexibility of the industry.

\FloatBarrier
\subsection*{Further plots and tables of the sensitivity settings}

\begin{figure}[!ht]
    \centering
    \includegraphics[width=.7\textwidth]{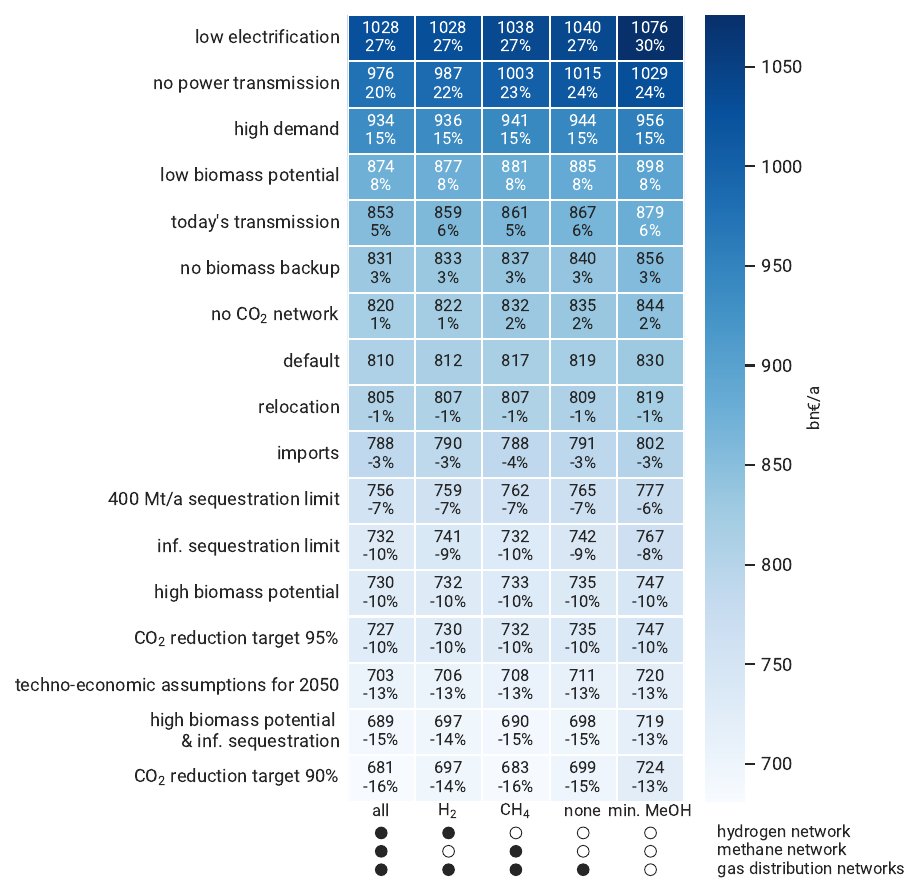}
    \caption{\textbf{Cost sensitivity for the different settings and the scenarios.} The percentage values show the relative cost increase (+) or decrease (-) compared to the primary setting with \co network.}
    \label{fig:sensitivity_total_cost}
\end{figure}

\begin{figure}[!ht]
    \centering
    \includegraphics[width=.7\textwidth]{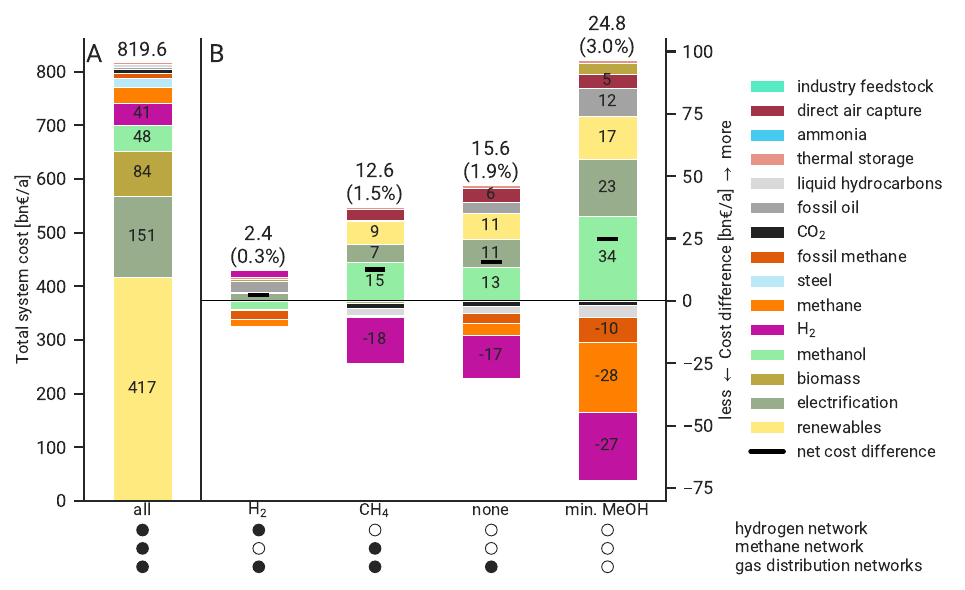}
    \caption{\textbf{Comparison of total system costs for the different scenarios in the setting without \co network.} The panel A shows the absolute cost of the `all networks' scenario. The panel B shows the cost increases and decreases of the other scenarios by component relative to the `all networks' scenario. The net absolute and relative cost difference is shown at the top of each bar. A breakdown of the cost groups is given in~\cref{tab:cost_groups}.}
    \label{fig:no_co2_network_total_cost}
\end{figure}
\begin{figure}[!ht]
    \centering
    \includegraphics[width=0.9\textwidth]{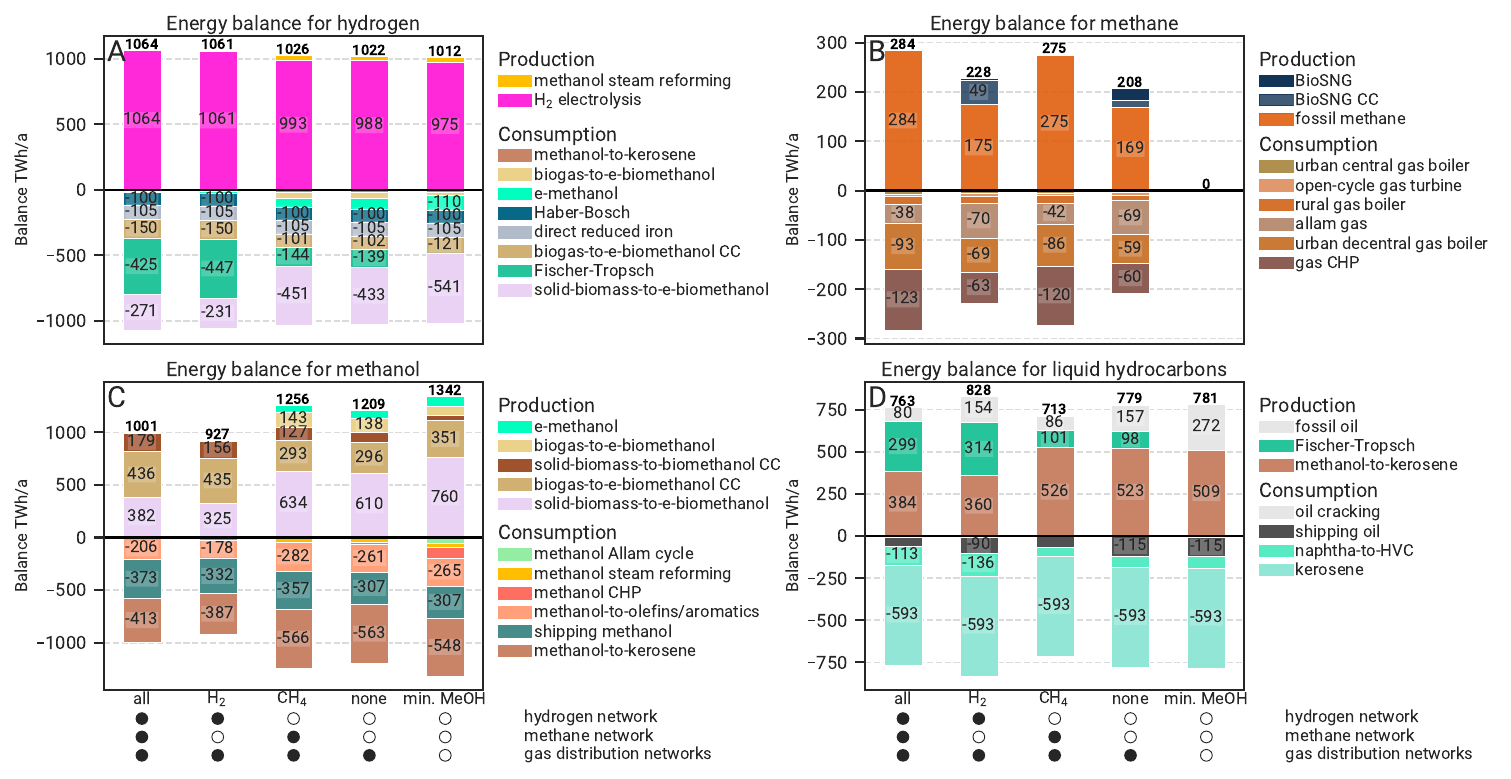}
    \caption{\textbf{Energy balances of the different scenarios for hydrogen, methane, methanol and oil in the setting without \co network.} The positive values show supply and the negative values show consumption. The bold number above each bar shows the total supply or consumption in TWh/a.}
    \label{fig:no_co2_networks_balances}
\end{figure}

\begin{figure}[!ht]
    \centering
    \includegraphics[width=.7\textwidth]{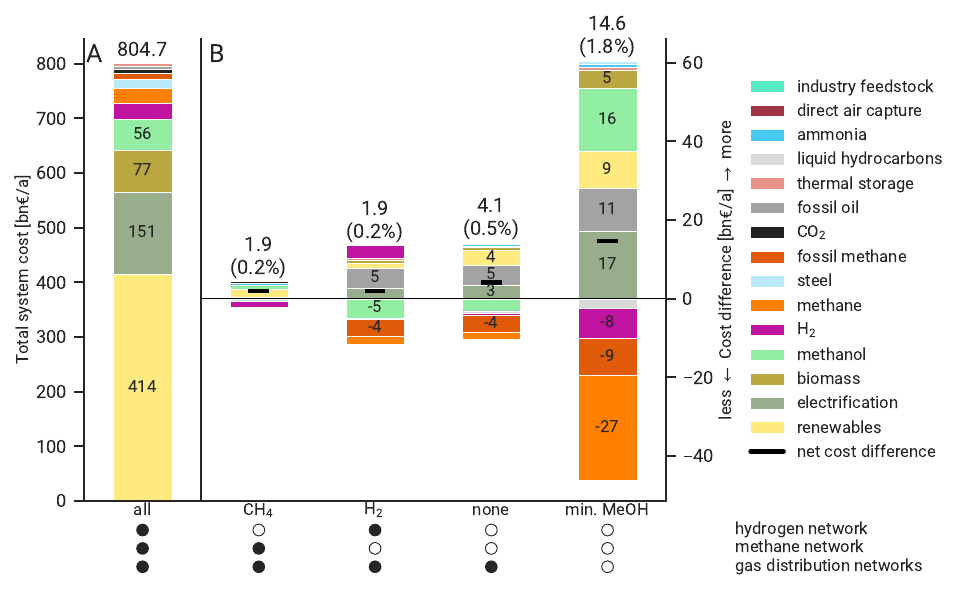}
    \caption{\textbf{Comparison of total system costs for the different scenarios in the setting with industry relocation.} The panel A shows the absolute cost of the `all networks' scenario. The panel B shows the cost increases and decreases of the other scenarios by component relative to the `all networks' scenario. The net absolute and relative cost difference is shown at the top of each bar. A breakdown of the cost groups is given in~\cref{tab:cost_groups}.}
    \label{fig:relocation_total_cost}
\end{figure}
\begin{figure}[!ht]
    \centering
    \includegraphics[width=0.9\textwidth]{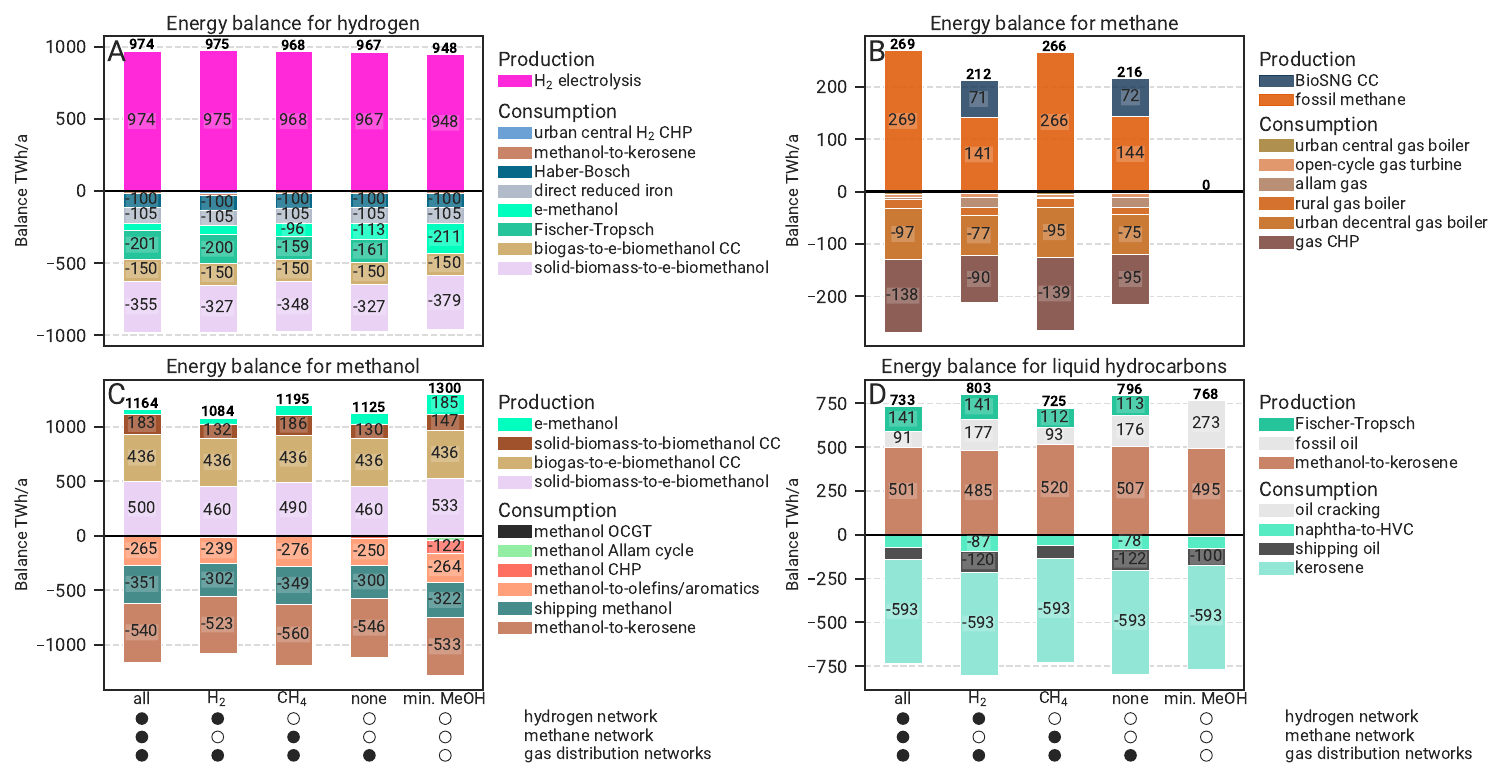}
    \caption{\textbf{Energy balances of the different scenarios for hydrogen, methane, methanol and oil in the setting with industry relocation.} The positive values show supply and the negative values show consumption. The bold number above each bar shows the total supply or consumption in TWh/a.}
    \label{fig:relocation_balances}
\end{figure}

\begin{figure}[!ht]
    \centering
    \includegraphics[width=.7\textwidth]{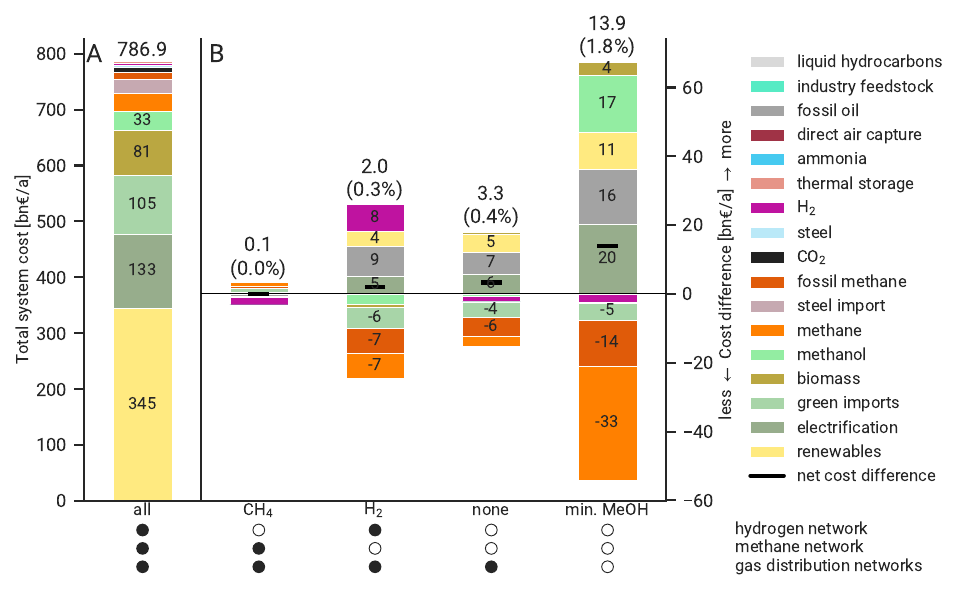}
    \caption{\textbf{Comparison of total system costs for the different scenarios in the setting with green imports and industry relocation.} The panel A shows the absolute cost of the `all networks' scenario. The panel B shows the cost increases and decreases of the other scenarios by component relative to the `all networks' scenario. The net absolute and relative cost difference is shown at the top of each bar. A breakdown of the cost groups is given in~\cref{tab:cost_groups}.}
    \label{fig:imports_total_cost}
\end{figure}
\begin{figure}[!ht]
    \centering
    \includegraphics[width=0.9\textwidth]{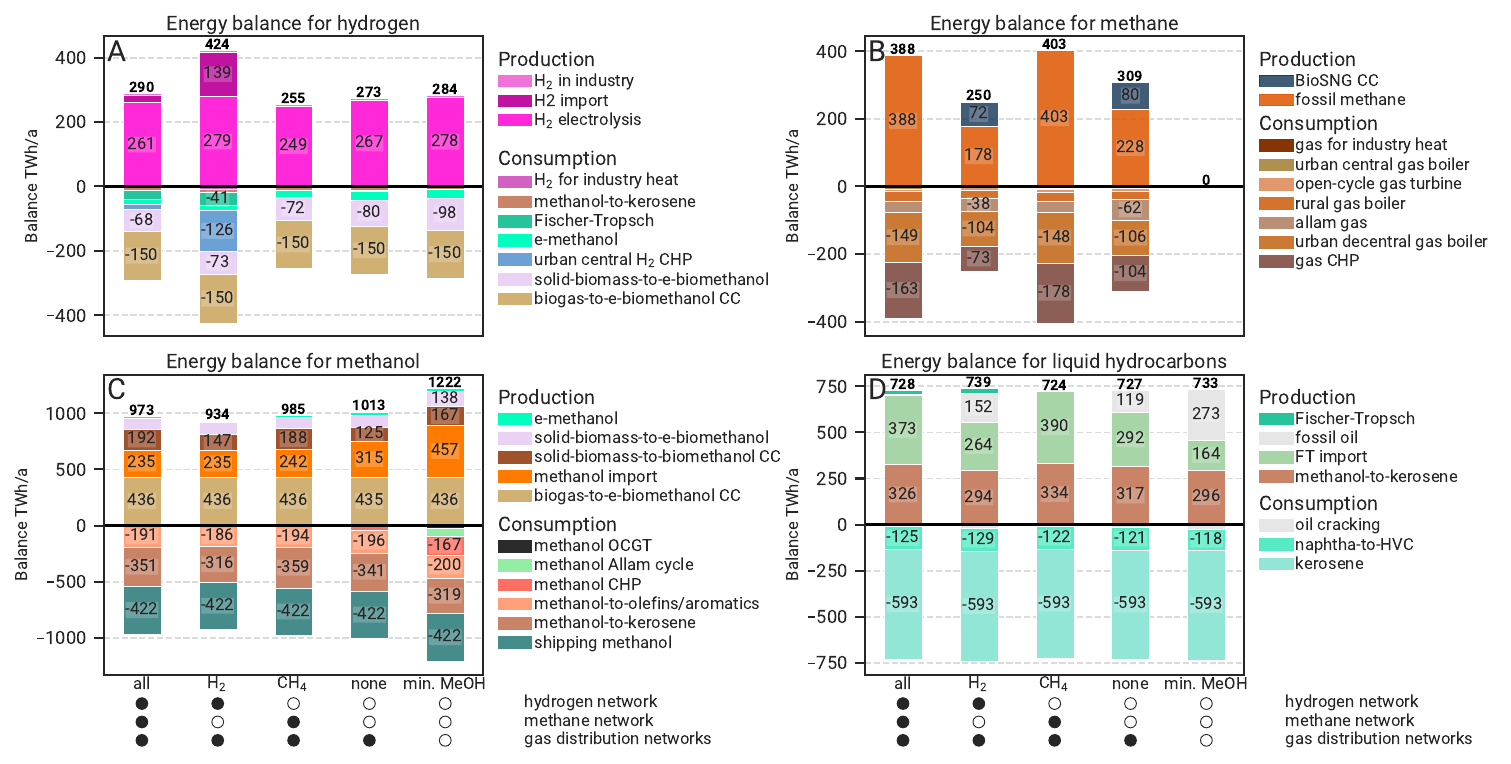}
    \caption{\textbf{Energy balances of the different scenarios for hydrogen, methane, methanol and oil in the setting with green imports and industry relocation.} The positive values show supply and the negative values show consumption. The bold number above each bar shows the total supply or consumption in TWh/a.}
    \label{fig:imports_balances}
\end{figure}

\begin{figure}[!ht]
    \centering
    \includegraphics[width=.7\textwidth]{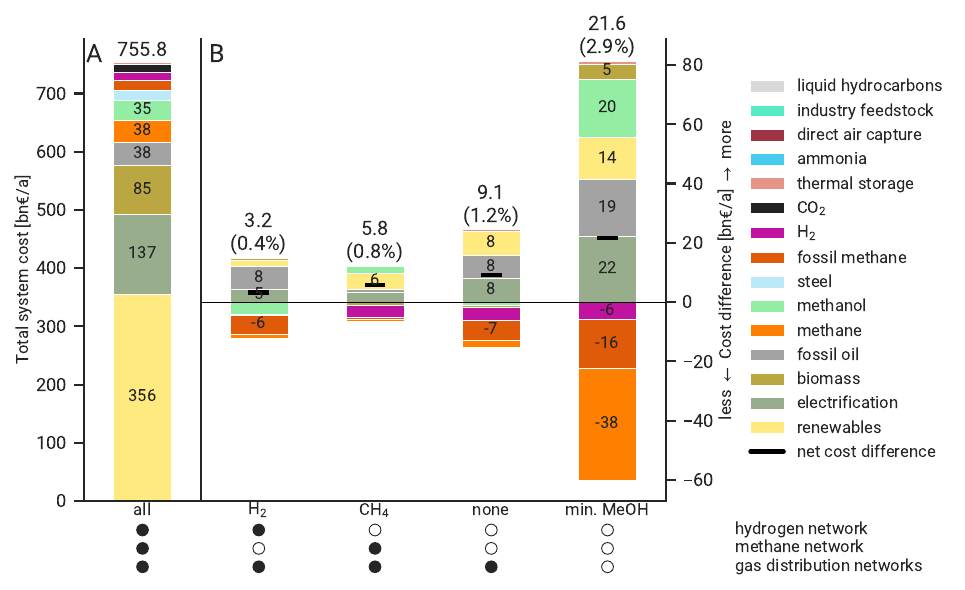}
    \caption{\textbf{Comparison of total system costs for the different scenarios in the setting with sequestration limit of 400~Mt$_\text{CO2}$.} The panel A shows the absolute cost of the `all networks' scenario. The panel B shows the cost increases and decreases of the other scenarios by component relative to the `all networks' scenario. The net absolute and relative cost difference is shown at the top of each bar. A breakdown of the cost groups is given in~\cref{tab:cost_groups}.}
    \label{fig:seq_400_total_cost}
\end{figure}
\begin{figure}[!ht]
    \centering
    \includegraphics[width=0.9\textwidth]{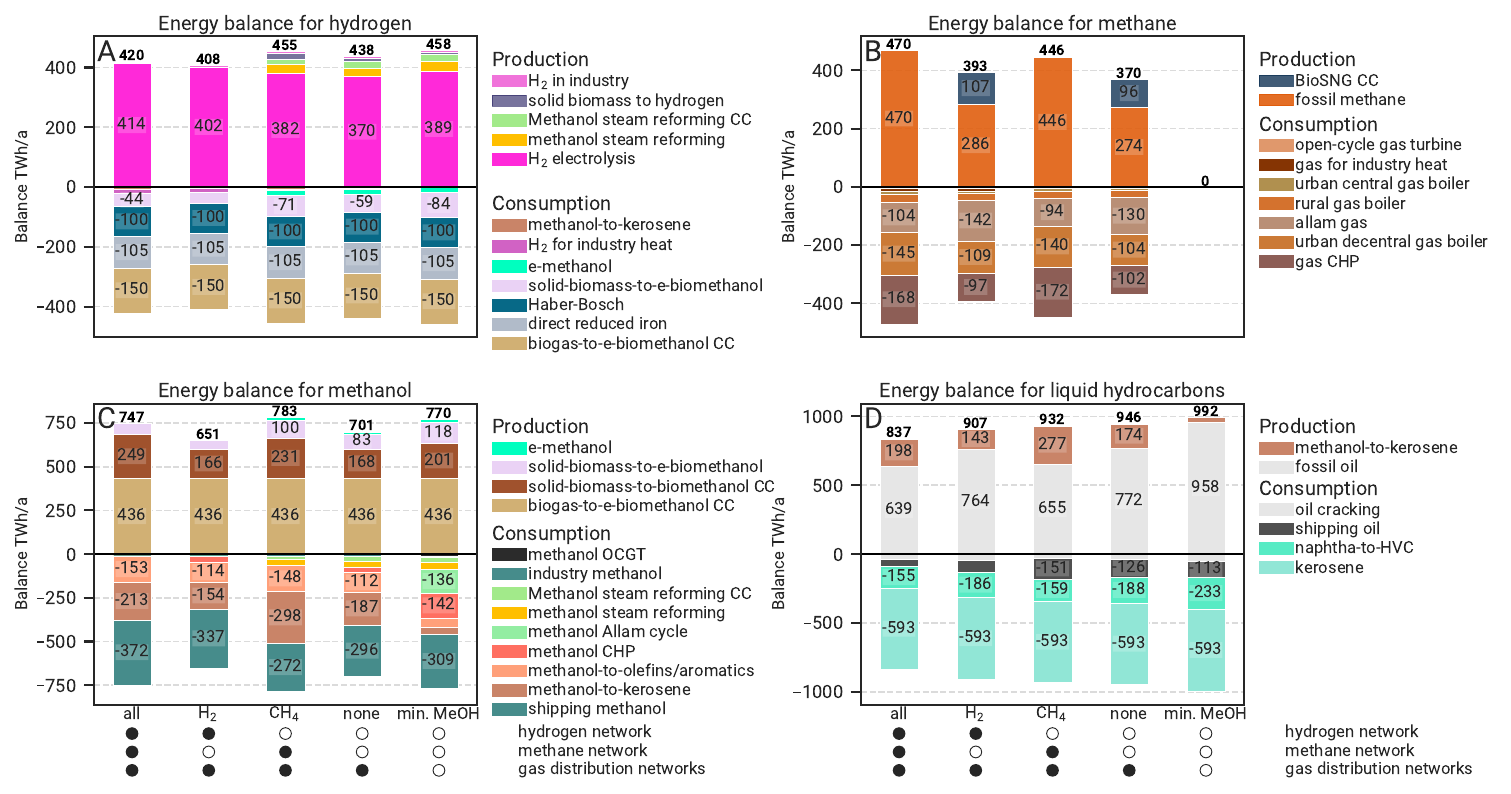}
    \caption{\textbf{Energy balances of the different scenarios for hydrogen, methane, methanol and oil in the setting with sequestration limit of 400~Mt$_\text{CO2}$.} The positive values show supply and the negative values show consumption. The bold number above each bar shows the total supply or consumption in TWh/a.}
    \label{fig:seq_400_balances}
\end{figure}

\begin{figure}[!ht]
    \centering
    \includegraphics[width=.7\textwidth]{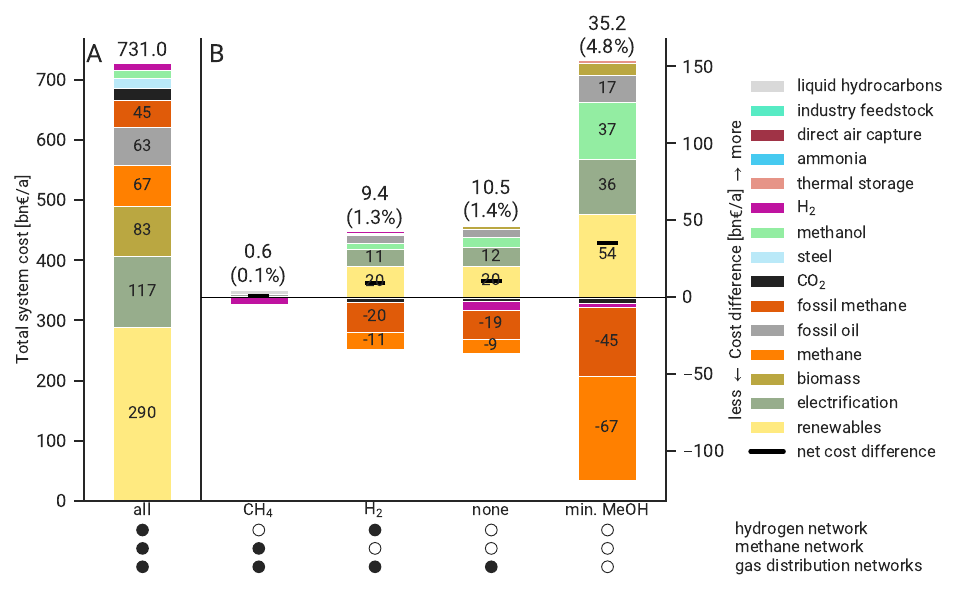}
    \caption{\textbf{Comparison of total system costs for the different scenarios in the setting infinite sequestration limit.} The panel A shows the absolute cost of the `all networks' scenario. The panel B shows the cost increases and decreases of the other scenarios by component relative to the `all networks' scenario. The net absolute and relative cost difference is shown at the top of each bar. A breakdown of the cost groups is given in~\cref{tab:cost_groups}.}
    \label{fig:inf_seq_total_cost}
\end{figure}
\begin{figure}[!ht]
    \centering
    \includegraphics[width=0.9\textwidth]{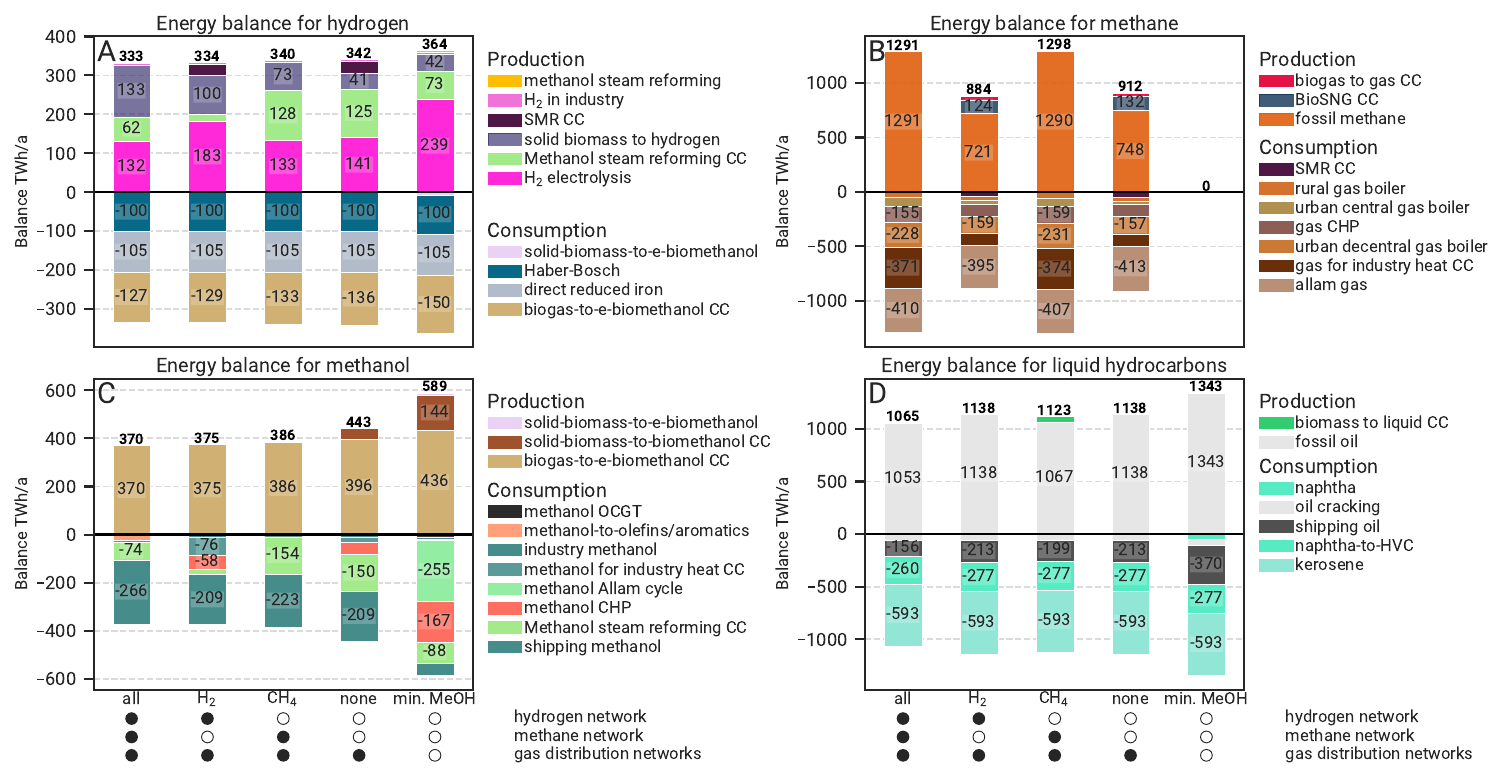}
    \caption{\textbf{Energy balances of the different scenarios for hydrogen, methane, methanol and oil in the setting with infinite sequestration limit} The positive values show supply and the negative values show consumption. The bold number above each bar shows the total supply or consumption in TWh/a.}
    \label{fig:inf_seq_balances}
\end{figure}

\begin{figure}[!ht]
    \centering
    \includegraphics[width=.7\textwidth]{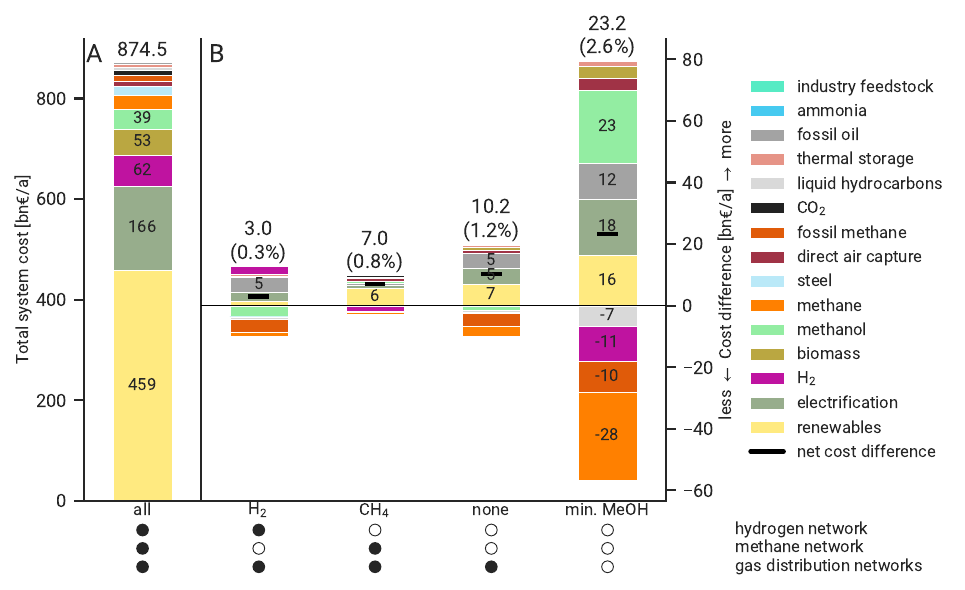}
    \caption{\textbf{Comparison of total system costs for the different scenarios in the setting with low biomass potential of 840~TWh/a.} The panel A shows the absolute cost of the `all networks' scenario. The panel B shows the cost increases and decreases of the other scenarios by component relative to the `all networks' scenario. The net absolute and relative cost difference is shown at the top of each bar. A breakdown of the cost groups is given in~\cref{tab:cost_groups}.}
    \label{fig:low_bio_total_cost}
\end{figure}
\begin{figure}[!ht]
    \centering
    \includegraphics[width=0.9\textwidth]{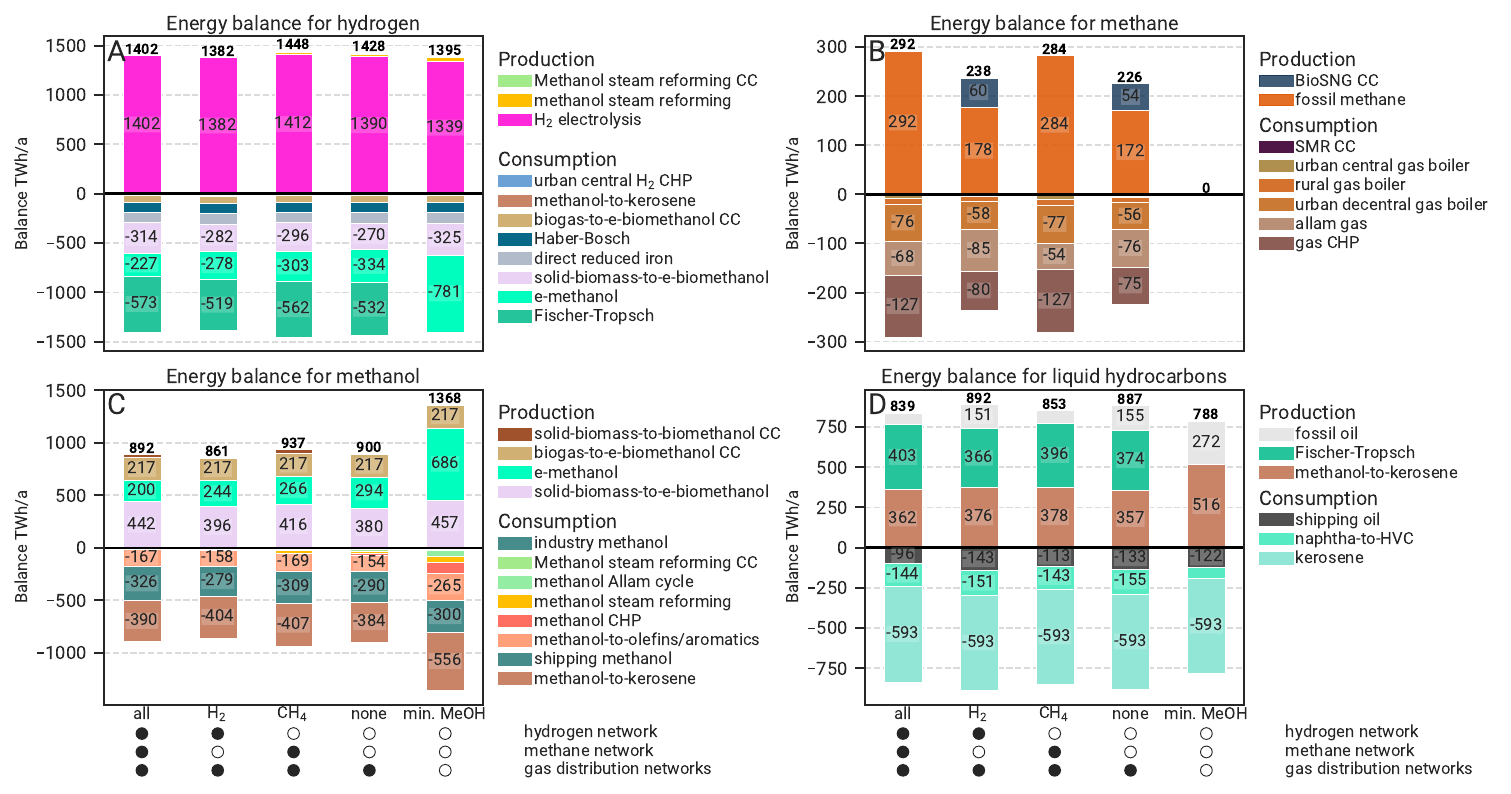}
    \caption{\textbf{Energy balances of the different scenarios for hydrogen, methane, methanol and oil in the setting with low biomass potential of 840~TWh/a.} The positive values show supply and the negative values show consumption. The bold number above each bar shows the total supply or consumption in TWh/a.}
    \label{fig:low_bio_balances}
\end{figure}

\begin{figure}[!ht]
    \centering
    \includegraphics[width=.7\textwidth]{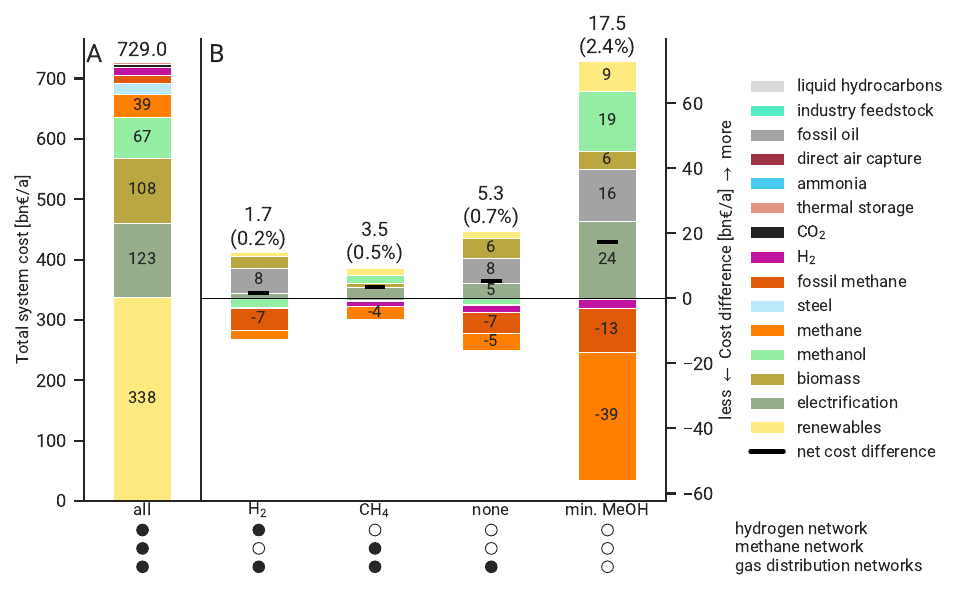}
    \caption{\textbf{Comparison of total system costs for the different scenarios in the setting with high biomass potential of 3300~TWh/a.} The panel A shows the absolute cost of the `all networks' scenario. The panel B shows the cost increases and decreases of the other scenarios by component relative to the `all networks' scenario. The net absolute and relative cost difference is shown at the top of each bar. A breakdown of the cost groups is given in~\cref{tab:cost_groups}.}
    \label{fig:high_bio_total_cost}
\end{figure}
\begin{figure}[!ht]
    \centering
    \includegraphics[width=0.9\textwidth]{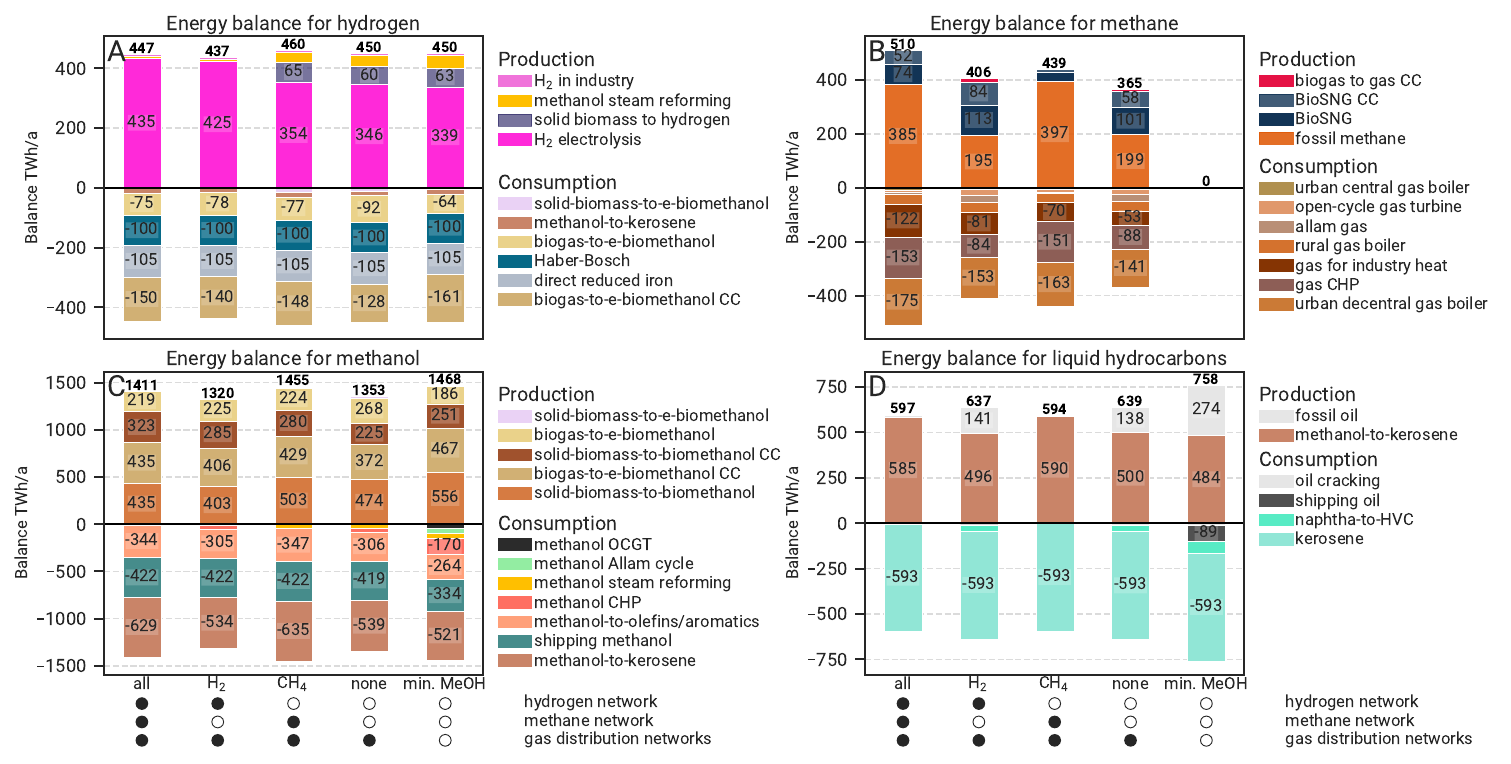}
    \caption{\textbf{Energy balances of the different scenarios for hydrogen, methane, methanol and oil in the setting with high biomass potential of 3300~TWh/a.} The positive values show supply and the negative values show consumption. The bold number above each bar shows the total supply or consumption in TWh/a.}
    \label{fig:high_bio_balances}
\end{figure}

\begin{figure}[!ht]
    \centering
    \includegraphics[width=.7\textwidth]{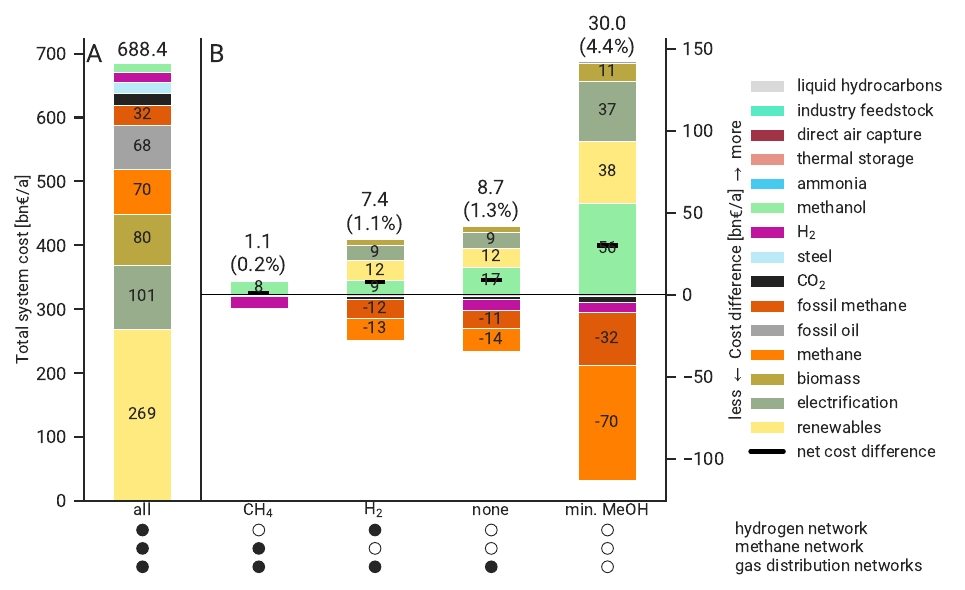}
    \caption{\textbf{Comparison of total system costs for the different scenarios in the setting with high biomass potential of 3300~TWh/a and infinite sequestration.} The panel A shows the absolute cost of the `all networks' scenario. The panel B shows the cost increases and decreases of the other scenarios by component relative to the `all networks' scenario. The net absolute and relative cost difference is shown at the top of each bar. A breakdown of the cost groups is given in~\cref{tab:cost_groups}.}
    \label{fig:high_bio_inf_seq_total_cost}
\end{figure}
\begin{figure}[!ht]
    \centering
    \includegraphics[width=0.9\textwidth]{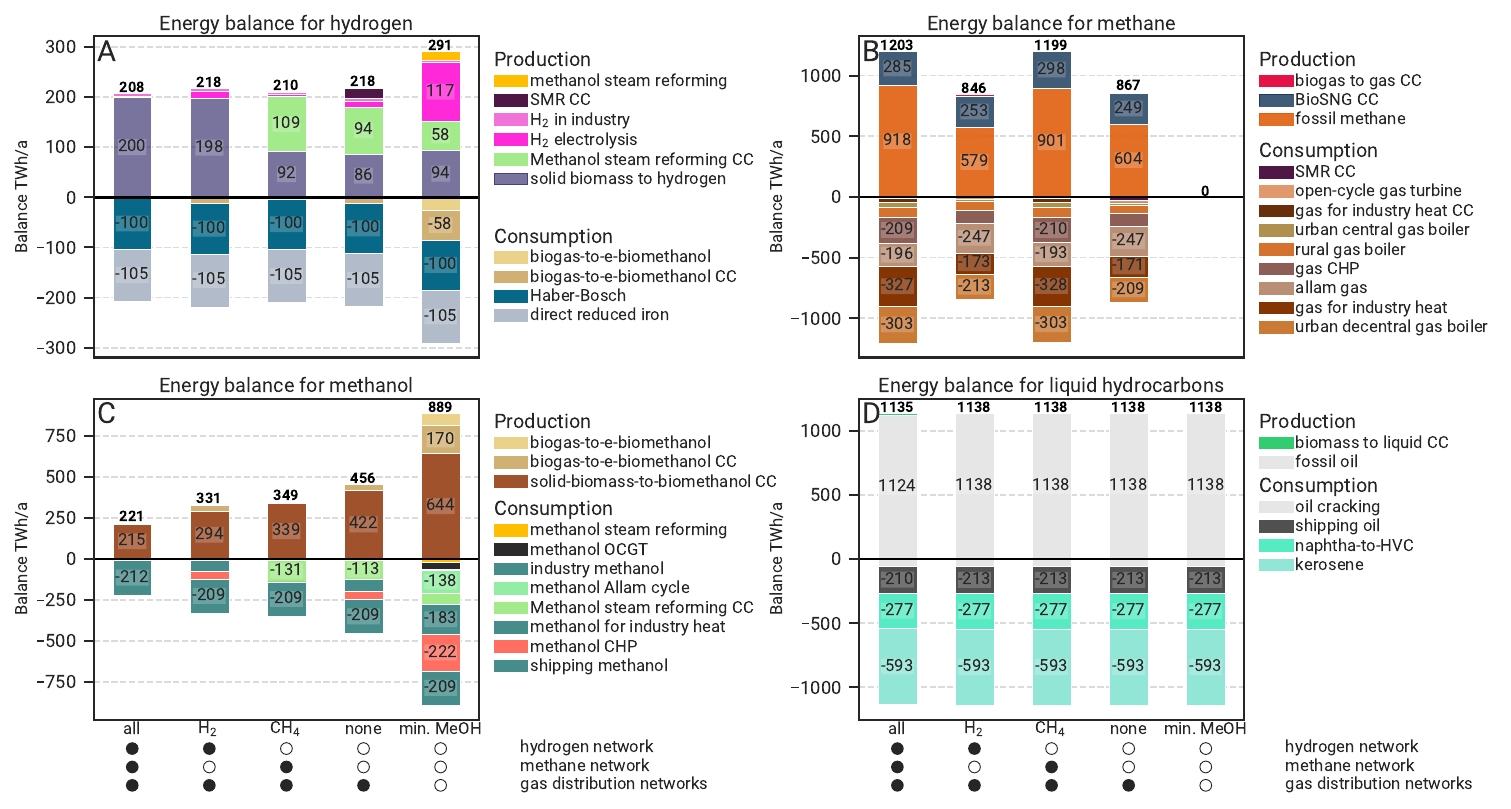}
    \caption{\textbf{Energy balances of the different scenarios for hydrogen, methane, methanol and oil in the setting with high biomass potential of 3300~TWh/a and infinite sequestration.} The positive values show supply and the negative values show consumption. The bold number above each bar shows the total supply or consumption in TWh/a.}
    \label{fig:high_bio_inf_seq_balances}
\end{figure}

\begin{figure}[!ht]
    \centering
    \includegraphics[width=.7\textwidth]{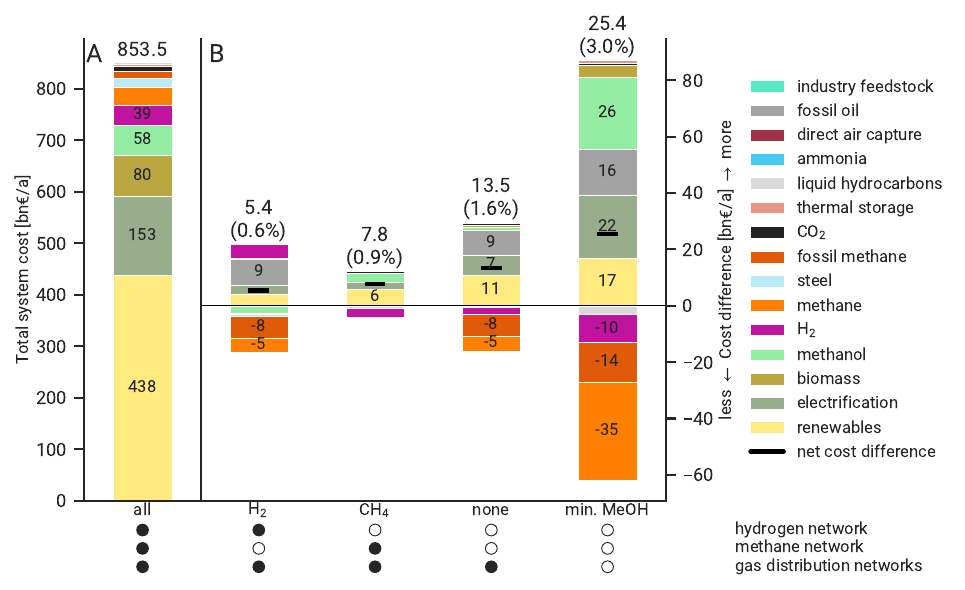}
    \caption{\textbf{Comparison of total system costs for the different scenarios in the setting using today's power transmission capacities.} The panel A shows the absolute cost of the `all networks' scenario. The panel B shows the cost increases and decreases of the other scenarios by component relative to the `all networks' scenario. The net absolute and relative cost difference is shown at the top of each bar. A breakdown of the cost groups is given in~\cref{tab:cost_groups}.}
    \label{fig:todays_transmission_total_cost}
\end{figure}
\begin{figure}[!ht]
    \centering
    \includegraphics[width=0.9\textwidth]{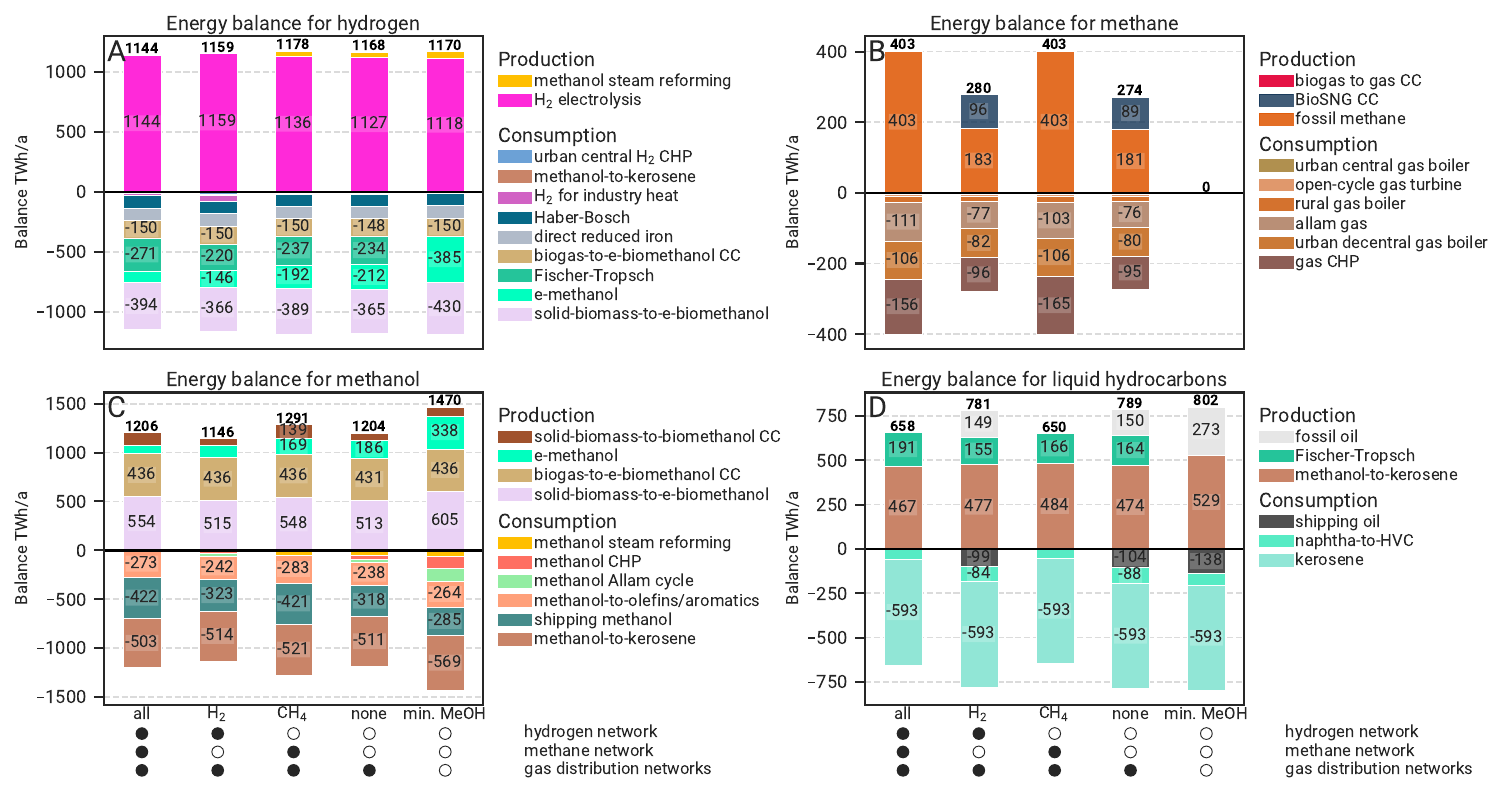}
    \caption{\textbf{Energy balances of the different scenarios for hydrogen, methane, methanol and oil in the setting using today's power transmission capacities.} The positive values show supply and the negative values show consumption. The bold number above each bar shows the total supply or consumption in TWh/a.}
    \label{fig:todays_transmission_balances}
\end{figure}

\begin{figure}[!ht]
    \centering
    \includegraphics[width=.7\textwidth]{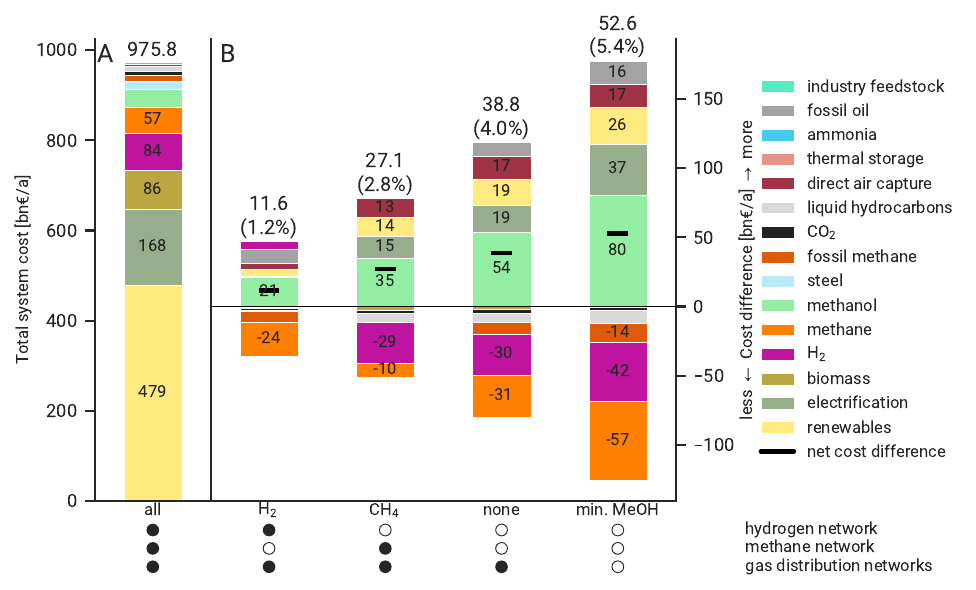}
    \caption{\textbf{Comparison of total system costs for the different scenarios in the setting without power transmission.} The panel A shows the absolute cost of the `all networks' scenario. The panel B shows the cost increases and decreases of the other scenarios by component relative to the `all networks' scenario. The net absolute and relative cost difference is shown at the top of each bar. A breakdown of the cost groups is given in~\cref{tab:cost_groups}.}
    \label{fig:no_grid_total_cost}
\end{figure}
\begin{figure}[!ht]
    \centering
    \includegraphics[width=0.9\textwidth]{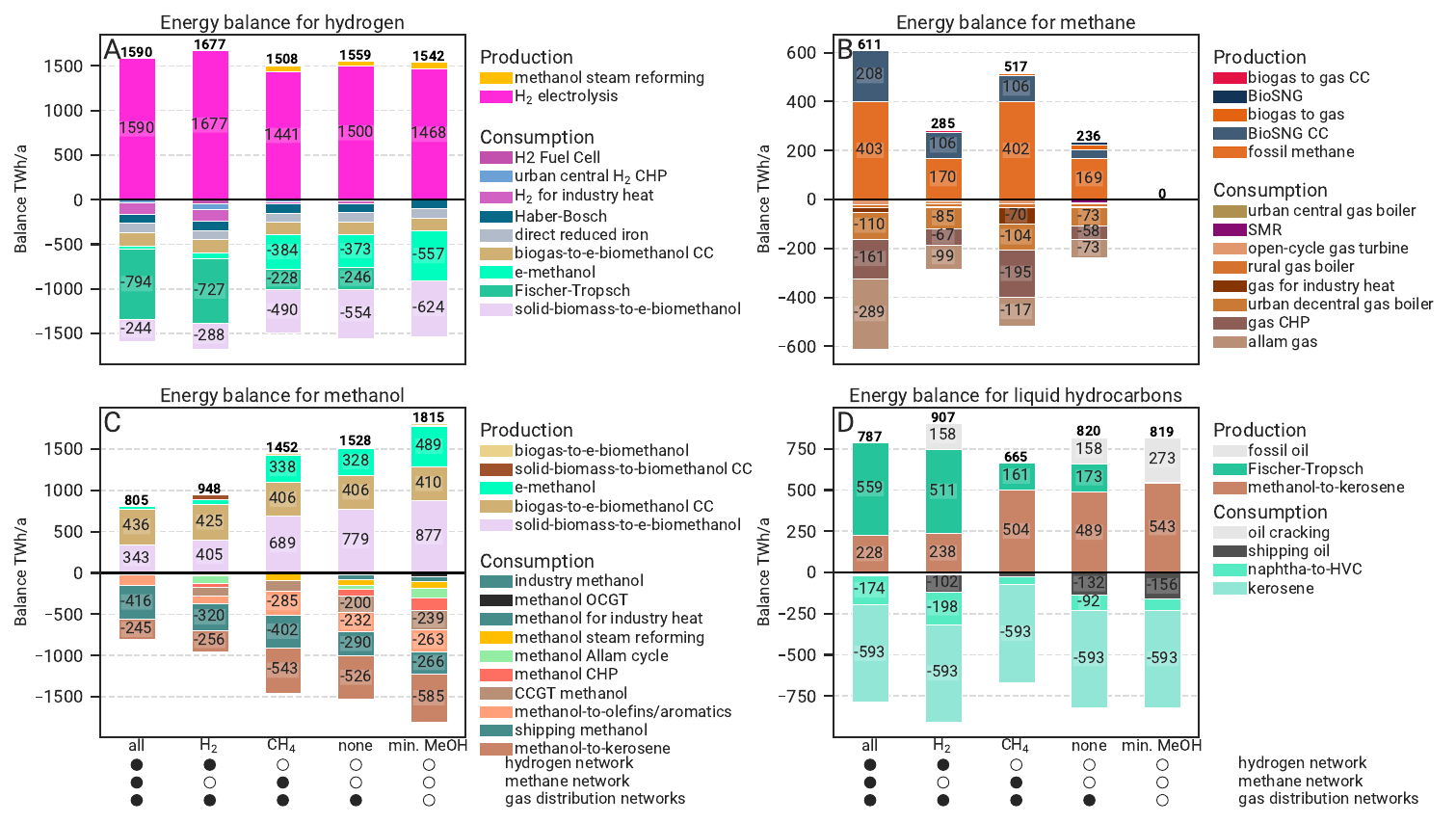}
    \caption{\textbf{Energy balances of the different scenarios for hydrogen, methane, methanol and oil in the setting without power transmission.} The positive values show supply and the negative values show consumption. The bold number above each bar shows the total supply or consumption in TWh/a.}
    \label{fig:no_grid_balances}
\end{figure}

\begin{figure}[!ht]
    \centering
    \includegraphics[width=.7\textwidth]{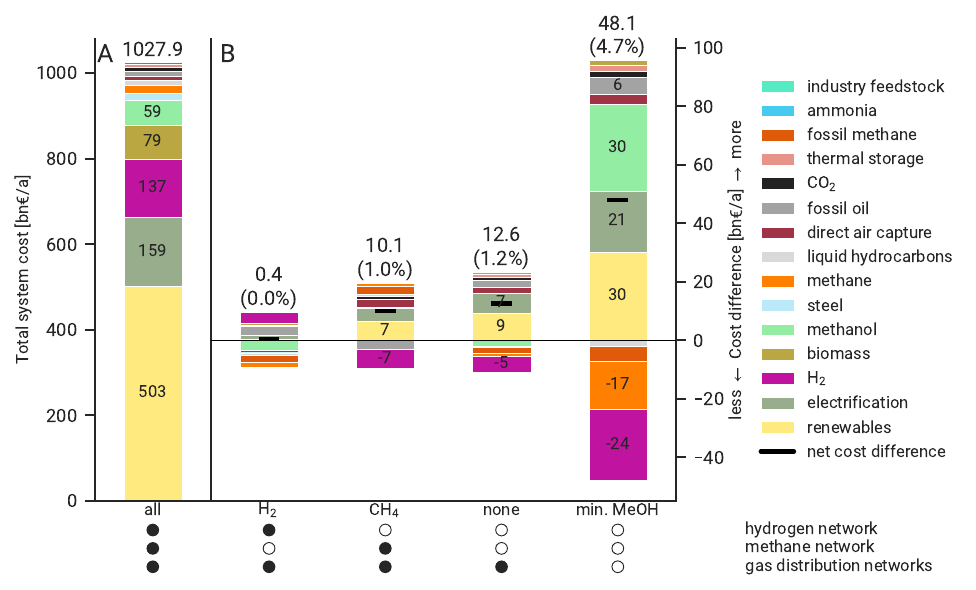}
    \caption{\textbf{Comparison of total system costs for the different scenarios in the setting with low electrification.} The panel A shows the absolute cost of the `all networks' scenario. The panel B shows the cost increases and decreases of the other scenarios by component relative to the `all networks' scenario. The net absolute and relative cost difference is shown at the top of each bar. A breakdown of the cost groups is given in~\cref{tab:cost_groups}.}
    \label{fig:low_electrification_total_cost}
\end{figure}
\begin{figure}[!ht]
    \centering
    \includegraphics[width=0.9\textwidth]{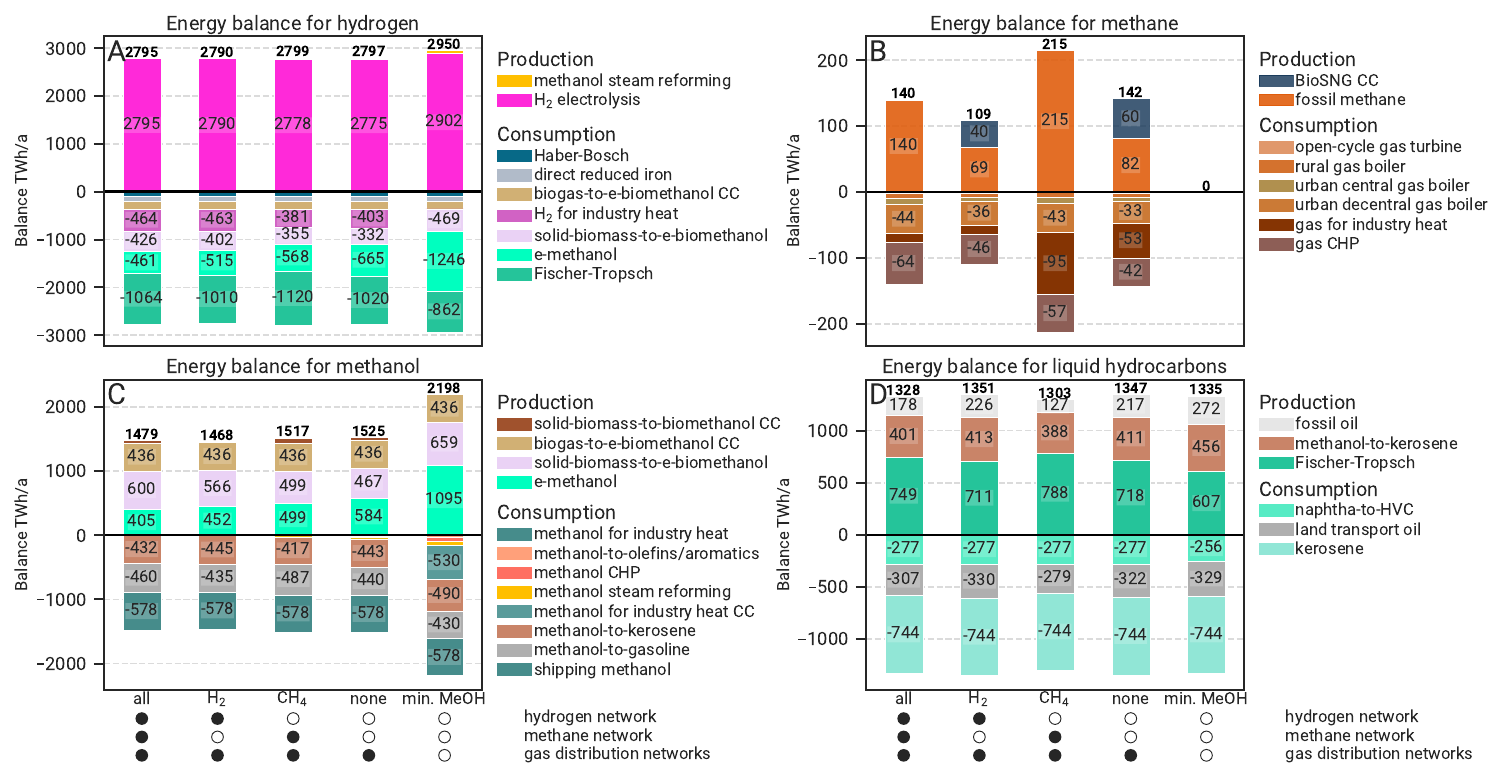}
    \caption{\textbf{Energy balances of the different scenarios for hydrogen, methane, methanol and oil in the setting with low electrification.} The positive values show supply and the negative values show consumption. The bold number above each bar shows the total supply or consumption in TWh/a.}
    \label{fig:low_electrification_balances}
\end{figure}

\begin{figure}[!ht]
    \centering
    \includegraphics[width=.7\textwidth]{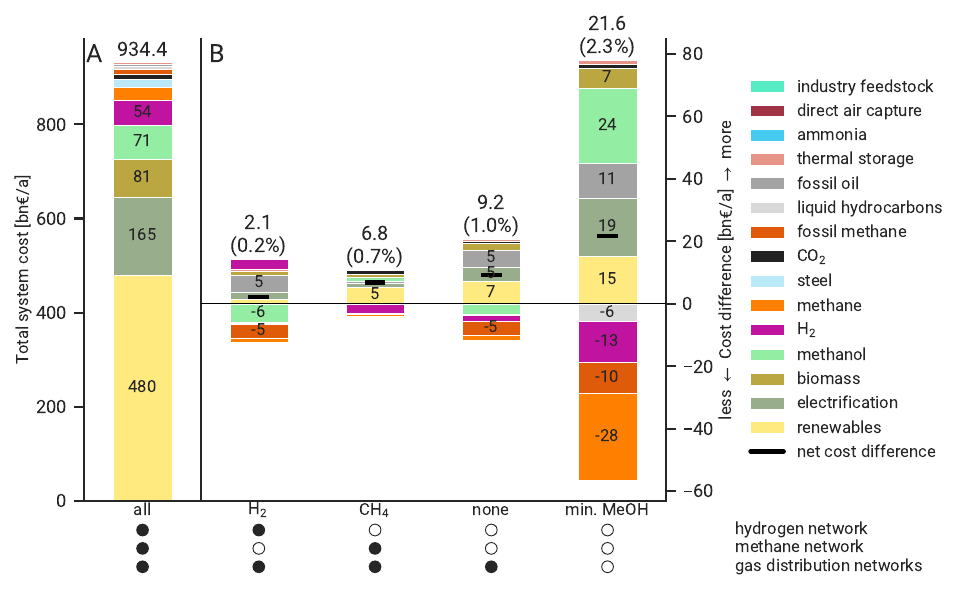}
    \caption{\textbf{Comparison of total system costs for the different scenarios in the setting with 30\% higher demand in road, aviation and shipping transport as well as for high value chemicals.} The panel A shows the absolute cost of the `all networks' scenario. The panel B shows the cost increases and decreases of the other scenarios by component relative to the `all networks' scenario. The net absolute and relative cost difference is shown at the top of each bar. A breakdown of the cost groups is given in~\cref{tab:cost_groups}.}
    \label{fig:high_demand_total_cost}
\end{figure}
\begin{figure}[!ht]
    \centering
    \includegraphics[width=0.9\textwidth]{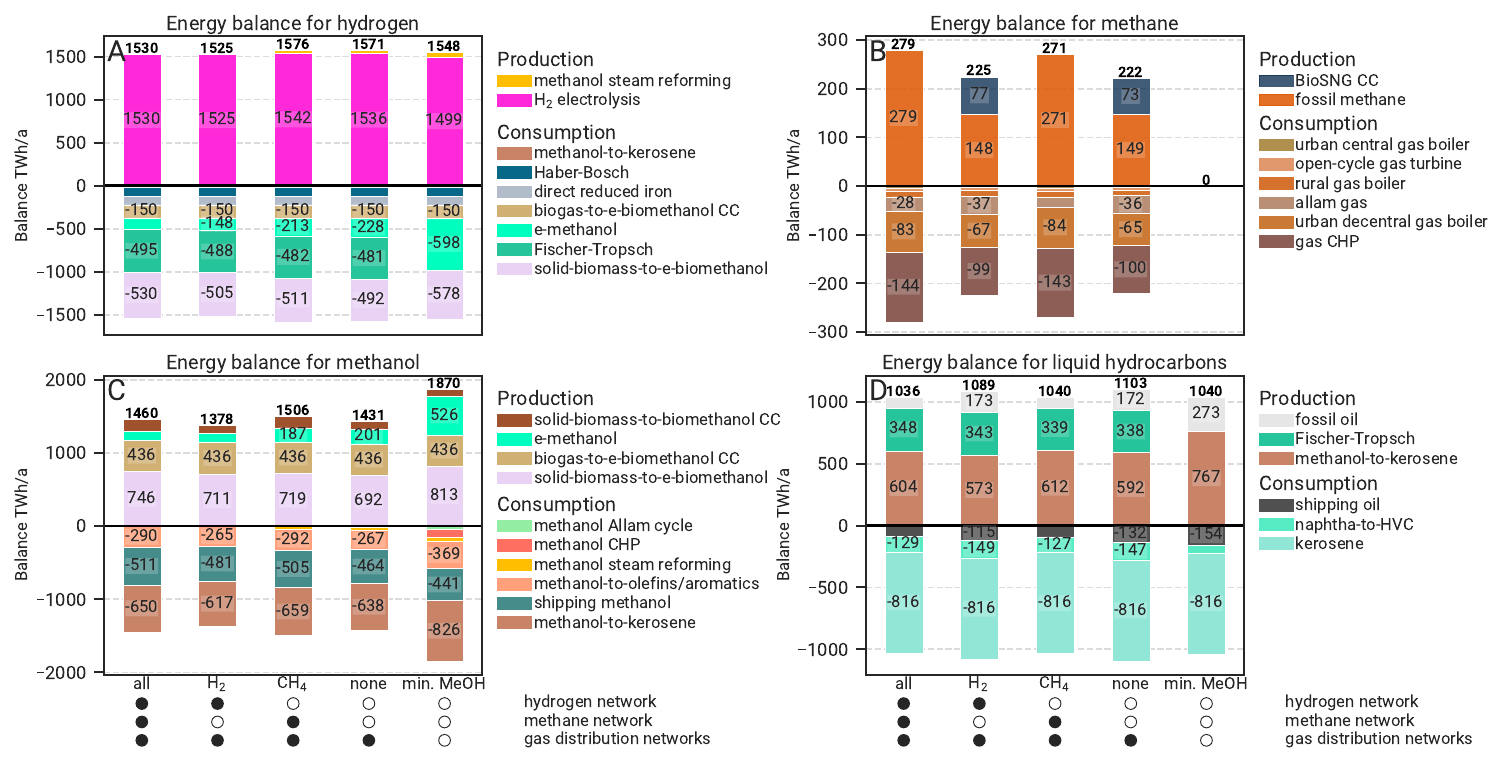}
    \caption{\textbf{Energy balances of the different scenarios for hydrogen, methane, methanol and oil in the setting with 30\% higher demand in road, aviation and shipping transport as well as for high value chemicals.} The positive values show supply and the negative values show consumption. The bold number above each bar shows the total supply or consumption in TWh/a.}
    \label{fig:high_demand_balances}
\end{figure}

\begin{figure}[!ht]
    \centering
    \includegraphics[width=.7\textwidth]{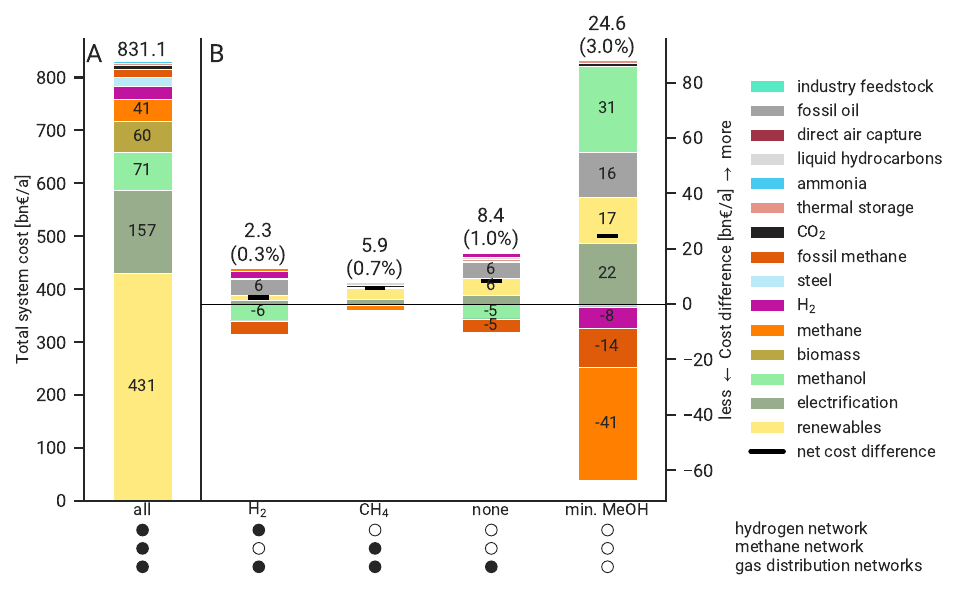}
    \caption{\textbf{Comparison of total system costs for the different scenarios in the setting without biomass in power and heat backup.} The panel A shows the absolute cost of the `all networks' scenario. The panel B shows the cost increases and decreases of the other scenarios by component relative to the `all networks' scenario. The net absolute and relative cost difference is shown at the top of each bar. A breakdown of the cost groups is given in~\cref{tab:cost_groups}.}
    \label{fig:no_bio_total_cost}
\end{figure}
\begin{figure}[!ht]
    \centering
    \includegraphics[width=0.9\textwidth]{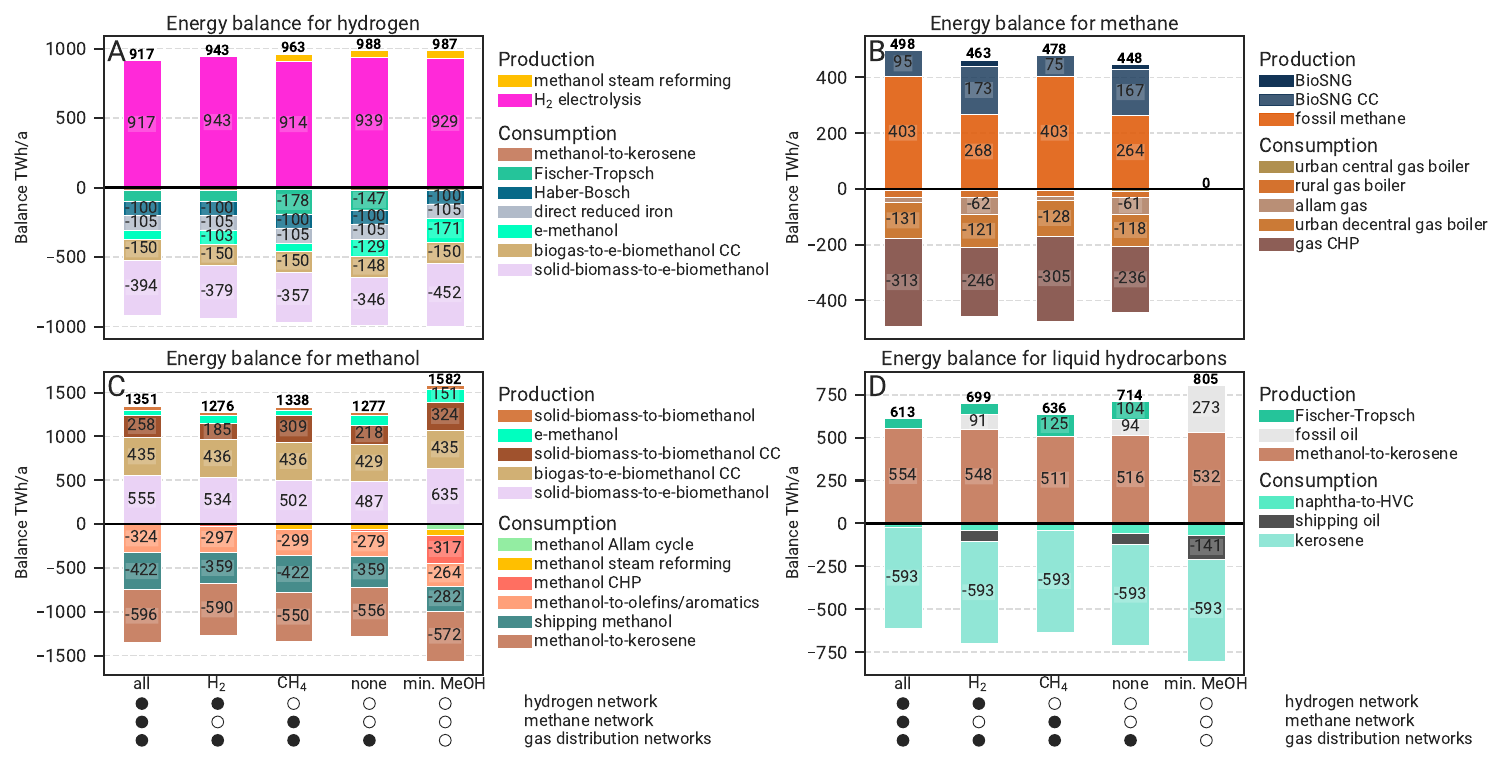}
    \caption{\textbf{Energy balances of the different scenarios for hydrogen, methane, methanol and oil in the setting without biomass in power and heat backup.} The positive values show supply and the negative values show consumption. The bold number above each bar shows the total supply or consumption in TWh/a.}
    \label{fig:no_bio_energy_balances}
\end{figure}

\begin{figure}[!ht]
    \centering
    \includegraphics[width=.7\textwidth]{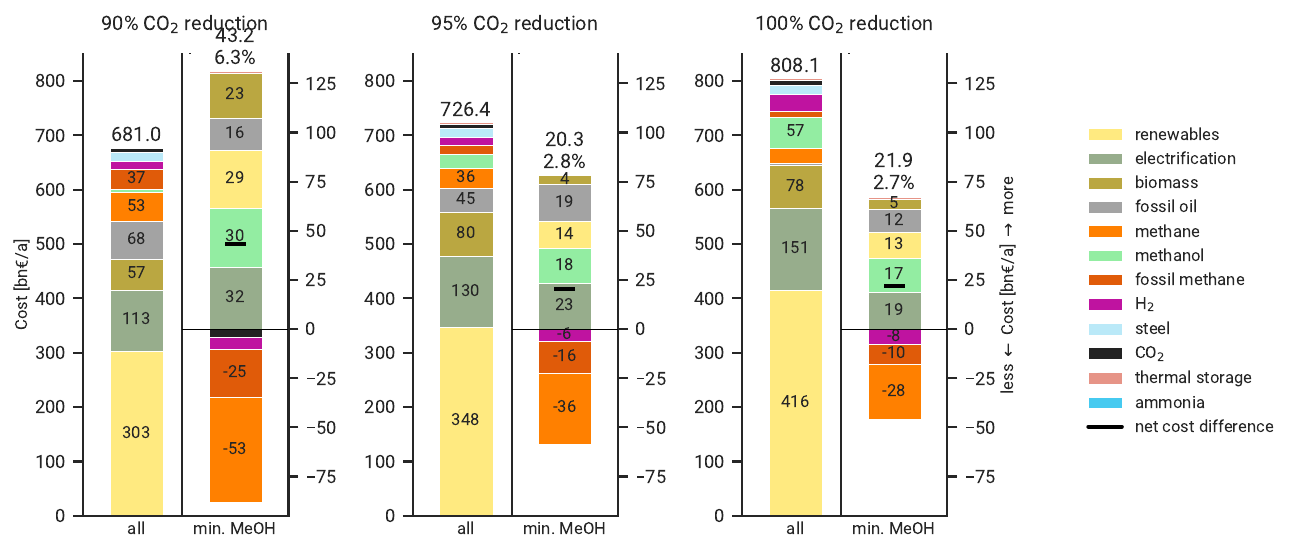}
    \caption{\textbf{Comparison of total system costs for the `all networks' and `minimal methanol backstop' scenarios in the setting with \co reduction targets of 90, 95 and 100\%~(default) compared to 1990 emissions.} The left panels show the absolute cost of the `all networks' scenario. The right panels show the cost increases and decreases of the `minimal methanol backstop' scenario by component relative to the `all networks' scenario. The net absolute and relative cost difference is shown at the top of each bar. A breakdown of the cost groups is given in~\cref{tab:cost_groups}.}
    \label{fig:reduced_aims_total_cost}
\end{figure}
\begin{figure}[!ht]
    \centering
    \includegraphics[width=0.9\textwidth]{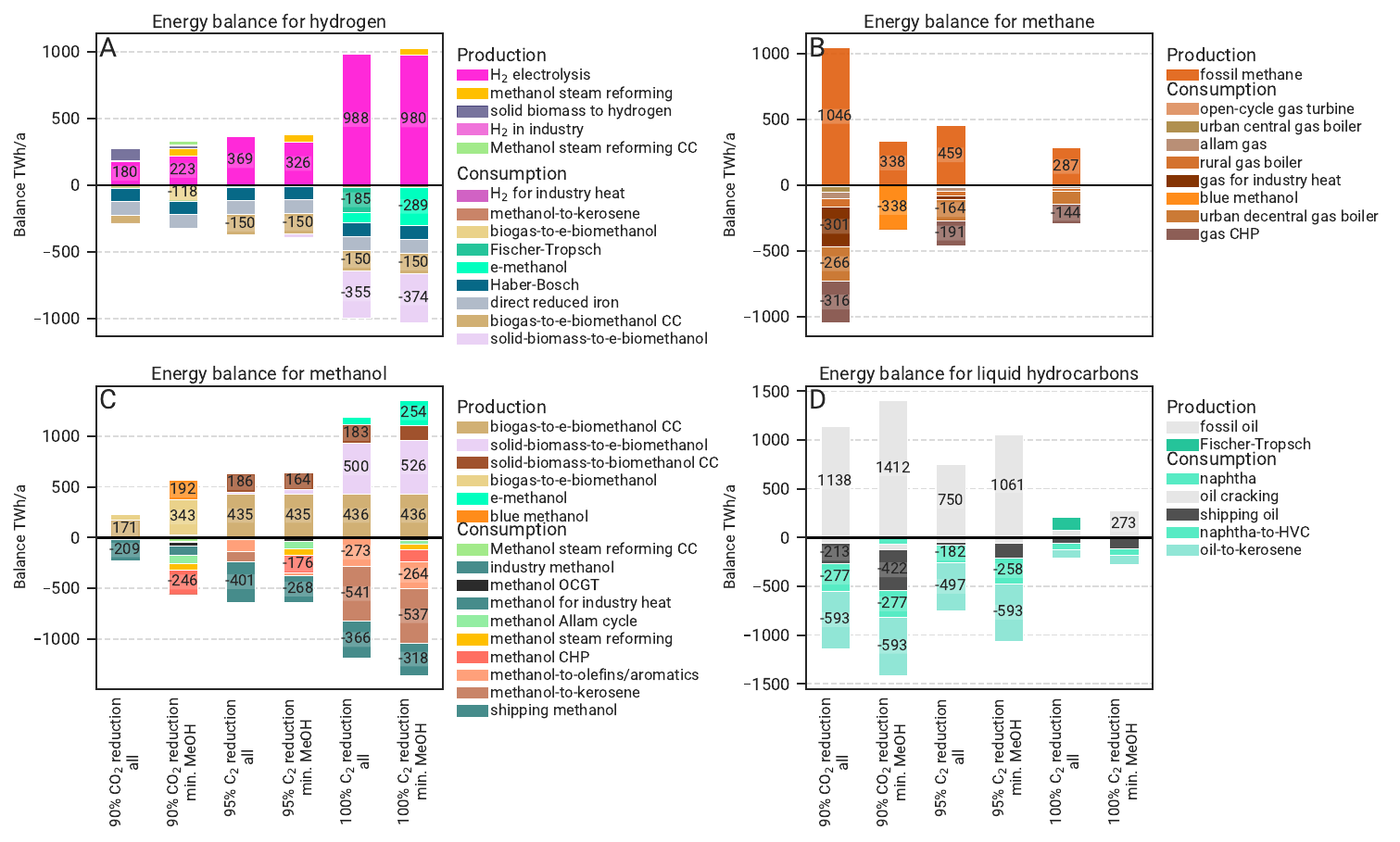}
    \caption{\textbf{Energy balances of the different scenarios for hydrogen, methane, methanol and oil in the setting with \co reduction targets of 90, 95 and 100\%~(default) compared to 1990 emissions.} The positive values show supply and the negative values show consumption. The bold number above each bar shows the total supply or consumption in TWh/a.}
    \label{fig:reduced_aims_energy_balances}
\end{figure}

\begin{figure}[!ht]
    \centering
    \includegraphics[width=.7\textwidth]{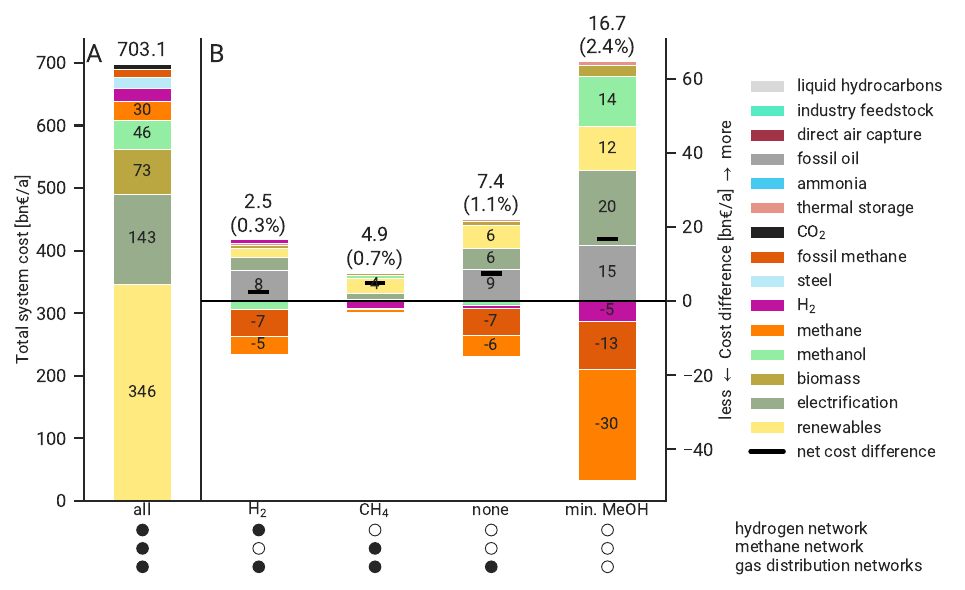}
    \caption{\textbf{Comparison of total system costs for the different scenarios in the setting using cost assumptions for 2050.} The panel A shows the absolute cost of the `all networks' scenario. The panel B shows the cost increases and decreases of the other scenarios by component relative to the `all networks' scenario. The net absolute and relative cost difference is shown at the top of each bar. A breakdown of the cost groups is given in~\cref{tab:cost_groups}.}
    \label{fig:2050_cost_total_cost}
\end{figure}
\begin{figure}[!ht]
    \centering
    \includegraphics[width=0.9\textwidth]{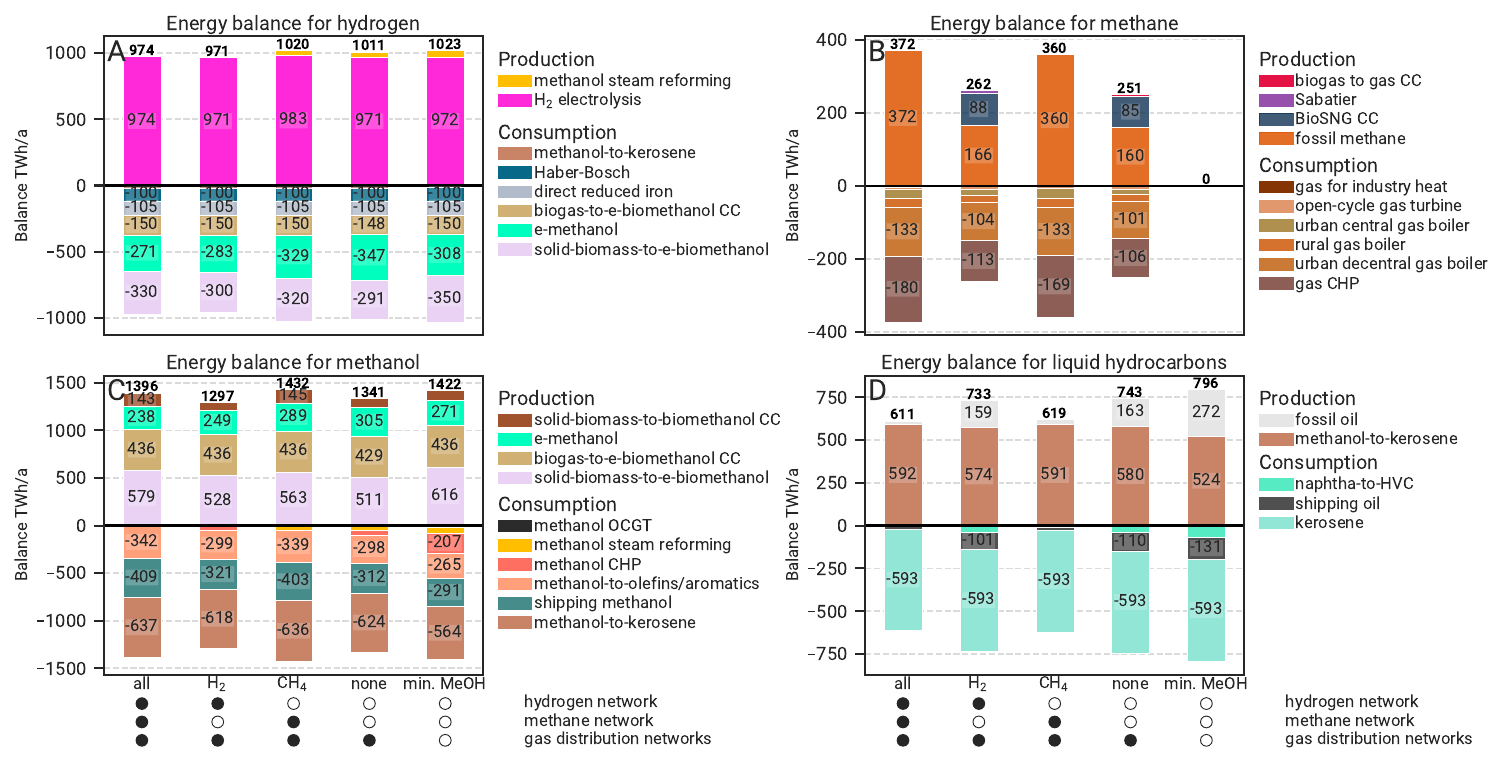}
    \caption{\textbf{Energy balances of the different scenarios for hydrogen, methane, methanol and oil in the setting using cost assumptions for 2050.} The positive values show supply and the negative values show consumption. The bold number above each bar shows the total supply or consumption in TWh/a.}
    \label{fig:2050_cost_energy_balances}
\end{figure}

\begin{figure}
    \centering
    \includegraphics[width=0.9\textwidth]{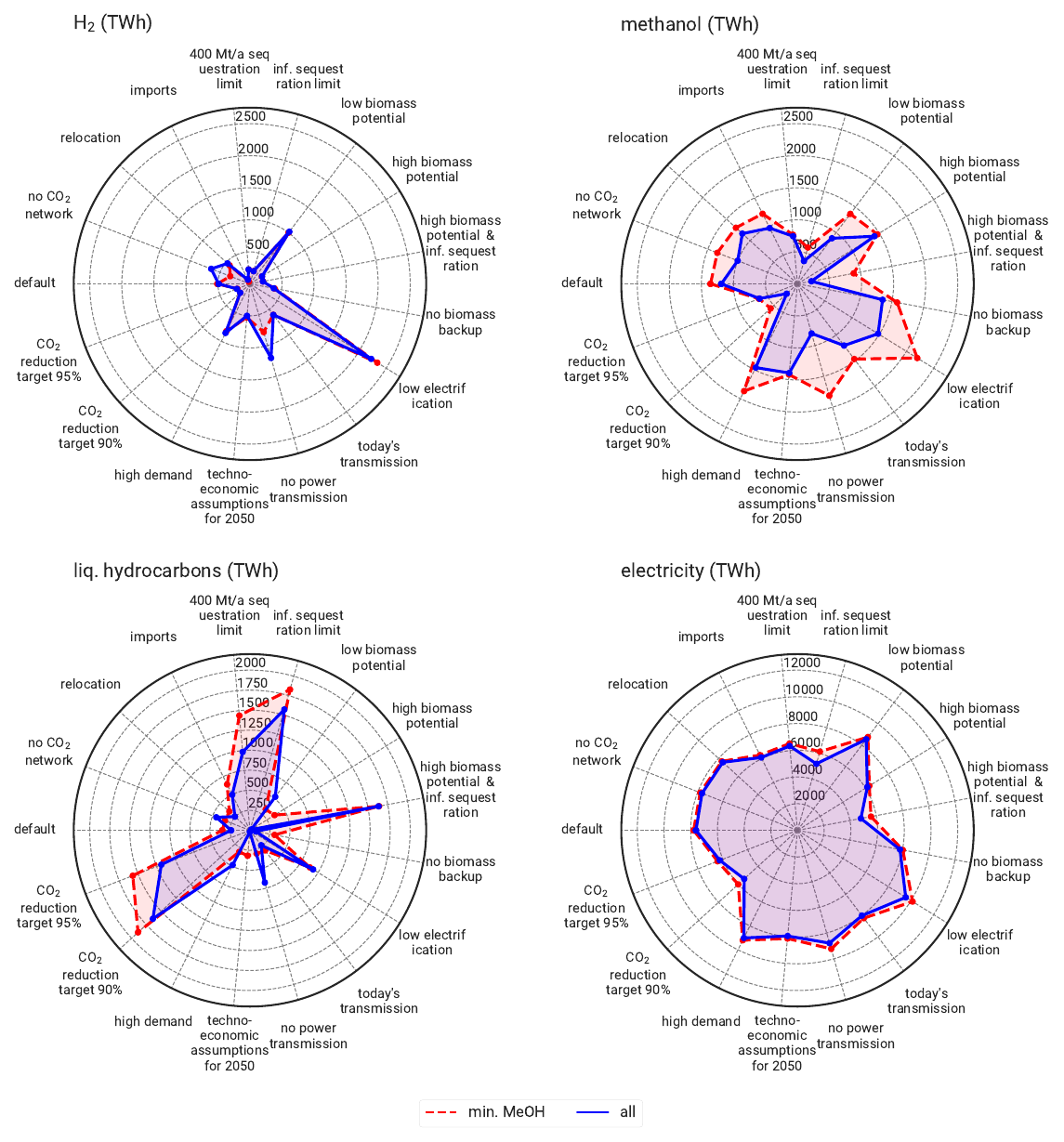}
    \caption{\textbf{Sensitivity of energy carrier consumption for different sensitivity settings.} The spider web charts show the variation in energy carrier consumption for the `all networks' and the `minimal methanol backstop' scenario in the different sensitivity settings.}
    \label{fig:sensitivity_energy_carrier_consumption}
\end{figure}

\begin{table}
\label{tab:trade_cmp}
\begin{threeparttable}
\caption{\textbf{Trade and consumption comparison in TWh/a for the setting using today's power transmission capacities and real-world data of 2024.} We depict the top 8 countries with the highest net trading volumes. The left side shows the results of the minimal methanol backstop and the right side shows the data from ENTSO-E for 2024\tnote{1}\tnote{2}.}
\begin{tabular}{lrrrr|rrrr}
\toprule
 & \multicolumn{4}{r}{min. MeOH} & \multicolumn{4}{r}{ENTSO-E 2024} \\
 & Import & Export & Net-Import & Consumption & Export & Import & Net-Import & Consumption \\
Country &  &  &  &  &  &  &  &  \\
\midrule
DE & 207 & 89 & 117 & 1200 & 40 & 65 & 25 & 465 \\
CH & 37 & 116 & -79 & 117 & 32 & 20 & -12 & 60 \\
AT & 31 & 68 & -37 & 191 & 19 & 14 & -5 & 59 \\
DK & 34 & 64 & -31 & 251 & 20 & 24 & 4 & 37 \\
ES & 20 & 49 & -29 & 705 & 19 & 12 & -8 & 232 \\
GR & 4 & 30 & -27 & 113 & 6 & 6 & 0 & 50 \\
RS & 42 & 18 & 24 & 41 & 6 & 7 & 1 & 34 \\
BE & 46 & 24 & 22 & 219 & 14 & 25 & 11 & 81 \\
\bottomrule
\end{tabular}
\begin{tablenotes}
\footnotesize
\item[1] \url{https://www.entsoe.eu/publications/data/power-stats/2024/physical_energy_power_flows_2024.csv}
\item[2] \url{https://www.entsoe.eu/publications/data/power-stats/2024/monthly_hourly_load_values_2024.csv}
\end{tablenotes}
\end{threeparttable}
\end{table}

\FloatBarrier

\subsection*{Methanol transport simulation}
In the transport simulation, we want to derive the cost for transport and distribution of methanol for the main analysis. There we do not model a dedicated transport infrastructure for methanol but assume uniform cost of 3€/MWh$_{MeOH}$.
This value is the consumption-weighted average cost of the transport simulation which corresponds to the `minimal methanol backstop' scenario with \co network. In the simulation, we consider transport of methanol by truck at 80€/t$_\text{MeOH}$/1000~km (14.6~€/MWh/1000~km)~\citeS{diasEnergyEconomicCosts2020S}, neglecting potentially less costly alternatives such as transport by pipeline, train, barge or ship.
\begin{figure}[!ht]
    \centering
    \includegraphics[width=.7\textwidth]{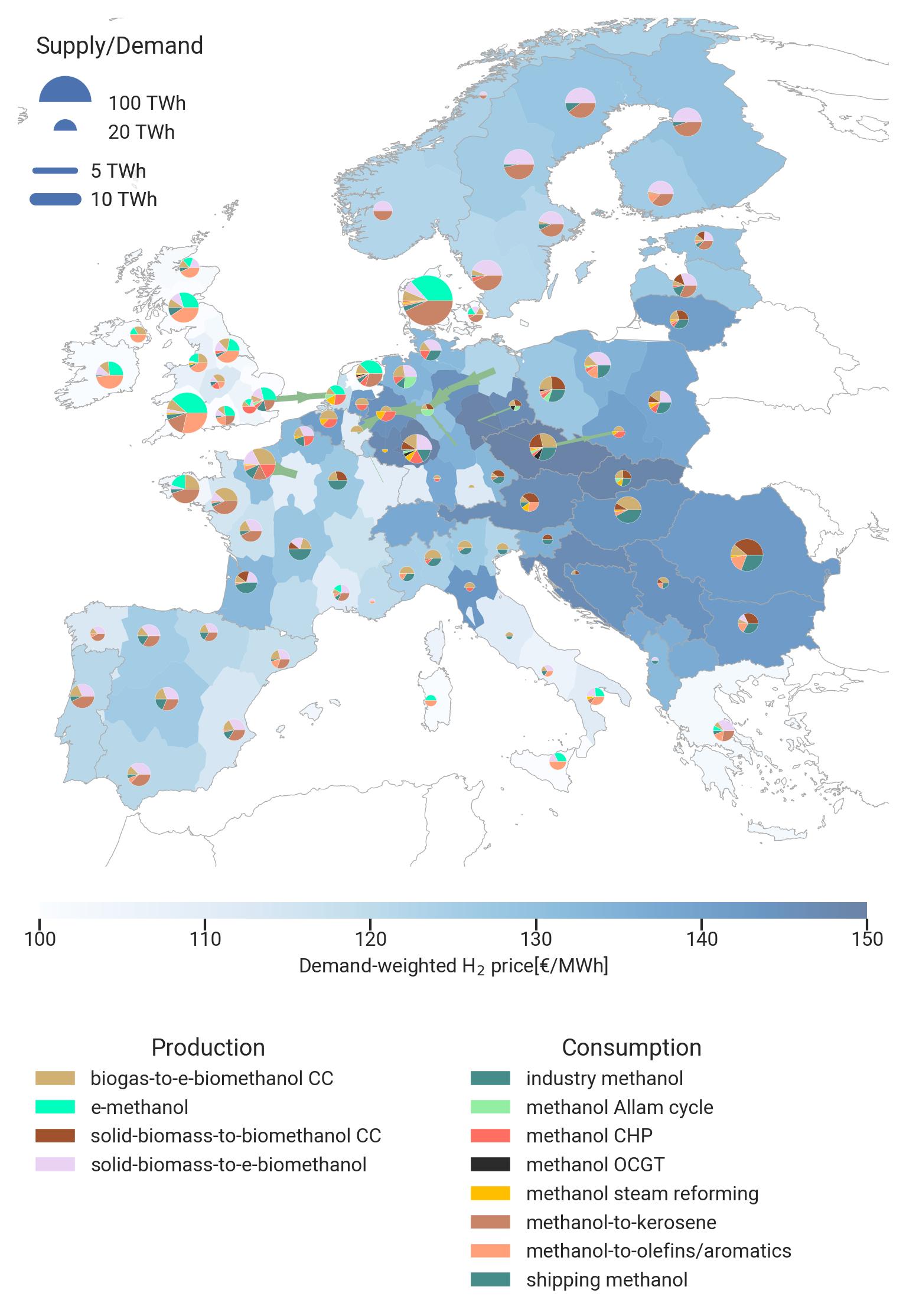}
    \caption{Methanol map of the min. MeOH economy scenario with methanol cost transport cost of 80€/t$_\text{MeOH}$/1000~km.}
    \label{fig:methanol_transport_map}
\end{figure}

\begin{figure}[!ht]
    \centering
    \includegraphics[width=.7\textwidth]{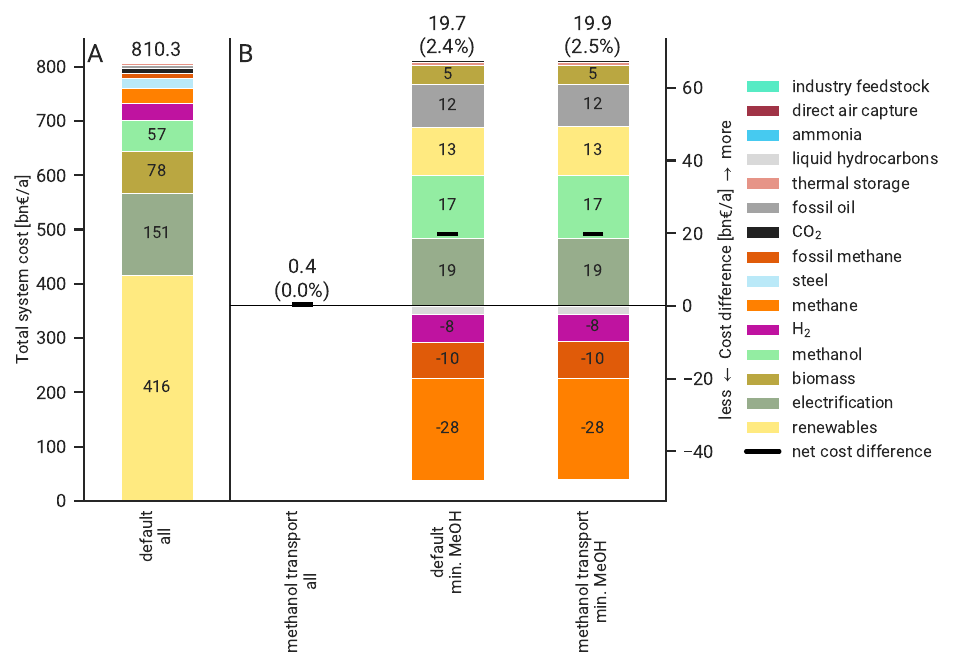}
    \caption{\textbf{Comparison of total system costs of the default and the transport setting.} The panel A shows the absolute cost of the `all networks' scenario. The panel B shows the cost increases and decreases of the other scenarios and settings by component relative to the `all networks' scenario. The net absolute and relative cost difference is shown at the top of each bar. A breakdown of the cost groups is given in~\cref{tab:cost_groups}.}
    \label{fig:methanol_transport_total_cost}
\end{figure}

\FloatBarrier

\bibliographyS{references}

\end{document}